\documentclass[prd,preprint,tightenlines,floatfix,
nofootinbib,eqsecnum,superscriptaddress]{revtex4-1}

\usepackage[T1]{fontenc}		

\usepackage{amsmath,amsfonts,amssymb,amstext,mathrsfs}
\usepackage{mathpazo}

\usepackage{mathpazo} 
\usepackage[scaled]{helvet} 
\usepackage{courier} 
\normalfont
\usepackage[T1]{fontenc}

\usepackage[dvips]{graphicx}
\usepackage{epsf,float}
\usepackage{revsymb}

\usepackage{dcolumn}
\usepackage{braket}
\usepackage{color,xcolor}
\usepackage{graphicx}
\usepackage{subfigure}
\usepackage{multirow}
\usepackage{tabularx}
\usepackage{pstricks}
\usepackage[section]{placeins}
\usepackage{booktabs}
\usepackage{array}

\usepackage{hyperref}

\newcommand{\Pom}{\mathbb{P}}

\newcommand{\Reg}{\mathbb{R}}

\newcommand{\bdPt}{\mbox{\boldmath $dP_{t}$}}

\newcommand{\bpta}{\mbox{\boldmath $p_{t,1}$}}
\newcommand{\bptb}{\mbox{\boldmath $p_{t,2}$}}

\newcommand{\bq}{\mbox{\boldmath $q$}}
\newcommand{\bqa}{\mbox{\boldmath $q_{1}$}}

\newcommand{\bp}{\mbox{\boldmath $p$}}

\newcommand{\bpa}{\mbox{\boldmath $p_{a}$}}
\newcommand{\bpb}{\mbox{\boldmath $p_{b}$}}
\newcommand{\bpaa}{\mbox{\boldmath $p_{1}$}}
\newcommand{\bpbb}{\mbox{\boldmath $p_{2}$}}
\newcommand{\bpip}{\mbox{\boldmath $p_{3}$}}

\newcommand{\bpipm}{\mbox{\boldmath $p_{34}$}}

\newcommand{\bhpa}{\mbox{\boldmath $\hat{p}_{a}$}}
\newcommand{\bhpb}{\mbox{\boldmath $\hat{p}_{b}$}}
\newcommand{\bhpc}{\mbox{\boldmath $\hat{p}_{3}$}}

\newcommand{\be}{\mbox{\boldmath $e$}}

\begin{document}

\title{\boldmath 
Extracting the pomeron-pomeron-$f_{2}(1270)$ coupling in the $p p \to p p \pi^{+} \pi^{-}$ reaction through angular distributions of the pions}

\vspace{0.6cm}

\author{Piotr Lebiedowicz}
 \email{Piotr.Lebiedowicz@ifj.edu.pl}
\affiliation{Institute of Nuclear Physics Polish Academy of Sciences, Radzikowskiego 152, PL-31342 Krak\'ow, Poland}

\author{Otto Nachtmann}
 \email{O.Nachtmann@thphys.uni-heidelberg.de}
\affiliation{Institut f\"ur Theoretische Physik, Universit\"at Heidelberg,
Philosophenweg 16, D-69120 Heidelberg, Germany}

\author{Antoni Szczurek
\footnote{Also at \textit{College of Natural Sciences, 
Institute of Physics, University of Rzesz\'ow, 
Pigonia 1, PL-35310 Rzesz\'ow, Poland}.}}
\email{Antoni.Szczurek@ifj.edu.pl}
\affiliation{Institute of Nuclear Physics Polish Academy of Sciences, Radzikowskiego 152, PL-31342 Krak\'ow, Poland}

\begin{abstract}
We discuss how to extract the pomeron-pomeron-$f_2(1270)$ ($\Pom \Pom f_2(1270)$) coupling within the tensor-pomeron model. The general $\Pom \Pom f_2(1270)$ coupling is a combination of seven basic couplings (tensorial structures). To study these tensorial structures we propose to measure the central-exclusive production of a $\pi^+ \pi^-$ pair in the invariant mass region of the $ f_2(1270)$ meson. An analysis of angular distributions in the $\pi^+ \pi^-$ rest system, using the Collins-Soper (CS) and the Gottfried-Jackson (GJ) frames, turns out to be particularly relevant for our purpose. For both frames the $\cos\theta_{\pi^{+}}$ and $\phi_{\pi^{+}}$ distributions are discussed. We find that the azimuthal angle distributions in these frames are fairly sensitive to the choice of the $\Pom \Pom f_2$ coupling. We show results for the resonance case alone as well as when the dipion continuum is included. We show the influence of the experimental cuts on the angular distributions in the context of dedicated experimental studies at RHIC and LHC energies. Absorption corrections are included for our final distributions.
\end{abstract}


\maketitle

\section{Introduction}

The pomeron ($\Pom$) is an essential object for understanding diffractive phenomena in high-energy physics. 
Within QCD the pomeron is a color singlet, predominantly gluonic, object.
The spin structure of the pomeron, in particular its coupling to hadrons,
is, however, not yet a matter of consensus.
In the tensor-pomeron model for soft high-energy scattering
formulated in \cite{Ewerz:2013kda} 
the pomeron exchange is effectively treated as the exchange of a rank-2 symmetric tensor.
The diffractive amplitude for a given process 
with soft pomeron exchange can then be formulated 
in terms of effective propagators and vertices
respecting the rules of quantum field theory.

It is rather difficult to obtain definitive statements on
the spin structure of the pomeron from unpolarised 
elastic proton-proton scattering.
On the other hand, the results
from polarised proton-proton scattering by the STAR Collaboration \cite{Adamczyk:2012kn} 
provide valuable information on this question.
Three hypotheses for the spin structure of the pomeron, tensor,
vector, and scalar, were discussed in \cite{Ewerz:2016onn}
in view of the experimental results from \cite{Adamczyk:2012kn}.
Only the tensor-ansatz for the pomeron was found to be compatible with the experiment.
Also some historical remarks on different views of the pomeron were made in~\cite{Ewerz:2016onn}.

In \cite{Britzger:2019lvc} further strong evidence against the hypothesis of a vector
character of the pomeron was given. It was shown there that a vector pomeron
necessarily decouples in elastic photon-proton scattering and in
the absorption cross sections of virtual photons on the proton, that is,
in the structure functions of deep inelastic lepton-nucleon scattering.
A tensor pomeron, on the other hand, has no such problems and
tensor-pomeron exchanges, soft and hard, give an excellent description
of the absorption cross sections for real and virtual photons
on the proton at high energies.

In the last few years we have undertaken a scientific program to analyse 
the production of light mesons in the tensor-pomeron and vector-odderon model
in several exclusive reactions: $p p \to p p M$~\cite{Lebiedowicz:2013ika}, 
$p p \to p p \pi^{+}\pi^{-}$~\cite{Lebiedowicz:2014bea,Lebiedowicz:2016ioh}, 
$p p \to p n \rho^{0} \pi^{+}$ ($p p \rho^{0} \pi^{0}$)~\cite{Lebiedowicz:2016ryp},
$p p \to p p K^{+}K^{-}$ \cite{Lebiedowicz:2018eui},
$p p \to p p (\sigma \sigma, \rho^{0} \rho^{0} \to \pi^{+}\pi^{-}\pi^{+}\pi^{-})$~\cite{Lebiedowicz:2016zka},
$p p \to p p p \bar{p}$~\cite{Lebiedowicz:2018sdt},
$p p \to p p (\phi \phi \to K^{+}K^{-}K^{+}K^{-})$~\cite{Lebiedowicz:2019jru},
and $p p \to p p (\phi \to K^{+}K^{-}, \mu^{+}\mu^{-})$~\cite{Lebiedowicz:2019boz}. 
Some azimuthal angle correlations between the outgoing protons 
can verify the $\Pom \Pom M$ couplings for scalar 
$f_{0}(980)$, $f_{0}(1370)$, $f_{0}(1500)$, $f_{0}(1710)$ 
and pseudoscalar $\eta$, $\eta'(958)$ mesons~\cite{Lebiedowicz:2013ika,Lebiedowicz:2018eui}. 
The couplings, being of nonperturbative nature, are difficult 
to obtain from first principles of QCD. 
The corresponding coupling constants were fitted 
to differential distributions of the WA102 Collaboration
\cite{Barberis:1998ax,Barberis:1999cq,Barberis:1999zh} and to the results of \cite{Kirk:2000ws}.
As was shown in \cite{Lebiedowicz:2013ika,Lebiedowicz:2018eui},
the tensorial $\Pom \Pom f_{0}$, $\Pom \Pom \eta$, and $\Pom \Pom \eta'$ vertices
correspond to the sum of two lowest orbital angular momentum - spin couplings, 
except for the $f_{0}(1370)$ meson.
The tensor meson case is a bit complicated as there are,
in our approach, seven (!) 
possible pomeron-pomeron-$f_{2}(1270)$ couplings in principle;
see the list of possible $\Pom \Pom f_{2}$ couplings in Appendix~A of \cite{Lebiedowicz:2016ioh} and in Sec.~\ref{sec:formalism} below.

It was shown in \cite{Barberis:1996iq,Barberis:1999cq} 
that the cross section for the undisputed $q\bar{q}$ tensor mesons,
$f_{2}(1270)$, $f'_{2}(1525)$, peaks at $\phi_{pp} = \pi$
and is suppressed at small $\rm{dP_{t}}$
in contrast to the tensor glueball candidate $f_{2}(1950)$; see e.g. \cite{Barberis:2000em}.
Here, $\phi_{pp}$ is the azimuthal angle
between the transverse momentum vectors $\bpta$, $\bptb$ of the outgoing protons and
$\rm{dP_{t}}$ (the so-called 'glueball-filter variable' \cite{Close:1997pj}) 
is defined by their difference $\bdPt = \bptb - \bpta$, $\rm{dP_{t}} = |\bdPt|$.
In \cite{Lebiedowicz:2016ioh} we gave some arguments from studying 
the $\phi_{pp}$ and $\rm{dP_{t}}$ distributions
that one particular coupling $\Pom \Pom f_{2}$ (denoted by $j=2$)
may be preferred.
We roughly reproduced the experimental data
obtained by the WA102 Collaboration \cite{Barberis:1999cq} and  
by the ABCDHW Collaboration \cite{Breakstone:1990at} with this coupling.
It was demonstrated in \cite{Lebiedowicz:2016ioh}
that the relative contribution of resonant $f_{2}(1270)$ 
and dipion continuum strongly depends on 
the cut on four-momentum transfer squared $t_{1,2}$ in a given experiment.
However, we must remember that at low energies
also the secondary (especially $f_{2 \Reg}$) exchanges may play an important role.

Now, we ask the question whether and how the $\Pom \Pom f_{2}$ couplings
can be studied in central-exclusive processes.
In the present work we discuss such a possibility: analysis of 
angular distributions of pions from the decay of $f_{2}$,
in two systems of reference,
the Collins-Soper~(CS) and the Gottfried-Jackson (GJ) systems. 
We will consider diffractive production 
of the $f_{2}(1270)$ resonance which is expected
to be abundantly produced in the $p p \to p p \pi^+ \pi^-$ reaction;
see e.g. \cite{Lebiedowicz:2016ioh}.
We will try to analyse whether such a study could shed light 
on the $\Pom \Pom f_2(1270)$ couplings.
In \cite{Austregesilo:2013yxa,Austregesilo:2014oxa,Austregesilo:2016sss} 
the central exclusive production of two-pseudoscalar mesons in $pp$ collisions
at the COMPASS experiment at CERN SPS was reported.
There, preliminary data of pion angular distributions 
in the $\pi^+ \pi^-$ rest system using the GJ frame was shown.
We refer the reader to 
\cite{Adamczyk:2014ofa,Aaltonen:2015uva,Khachatryan:2017xsi,Sikora:2018cyk,CMS:2019qjb,Schicker:2019qcn}
for the latest measurements of central $\pi^+ \pi^-$ production 
in high-energy proton-(anti)proton collisions.
In the future the corresponding $\Pom \Pom f_{2}(1270)$ couplings
could be adjusted by comparison to precise experimental data 
from both RHIC and the LHC.

\section{Formalism}
\label{sec:formalism}

We study central exclusive production of $\pi^+ \pi^-$ 
in proton-proton collisions
\begin{eqnarray}
p(p_{a},\lambda_{a}) + p(p_{b},\lambda_{b}) \to
p(p_{1},\lambda_{1}) + \pi^{+}(p_{3}) + \pi^{-}(p_{4}) + p(p_{2},\lambda_{2}) \,,
\label{2to4_reaction}
\end{eqnarray}
where $p_{a,b}$, $p_{1,2}$ and $\lambda_{a,b}$, 
$\lambda_{1,2} \in \lbrace +1/2, -1/2 \rbrace$ 
denote the four-momenta and helicities of the protons, 
and $p_{3,4}$ denote the four-momenta of the charged pions, respectively.

\begin{figure}[!ht]
(a)\includegraphics[width=6.cm]{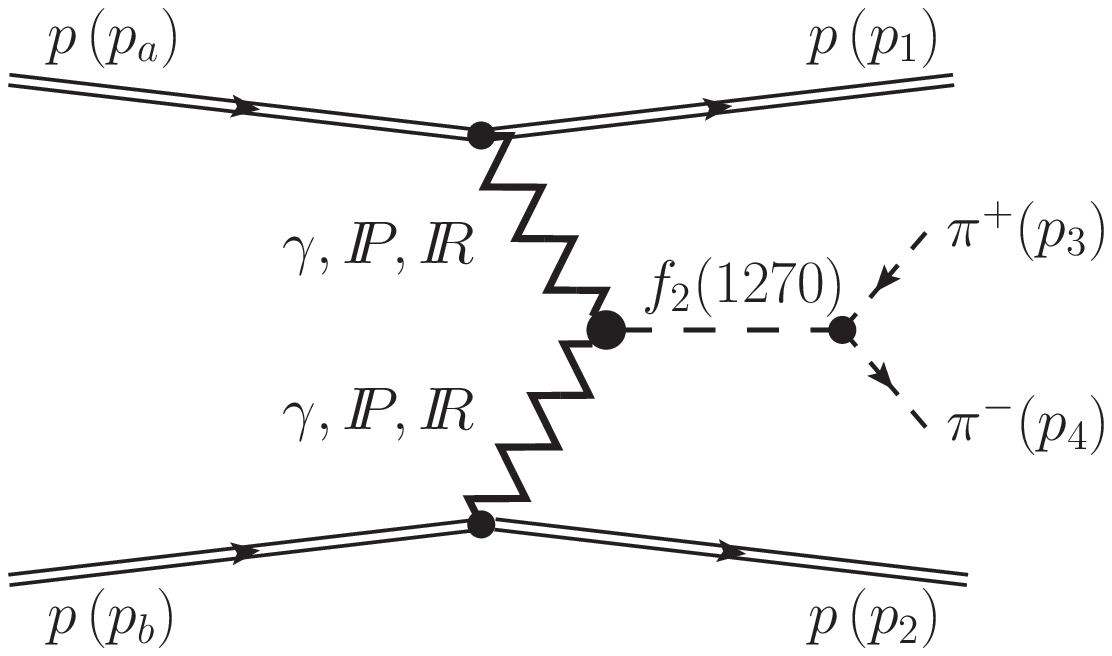}
(b)\includegraphics[width=4.6cm]{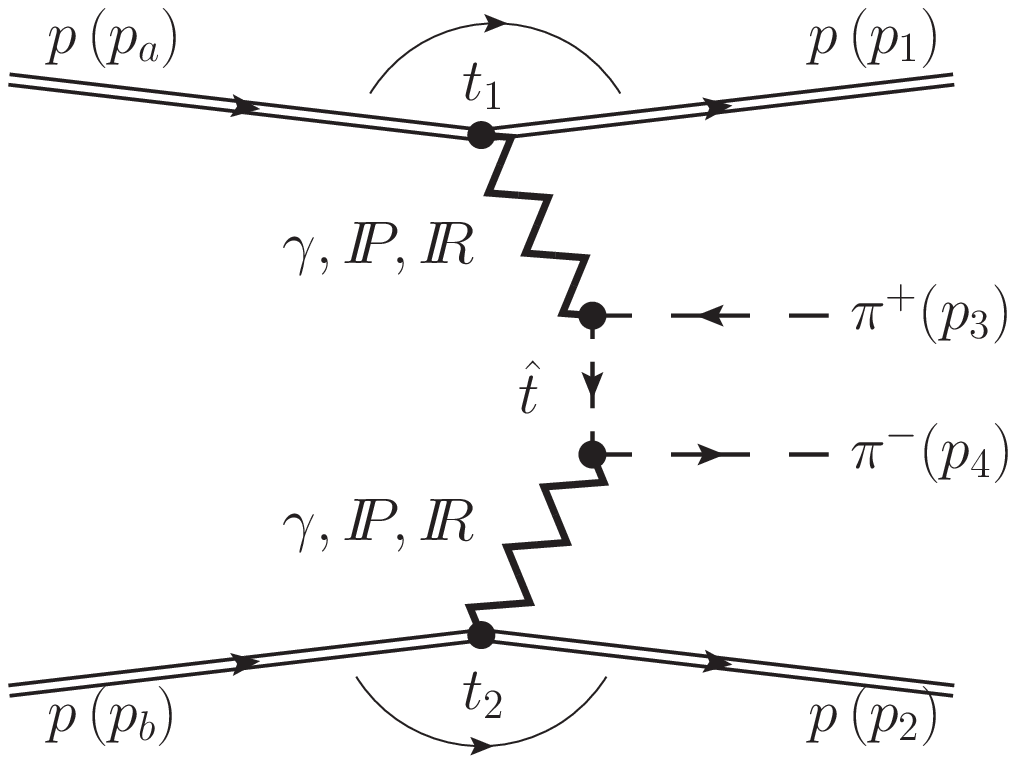}
   \includegraphics[width=4.6cm]{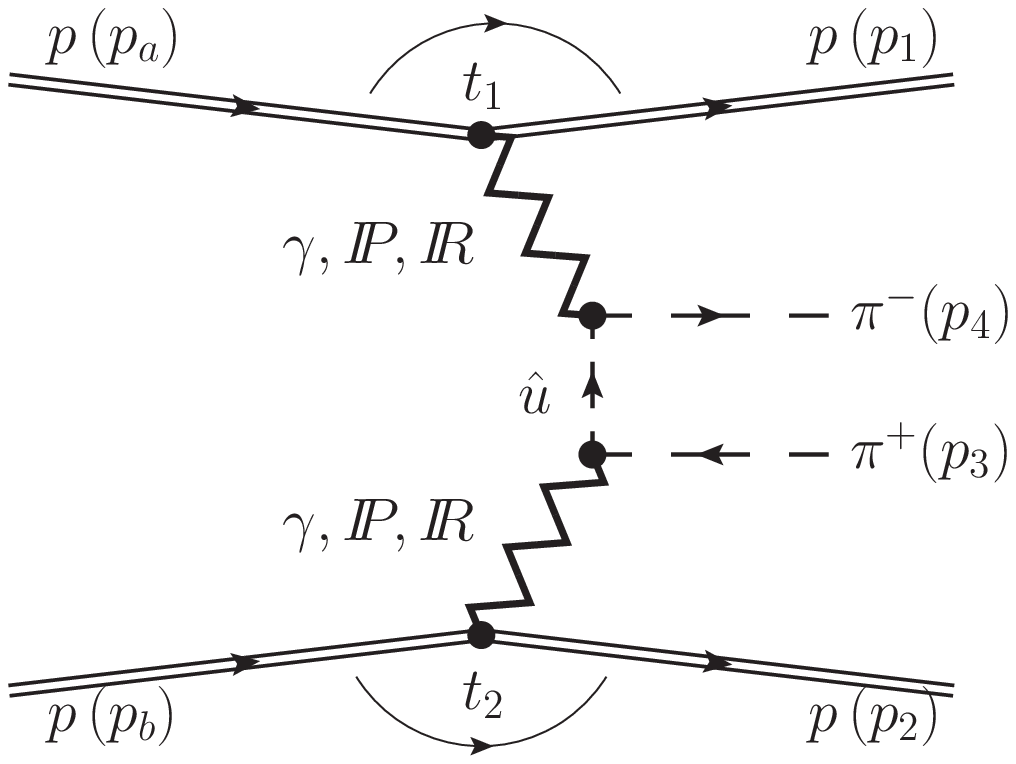}
\caption{\label{fig:diagrams}
\small
The Born diagrams for the $p p \to p p \pi^+ \pi^-$ reaction.
In (a) we have the $\pi^{+}\pi^{-}$ production via the $f_{2}(1270)$ resonance,
in (b) the continuum $\pi^{+}\pi^{-}$ production.
The exchange objects are the photon ($\gamma$), the pomeron ($\Pom$)
and the reggeons ($\Reg$).}
\end{figure}

We are, in the present article, mainly interested in the region
of the $\pi^+ \pi^-$ invariant mass in the $f_{2}(1270)$ region.
There we should take into account two main processes shown 
by the diagrams in Fig.~\ref{fig:diagrams}.
For the $f_{2}(1270)$ resonance (the diagram (a)) we consider only the $\Pom \Pom$ fusion.
The secondary reggeons $f_{2 \Reg}$, $a_{2 \Reg}$, $\omega_{\Reg}$, $\rho_{\Reg}$
should give small contributions at high energies.
We also neglect contributions involving the photon.
In the case of the non-resonant continuum (the diagrams (b)) 
we include in the calculations both $\Pom$ and $f_{2 \Reg}$-reggeon exchanges.
For an extensive discussion we refer to \cite{Lebiedowicz:2014bea,Lebiedowicz:2016ioh}.

The kinematic variables for the reaction (\ref{2to4_reaction}) are
\begin{eqnarray}
&&s = (p_{a} + p_{b})^{2}, 
\quad s_{34} = M_{\pi \pi}^{2} = (p_{3} + p_{4})^{2},\nonumber \\
&&q_1 = p_{a} - p_{1}, \quad q_2 = p_{b} - p_{2}, 
\quad  t_1 = q_{1}^{2}, \quad t_2 = q_{2}^{2}, \nonumber \\
&& p_{34} = q_{1} + q_{2} = p_{3} + p_{4},\nonumber \\
&&    s_{1} = (p_{a} + q_{2})^{2} = (p_{1} + p_{34})^{2},
\quad s_{2} = (p_{b} + q_{1})^{2} = (p_{2} + p_{34})^{2}\,.
\label{2to4_kinematic}
\end{eqnarray}

The $\Pom\Pom$-exchange (Born-level) amplitude for $\pi^{+} \pi^{-}$ production via
the tensor $f_{2}$-meson ($f_{2} \equiv f_{2}(1270)$) exchange
can be written as
\begin{equation}
\begin{split}
{\cal M}^{(\Pom \Pom \to f_{2}\to \pi^{+}\pi^{-})}_{\lambda_{a} \lambda_{b} \to \lambda_{1} \lambda_{2} \pi^{+}\pi^{-}} 
= & (-i)\,
\bar{u}(p_{1}, \lambda_{1}) 
i\Gamma^{(\Pom pp)}_{\mu_{1} \nu_{1}}(p_{1},p_{a}) 
u(p_{a}, \lambda_{a})\;
i\Delta^{(\Pom)\, \mu_{1} \nu_{1}, \alpha_{1} \beta_{1}}(s_{1},t_{1}) \\
& \times 
i\Gamma^{(\Pom \Pom f_{2})}_{\alpha_{1} \beta_{1},\alpha_{2} \beta_{2}, \rho \sigma}(q_{1},q_{2}) \;
i\Delta^{(f_{2})\,\rho \sigma, \alpha \beta}(p_{34})\;
i\Gamma^{(f_{2} \pi \pi)}_{\alpha \beta}(p_{3},p_{4})\\
& \times 
i\Delta^{(\Pom)\, \alpha_{2} \beta_{2}, \mu_{2} \nu_{2}}(s_{2},t_{2}) \;
\bar{u}(p_{2}, \lambda_{2}) 
i\Gamma^{(\Pom pp)}_{\mu_{2} \nu_{2}}(p_{2},p_{b}) 
u(p_{b}, \lambda_{b}) \,.
\end{split}
\label{amplitude_f2_pomTpomT}
\end{equation}
Here $\Delta^{(\Pom)}$ and $\Gamma^{(\Pom pp)}$ 
denote the effective propagator and proton vertex function, respectively, 
for the tensor-pomeron exchange.
For the explicit expressions, see Sect.~3 of \cite{Ewerz:2013kda}.
More details related to the amplitude (\ref{amplitude_f2_pomTpomT}) 
are given in \cite{Lebiedowicz:2016ioh}.
$\Delta^{(f_{2})}$ and $\Gamma^{(f_{2} \pi \pi)}$ denote the tensor-meson propagator
and the $f_{2} \pi \pi$ vertex, respectively.
As was mentioned in \cite{Ewerz:2013kda}
we cannot use a simple Breit-Wigner ansatz for the $f_{2}$ propagator
in conjunction with the $f_{2} \pi \pi$ vertex from (3.37), (3.38) of \cite{Ewerz:2013kda}
because the partial-wave unitarity relation is not satisfied.
We should use, therefore, a model for the $f_{2}$ propagator considered 
in Eqs. (3.6)--(3.8) and (5.19)--(5.22) of \cite{Ewerz:2013kda}.
The form factor $F^{(f_{2} \pi \pi)}(p_{34}^{2})$ for the $f_{2} \pi \pi$ vertex
and for the $f_{2}$ propagator
is taken to be the same as (\ref{Fpompommeson_ff_tensor}) below,
but with $\Lambda_{f_{2} \pi \pi}$ instead of $\Lambda_{\Pom \Pom f_{2}}$.

The main ingredient of the amplitude (\ref{amplitude_f2_pomTpomT}) 
is the pomeron-pomeron-$f_{2}$ vertex~\footnote{Here the label ``bare'' is used
for a vertex, as derived from a corresponding coupling Lagrangian
in Appendix~A of \cite{Lebiedowicz:2016ioh} without a form-factor function;
see (\ref{A11})--(\ref{A17}) below.}
\begin{eqnarray}
i\Gamma_{\mu \nu,\kappa \lambda,\rho \sigma}^{(\Pom \Pom f_{2})} (q_{1},q_{2}) =
\left( i\Gamma_{\mu \nu,\kappa \lambda,\rho \sigma}^{(\Pom \Pom f_{2})(1)} \mid_{\rm{bare}}
+ \sum_{j=2}^{7}i\Gamma_{\mu \nu,\kappa \lambda,\rho \sigma}^{(\Pom \Pom f_{2})(j)}(q_{1},q_{2}) \mid_{\rm{bare}} 
\right)
\tilde{F}^{(\Pom \Pom f_{2})}(q_{1}^{2},q_{2}^{2},p_{34}^{2}).\qquad
\label{vertex_pompomT}
\end{eqnarray}
Here $\tilde{F}^{(\Pom \Pom f_{2})}$
is a form factor for which we make a factorised ansatz (see (4.17) of \cite{Lebiedowicz:2016ioh})
\begin{eqnarray}
\tilde{F}^{(\Pom \Pom f_{2})}(q_{1}^{2},q_{2}^{2},p_{34}^{2}) = 
F_{M}(q_{1}^{2}) F_{M}(q_{2}^{2}) F^{(\Pom \Pom f_{2})}(p_{34}^{2})\,.
\label{Fpompommeson_tensor}
\end{eqnarray}
We are taking here the same form factor for each vertex with index $j$ ($j = 1, ..., 7$).
In principle, we could take a different form factor for each vertex.
We take
\begin{eqnarray}
&&F_{M}(t)=\frac{1}{1-t/\Lambda_{0}^{2}}\,,
\quad \Lambda_{0}^{2} = 0.5\;{\rm GeV}^{2}\,; 
\label{FM_t}\\
&&F^{(\Pom \Pom f_{2})}(p_{34}^{2}) = 
\exp{ \left( \frac{-(p_{34}^{2}-m_{f_{2}}^{2})^{2}}{\Lambda_{\Pom \Pom f_{2}}^{4}} \right)}\,,
\quad \Lambda_{\Pom \Pom f_{2}} = 1\;{\rm GeV}\,.
\label{Fpompommeson_ff_tensor}
\end{eqnarray}

The expressions for our bare vertices in (\ref{vertex_pompomT}), 
obtained from the coupling Lagrangians in Appendix~A of \cite{Lebiedowicz:2016ioh},
are as follows:
\begin{equation}
\begin{split}
i\Gamma_{\mu \nu,\kappa \lambda,\rho \sigma}^{(\Pom \Pom f_{2})(1)}=
2 i \,g^{(1)}_{\Pom \Pom f_{2}} M_{0}\, 
R_{\mu \nu \mu_{1} \nu_{1}}\,
R_{\kappa \lambda \alpha_{1} \lambda_{1}}\,
R_{\rho \sigma \rho_{1} \sigma_{1}}\,
g^{\nu_{1} \alpha_{1}}\,g^{\lambda_{1} \rho_{1}}\,g^{\sigma_{1} \mu_{1}}\,, 
\end{split}
\label{A11}
\end{equation}
\begin{equation}
\begin{split}
i\Gamma_{\mu \nu,\kappa \lambda,\rho \sigma}^{(\Pom \Pom f_{2})(2)} (q_{1},q_{2})=
-\frac{2i}{M_{0}} \,g^{(2)}_{\Pom \Pom f_{2}} \, 
\Bigl( &(q_{1} \cdot q_{2})\, R_{\mu \nu \rho_{1} \alpha}\, R_{\kappa \lambda \sigma_{1}}^{\quad \;\;\, \alpha}
- q_{1 \rho_{1}} \,q_{2}^{\mu_{1}} \,
  R_{\mu \nu \mu_{1} \alpha}\, R_{\kappa \lambda \sigma_{1}}^{\quad \;\;\, \alpha}\\
&- q_{1}^{\mu_{1}} \,q_{2 \sigma_{1}} \,
  R_{\mu \nu \rho_{1} \alpha}\, R_{\kappa \lambda \mu_{1}}^{\quad \;\;\, \alpha}
+ q_{1 \rho_{1}} \,q_{2 \sigma_{1}} \, R_{\mu \nu \kappa \lambda}
\Bigr) R_{\rho \sigma}^{\quad \rho_{1} \sigma_{1}}\,, 
\end{split}
\label{A12}
\end{equation}
\begin{equation}
\begin{split}
i\Gamma_{\mu \nu,\kappa \lambda,\rho \sigma}^{(\Pom \Pom f_{2})(3)} (q_{1},q_{2})=
-\frac{2i}{M_{0}} \,g^{(3)}_{\Pom \Pom f_{2}} \, 
\Bigl( &(q_{1} \cdot q_{2})\, R_{\mu \nu \rho_{1} \alpha}\, R_{\kappa \lambda \sigma_{1}}^{\quad \;\;\, \alpha}
+ q_{1 \rho_{1}} \,q_{2}^{\mu_{1}} \,
  R_{\mu \nu \mu_{1} \alpha}\, R_{\kappa \lambda \sigma_{1}}^{\quad \;\;\, \alpha}\\
&+ q_{1}^{\mu_{1}} \,q_{2 \sigma_{1}} \,
  R_{\mu \nu \rho_{1} \alpha}\, R_{\kappa \lambda \mu_{1}}^{\quad \;\;\, \alpha}
+ q_{1 \rho_{1}} \,q_{2 \sigma_{1}} \, R_{\mu \nu \kappa \lambda}
\Bigr) R_{\rho \sigma}^{\quad \rho_{1} \sigma_{1}}\,, 
\end{split}
\label{A13}
\end{equation}
\begin{equation}
\begin{split}
i\Gamma_{\mu \nu,\kappa \lambda,\rho \sigma}^{(\Pom \Pom f_{2})(4)} (q_{1},q_{2})=
-\frac{i}{M_{0}} \,g^{(4)}_{\Pom \Pom f_{2}} \, 
\Bigl( 
q_{1}^{\alpha_{1}} \,q_{2}^{\mu_{1}} \, 
R_{\mu \nu \mu_{1} \nu_{1}}\, R_{\kappa \lambda \alpha_{1} \lambda_{1}} + 
q_{2}^{\alpha_{1}} \,q_{1}^{\mu_{1}} \, 
R_{\mu \nu \alpha_{1} \lambda_{1}}\, R_{\kappa \lambda \mu_{1} \nu_{1}}
\Bigr) R^{\nu_{1} \lambda_{1}}_{\quad \; \; \; \rho \sigma}\,, 
\end{split}
\label{A14}
\end{equation}
\begin{equation}
\begin{split}
i\Gamma_{\mu \nu,\kappa \lambda,\rho \sigma}^{(\Pom \Pom f_{2})(5)} (q_{1},q_{2})=
-\frac{2i}{M_{0}^{3}} \,g^{(5)}_{\Pom \Pom f_{2}} \, 
\Bigl( &
q_{1}^{\mu_{1}} \,q_{2}^{\nu_{1}} \, 
R_{\mu \nu \nu_{1} \alpha}\, R_{\kappa \lambda \mu_{1}}^{\quad \;\;\, \alpha} + 
q_{1}^{\nu_{1}} \,q_{2}^{\mu_{1}} \, 
R_{\mu \nu \mu_{1} \alpha}\, R_{\kappa \lambda \nu_{1}}^{\quad \;\;\, \alpha} \\
&-2 (q_{1} \cdot q_{2})\, R_{\mu \nu \kappa \lambda}
\Bigr) 
q_{1 \alpha_{1}} \,q_{2 \lambda_{1}} \, 
R^{\alpha_{1} \lambda_{1}}_{\quad \; \; \; \rho \sigma}\,, 
\end{split}
\label{A15}
\end{equation}
\begin{equation}
\begin{split}
i\Gamma_{\mu \nu,\kappa \lambda,\rho \sigma}^{(\Pom \Pom f_{2})(6)} (q_{1},q_{2})=
\frac{i}{M_{0}^{3}} \,g^{(6)}_{\Pom \Pom f_{2}} \, 
\Bigl( &
q_{1}^{\alpha_{1}} \,q_{1}^{\lambda_{1}} \, q_{2}^{\mu_{1}} \, q_{2 \rho_{1}} \, 
R_{\mu \nu \mu_{1} \nu_{1}}\, R_{\kappa \lambda \alpha_{1} \lambda_{1}} \\
&+ 
q_{2}^{\alpha_{1}} \,q_{2}^{\lambda_{1}} \, q_{1}^{\mu_{1}} \, q_{1 \rho_{1}} \, 
R_{\mu \nu \alpha_{1} \lambda_{1}}\, R_{\kappa \lambda \mu_{1} \nu_{1}}
\Bigr) 
R^{\nu_{1} \rho_{1}}_{\quad \; \; \; \rho \sigma}\,, 
\end{split}
\label{A16}
\end{equation}
\begin{equation}
\begin{split}
i\Gamma_{\mu \nu,\kappa \lambda,\rho \sigma}^{(\Pom \Pom f_{2})(7)} (q_{1},q_{2})=
-\frac{2i}{M_{0}^{5}} \,g^{(7)}_{\Pom \Pom f_{2}} \, 
q_{1}^{\rho_{1}} \, q_{1}^{\alpha_{1}} \, q_{1}^{\lambda_{1}} \, 
q_{2}^{\sigma_{1}} \, q_{2}^{\mu_{1}} \, q_{2}^{\nu_{1}} \, 
R_{\mu \nu \mu_{1} \nu_{1}}\, R_{\kappa \lambda \alpha_{1} \lambda_{1}}\,
R_{\rho \sigma \rho_{1} \sigma_{1}}\,,
\end{split}
\label{A17}
\end{equation}
where 
\begin{eqnarray}
R_{\mu \nu \kappa \lambda} = \frac{1}{2} g_{\mu \kappa} g_{\nu \lambda} 
                           + \frac{1}{2} g_{\mu \lambda} g_{\nu \kappa}
                            -\frac{1}{4} g_{\mu \nu} g_{\kappa \lambda}\,.
\label{A2}
\end{eqnarray}
In (\ref{A11}) to (\ref{A17}) the Lorentz indices of the pomeron
with momentum $q_{1}$ are denoted by $\mu \nu$,
of the pomeron with momentum $q_{2}$ by $\kappa \lambda$,
and of the $f_{2}$ by $\rho \sigma$.
Furthermore, $M_{0} \equiv 1$~GeV and the $g^{(j)}_{\Pom \Pom f_{2}}$ 
($j = 1, ..., 7$) are dimensionless coupling constants.
The values of the coupling constants $g^{(j)}_{\Pom \Pom f_{2}}$ 
are not known and are not easy to be found from first principles
of QCD, as they are of nonperturbative origin.
At the present stage these coupling constants $g_{\Pom \Pom f_{2}}^{(j)}$ 
should be fitted to experimental data.

Considering the fictitious reaction of two ``real tensor pomerons''
annihilating to the $f_{2}$ meson, see Appendix~A of \cite{Lebiedowicz:2016ioh},
we find that we can associate the couplings (\ref{A11})--(\ref{A17}) 
with the following $(l,S)$ values 
\footnote{Here, $l$ and $S$ denote orbital angular momentum and total spin of two 
fictitious ``real pomerons''
in the rest system of the $f_{2}$ meson, respectively.}
$(0,2)$, $(2,0)-(2,2)$, $(2,0)+(2,2)$, $(2,4)$, $(4,2)$, $(4,4)$, $(6,4)$, respectively.

To give the full physical amplitudes
we should include absorptive corrections to the Born amplitudes.
For the details how to include the $pp$-rescattering corrections 
in the eikonal approximation for the four-body reaction
see e.g. Sec.~3.3 of \cite{Lebiedowicz:2014bea}.
Other rescattering corrections, 
such as possible pion-proton \cite{Lebiedowicz:2015eka,Ryutin:2019khx} 
and pion-pion \cite{Lebiedowicz:2011nb} interactions 
in the final state,
and also so-called ``enhanced'' corrections \cite{Harland-Lang:2013dia},
are neglected in the present calculations.
In practice we work with the amplitudes in the high-energy approximation;
see Eqs.~(3.19)--(3.21) and (4.23) of \cite{Lebiedowicz:2016ioh}.

We are interested in the angular distribution of the $\pi^{+}$ in the 
center-of-mass system of the $\pi^{+}\pi^{-}$ pair.
Various reference systems are commonly used; see e.g. \cite{Bolz:2014mya}
for a discussion of such systems for the $\gamma p \to \pi^{+}\pi^{-} p$ reaction.
For the Collins-Soper system \cite{Collins:1977iv,Soper:1982wc} 
for the reaction (\ref{2to4_reaction}) we set the unit vectors 
defining the axes as follows:
\begin{equation}
\begin{split}
& \be_{\textbf{3},\,\rm CS} = \frac{\bhpa-\bhpb}{|\bhpa-\bhpb|}\,,\\
& \be_{\textbf{2},\,\rm CS} = \frac{\bhpa \times \bhpb}{|\bhpa \times \bhpb|}\,,\\
& \be_{\textbf{1},\,\rm CS} = \frac{\bhpa+\bhpb}{|\bhpa+\bhpb|}\,.
\end{split}
\label{CS}
\end{equation}
These satisfy the condition 
$\be_{\textbf{1},\,\rm CS} = 
\be_{\textbf{2},\,\rm CS} \times \be_{\textbf{3},\,\rm CS}$.
Here $\bhpa = \bpa / |\bpa|$, $\bhpb = \bpb / |\bpb|$,
where $\bpa$, $\bpb$ are the three-momenta of the initial protons
in the $\pi^{+}\pi^{-}$ rest system.
There we have $\bpipm = 0$ and $\bpa + \bpb = \bpaa + \bpbb$.
Now we denote by $\theta_{\pi^{+},\,{\rm CS}}$ and $\phi_{\pi^{+},\,{\rm CS}}$
the polar and azimuthal angles of $\bhpc$ (the $\pi^{+}$ meson momentum)
relative to the coordinate axes (\ref{CS}).
We have then e.g.
\begin{equation}
\cos\theta_{\pi^{+},{\,\rm CS}} = \bhpc \cdot \be_{\textbf{3},\,\rm CS}\,,
\end{equation}
where $\bhpc = \bpip / |\bpip|$.

Alternatively, for the experiments that can measure
at least one of the outgoing protons,
the Gottfried-Jackson (GJ) system could be used as well.
For the GJ system \cite{Gottfried:1964nx} we set
\begin{equation}
\begin{split}
& \be_{\textbf{3},\,\rm GJ} = \frac{\bqa}{|\bqa|}\,,\\
& \be_{\textbf{2},\,\rm GJ} = \frac{\bq_{\textbf{1},\,\rm c.m.} \times \bq_{\textbf{2},\,\rm c.m.}}
                         {|\bq_{\textbf{1},\,\rm c.m.} \times \bq_{\textbf{2},\,\rm c.m.}|}\,,\\
& \be_{\textbf{1},\,\rm GJ} = \be_{\textbf{2},\,\rm GJ} \times \be_{\textbf{3},\,\rm GJ}\,.
\end{split}
\label{GJ}
\end{equation}
Here $\bqa$ is
the three-momentum of the pomeron (emitted by the proton with positive $p_{z}$)
in the $\pi^{+}\pi^{-}$ rest system.
The second axis of the GJ coordinate system 
is fixed by the normal to the production plane ($\Pom$-$\Pom$-$\pi^{+}\pi^{-}$ plane)
in the $pp$ center-of-mass (c.m.) system.
$\bq_{\textbf{1},\,\rm c.m.}$ and $\bq_{\textbf{2},\,\rm c.m.}$ are three-momenta defined in the $pp$ c.m. frame.

For some further remarks on this GJ system see Appendix~\ref{sec:appendixA}.

Having defined these angles
we can now examine the differential cross sections
$d^{2}\sigma/(d\cos\theta_{\pi^{+},\,{\rm CS}} \,d\phi_{\pi^{+},\,{\rm CS}})$,
$d\sigma/d\cos\theta_{\pi^{+},\,{\rm CS}}$, $d\sigma/d\phi_{\pi^{+},\,{\rm CS}}$,
and the corresponding distributions in the GJ system.

\section{Results}
\label{sec:results}

\begin{figure}[!ht]
\includegraphics[width=0.45\textwidth]{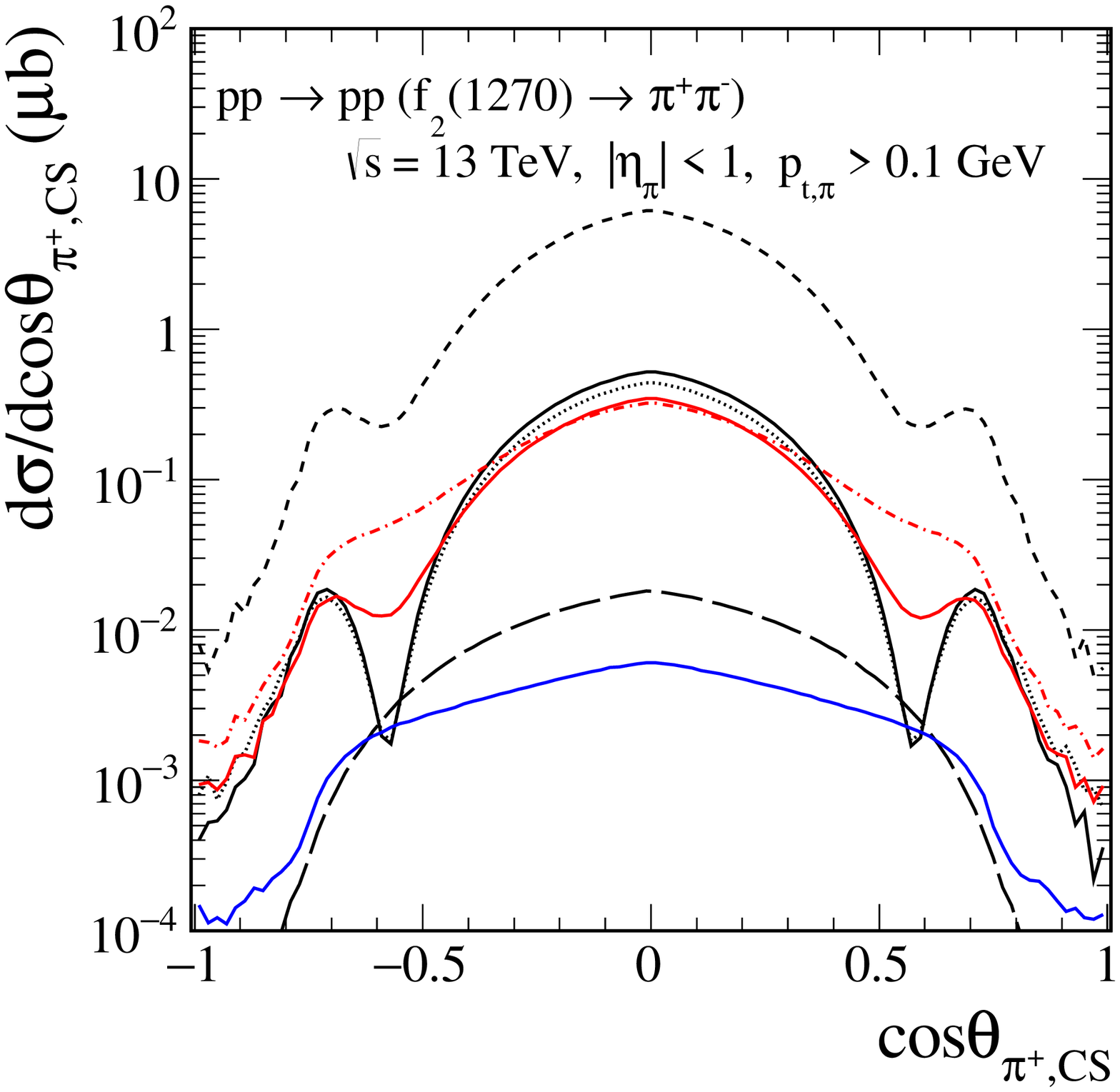}
\includegraphics[width=0.45\textwidth]{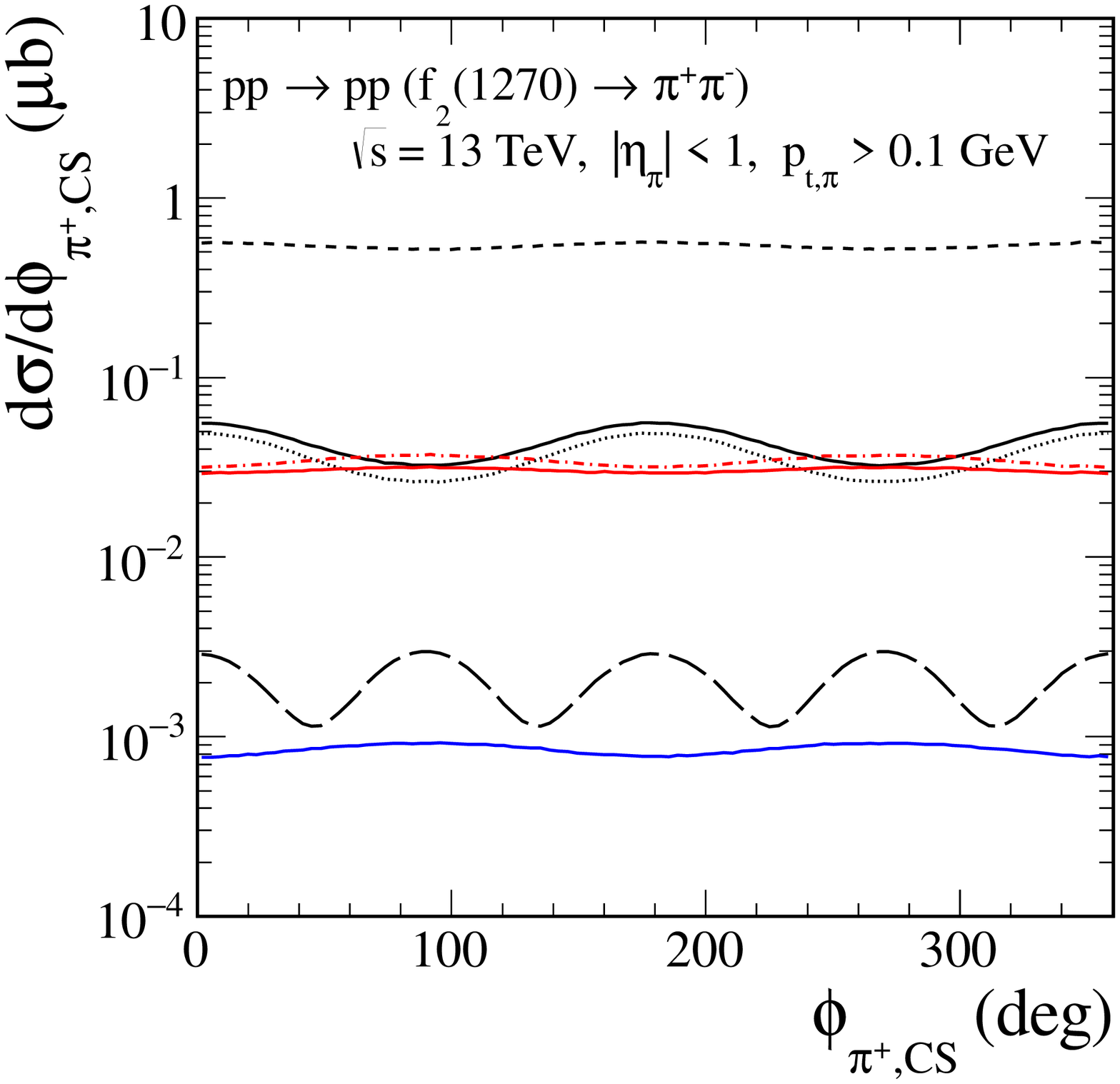}
\includegraphics[width=0.45\textwidth]{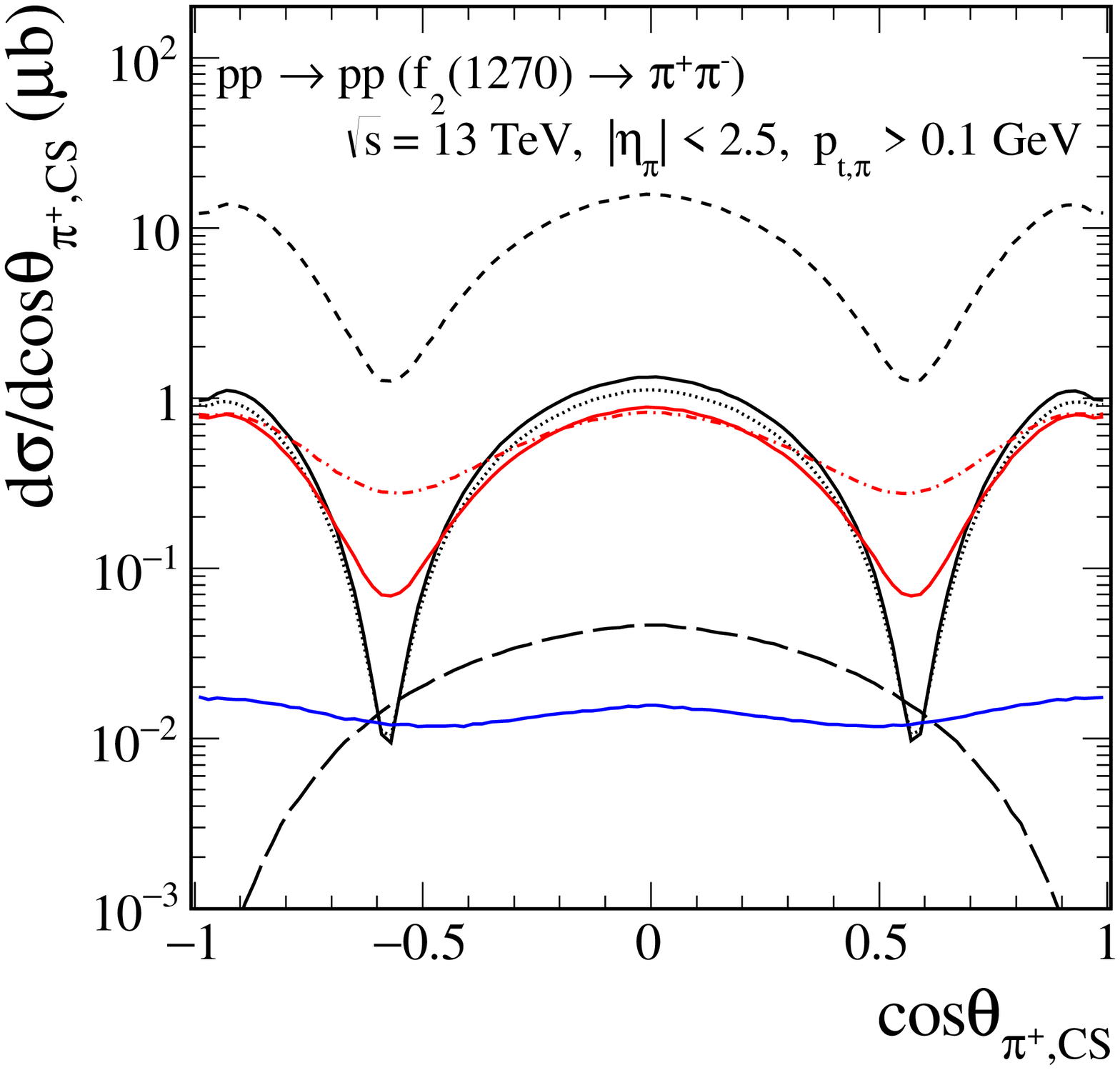}
\includegraphics[width=0.45\textwidth]{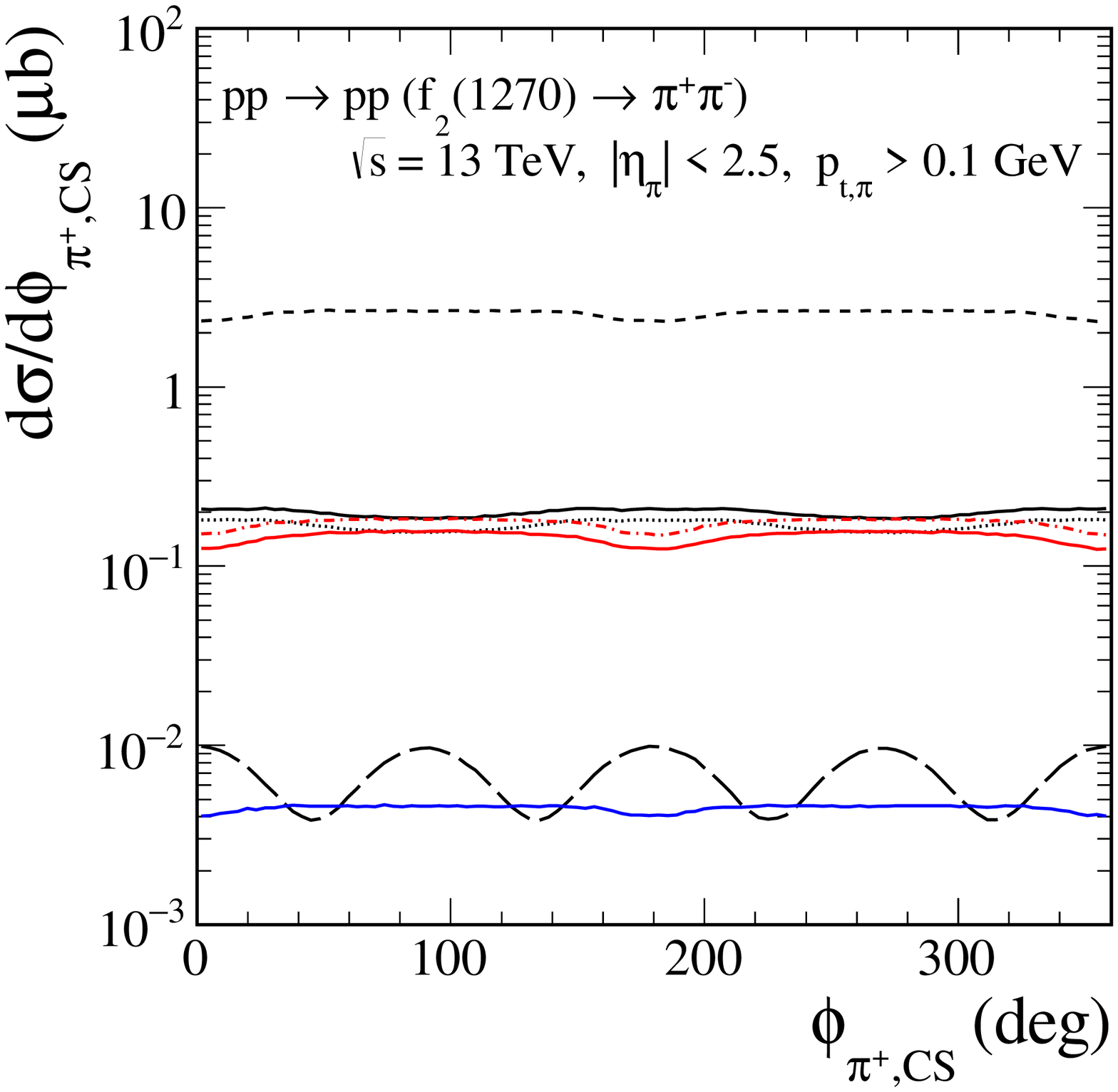}
\caption{\label{fig:1}
\small
The distributions in $\cos\theta_{\pi^{+},\,{\rm CS}}$ (the left panels) 
and in $\phi_{\pi^{+},\,{\rm CS}}$ (the right panels) for 
the $pp \to pp (f_{2}(1270) \to \pi^{+}\pi^{-})$ reaction.
The calculations were done for $\sqrt{s} = 13$~TeV 
with different cuts on $|\eta_{\pi}|$ and $p_{t, \pi} > 0.1$~GeV.
The individual contributions of the different $\Pom\Pom f_{2}$ couplings 
from (\ref{A11}) to (\ref{A17}) are shown:
$j=1$ (the black solid line), $j=2$ (the black long-dashed line),
$j=3$ (the black dashed line), $j=4$ (the black dotted line),
$j=5$ (the blue solid line), $j=6$ (the red solid line), 
and $j=7$ (the red dot-dashed line).
The results correspond to the arbitrary choice 
of coupling constants $g_{\Pom \Pom f_{2}}^{(j)} = 1.0$.
No absorption effects were included here.}
\end{figure}

\begin{figure}[!ht]
\includegraphics[width=0.45\textwidth]{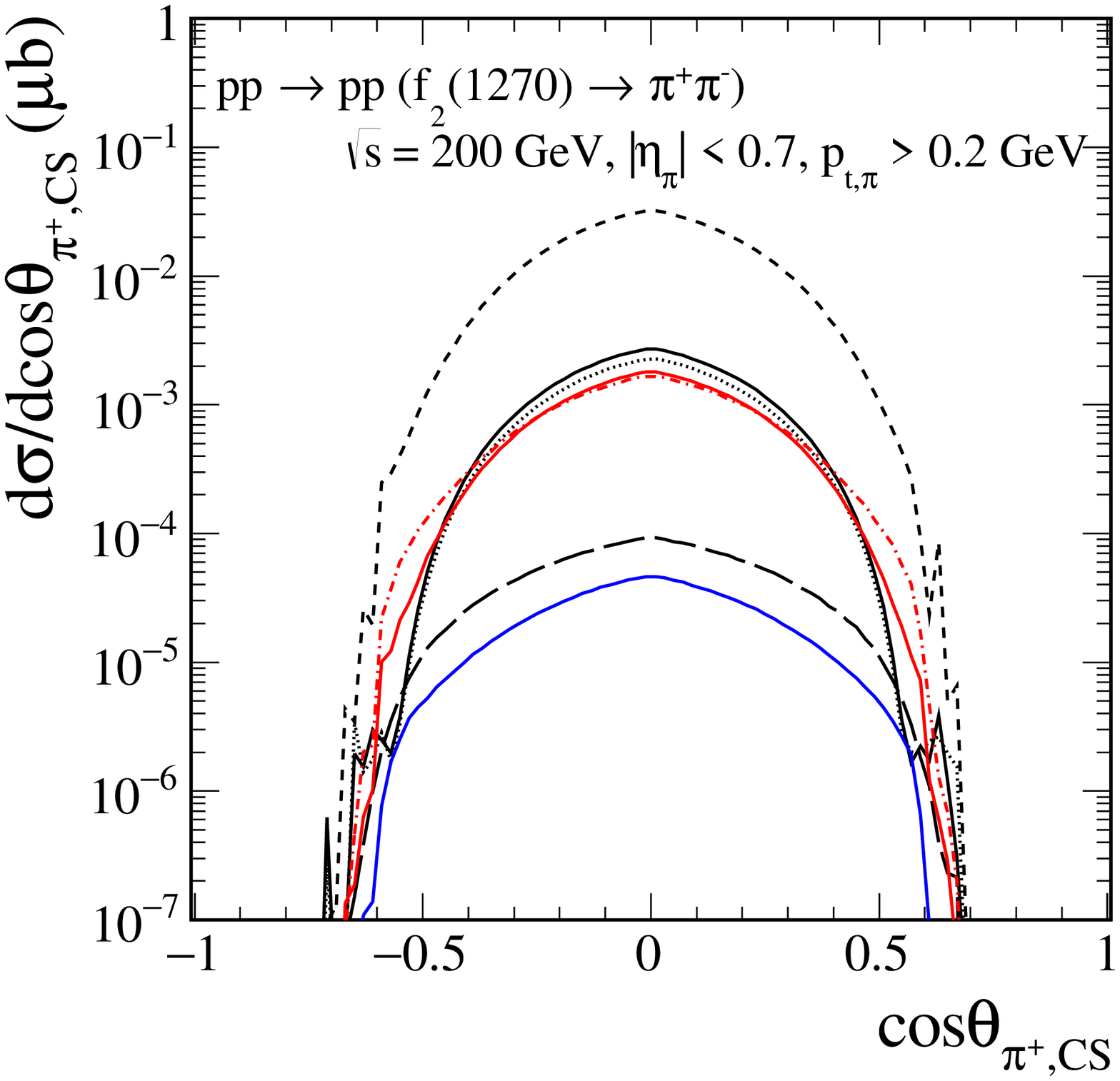}
\includegraphics[width=0.45\textwidth]{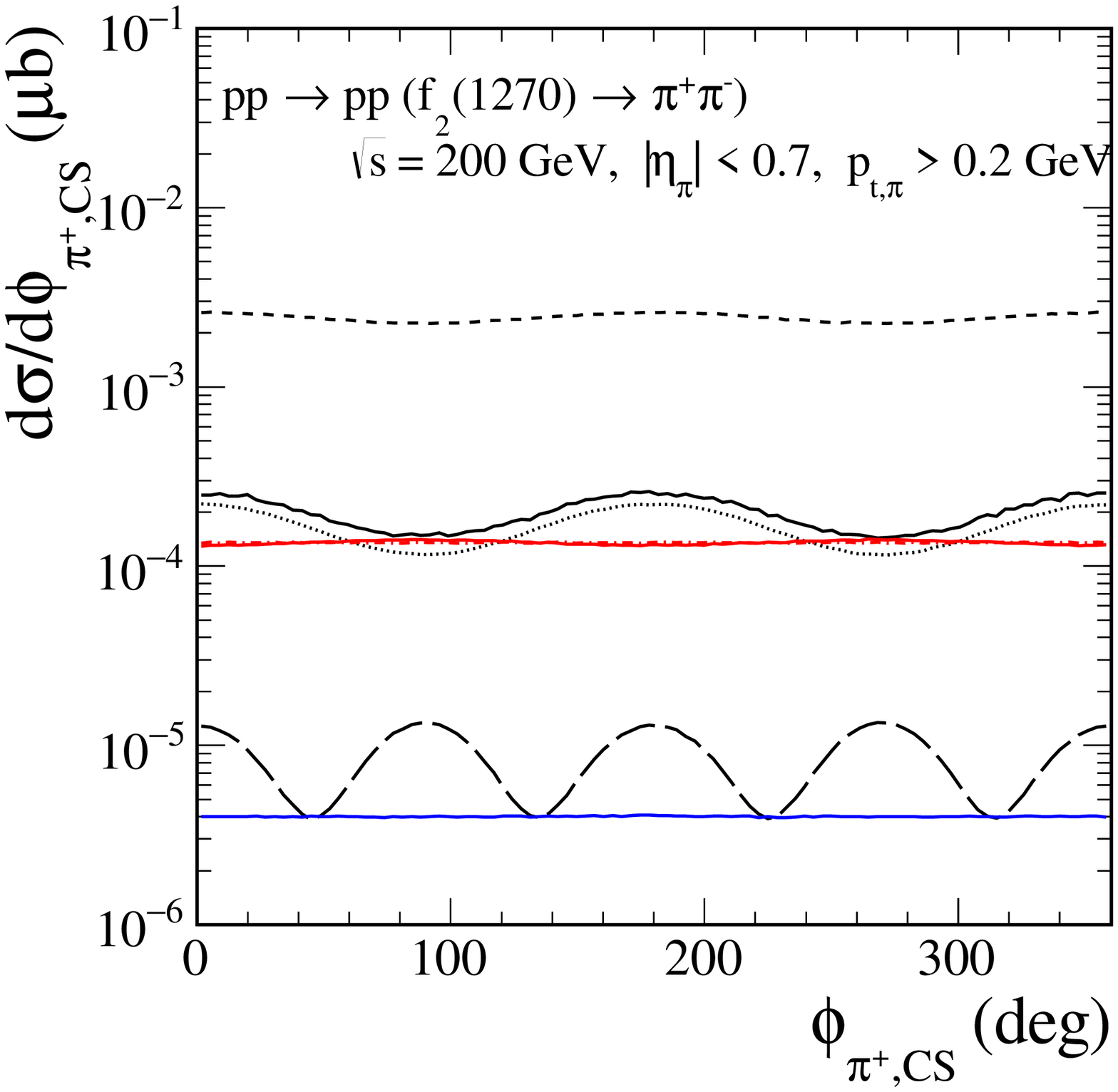}
\includegraphics[width=0.45\textwidth]{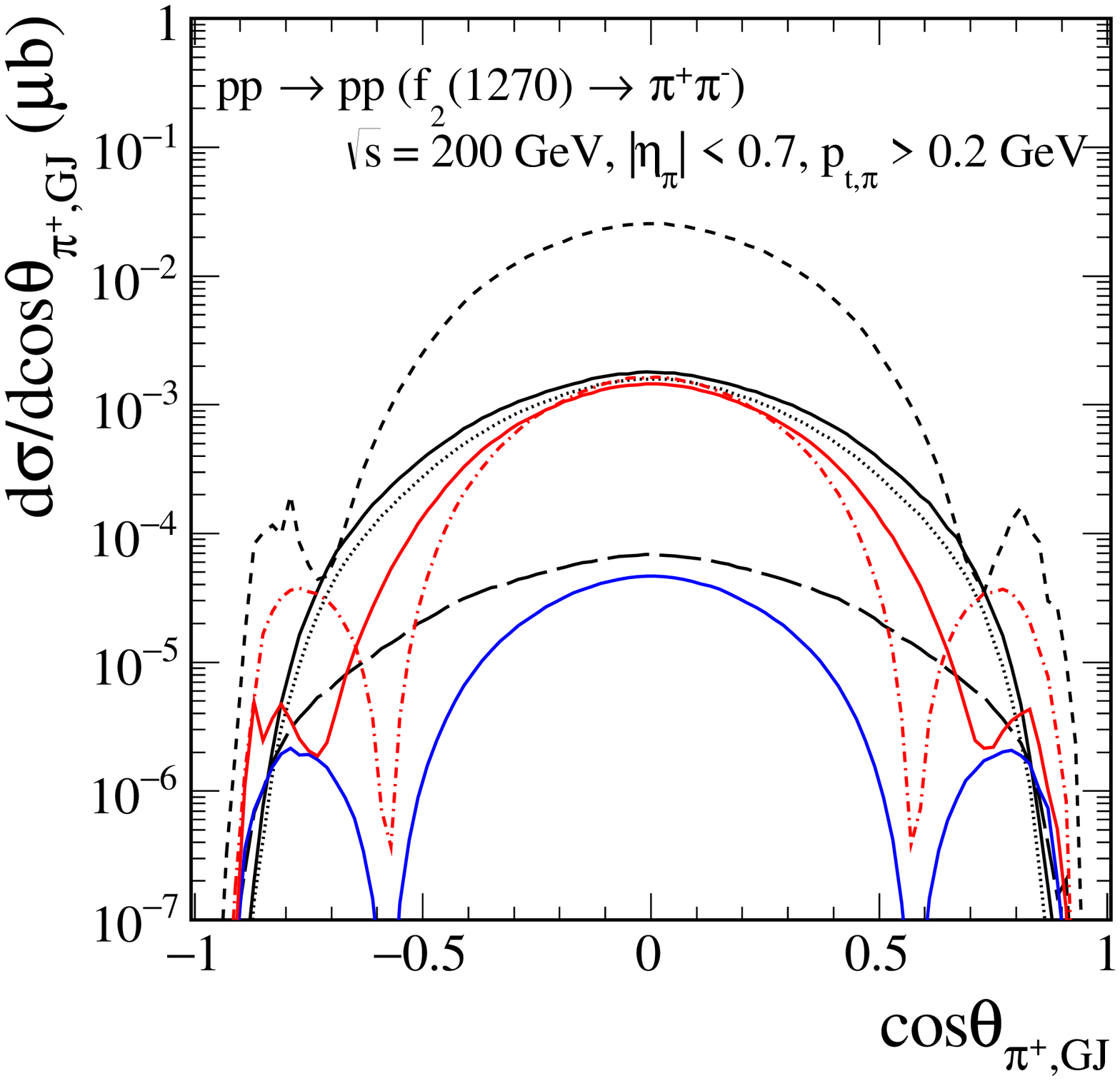}
\includegraphics[width=0.45\textwidth]{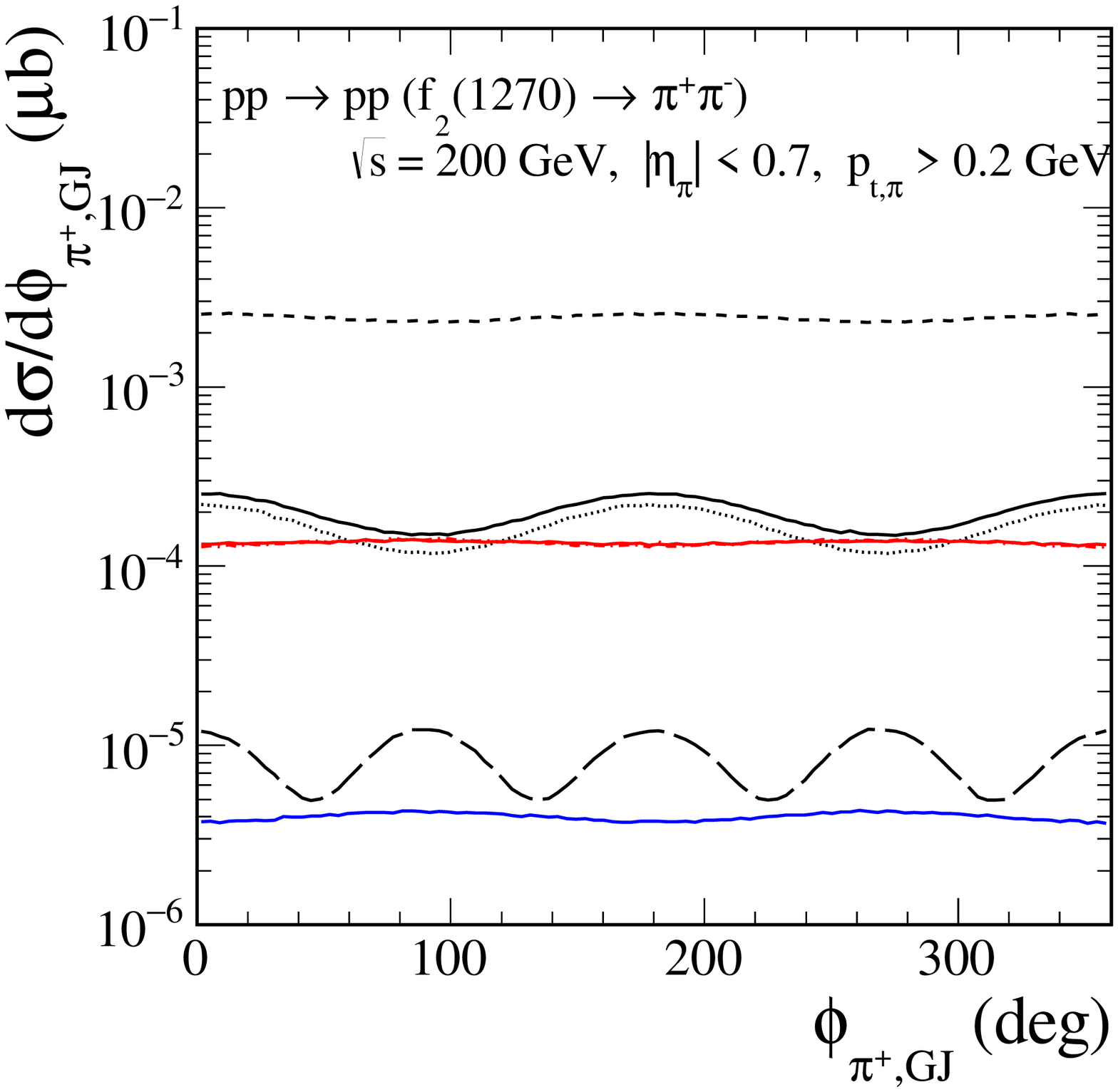}
\caption{\label{fig:STAR}
\small
The same as in Fig.~\ref{fig:1} but for $\sqrt{s} = 200$~GeV
and the STAR experimental cuts from \cite{Sikora:2018cyk}: 
$|\eta_{\pi}| < 0.7$, $p_{t, \pi} > 0.2$~GeV,
and with cuts on the leading protons (\ref{STAR_cuts}).
In the top panels, we show the pion angular distributions 
in the $\pi^+ \pi^-$ rest system using the CS frame (\ref{CS}).
In the bottom panels, we show the results using the GJ frame (\ref{GJ}).}
\end{figure}

\begin{figure}[!ht]
(a)\includegraphics[width=0.4\textwidth]{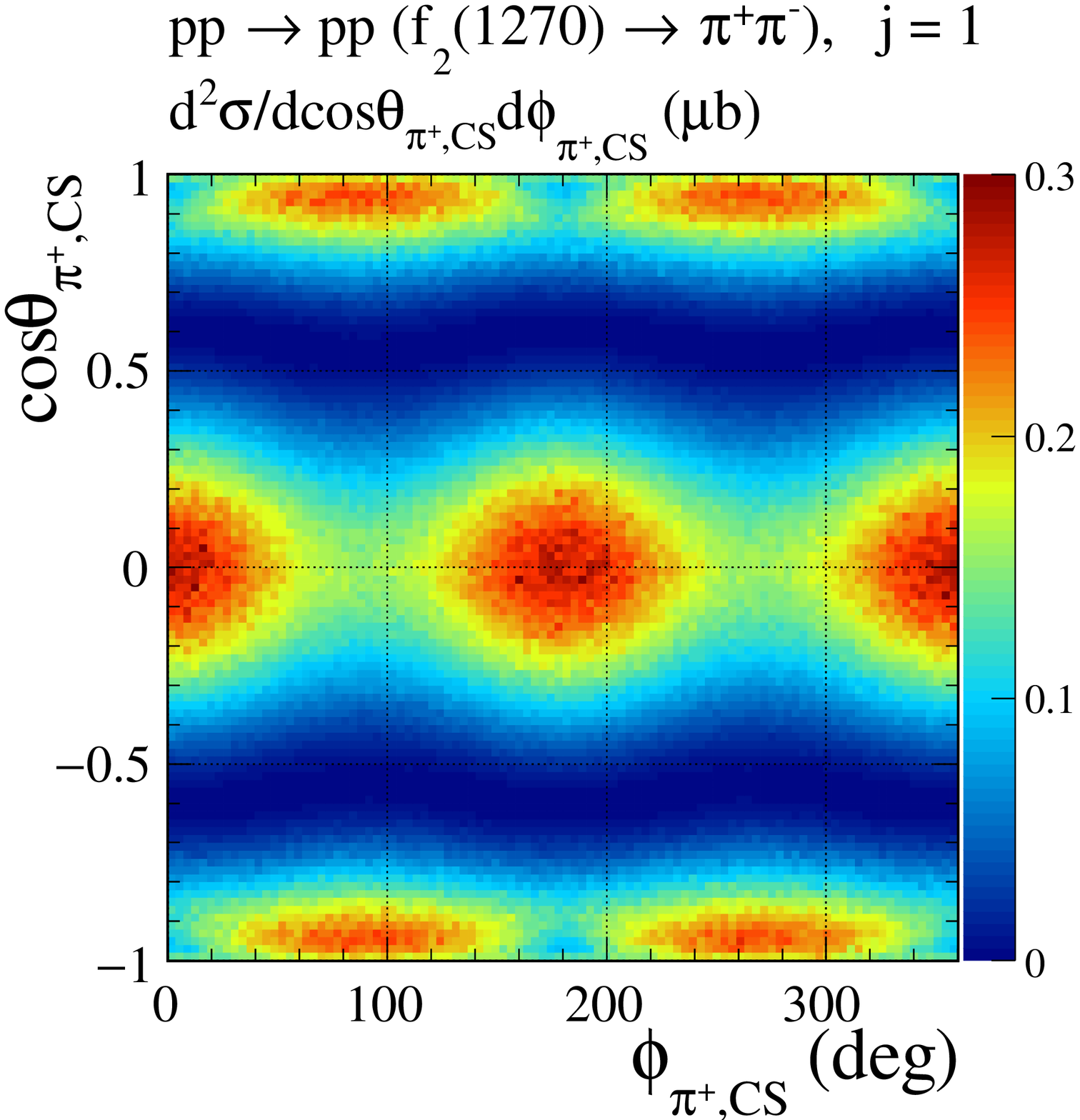}
(b)\includegraphics[width=0.4\textwidth]{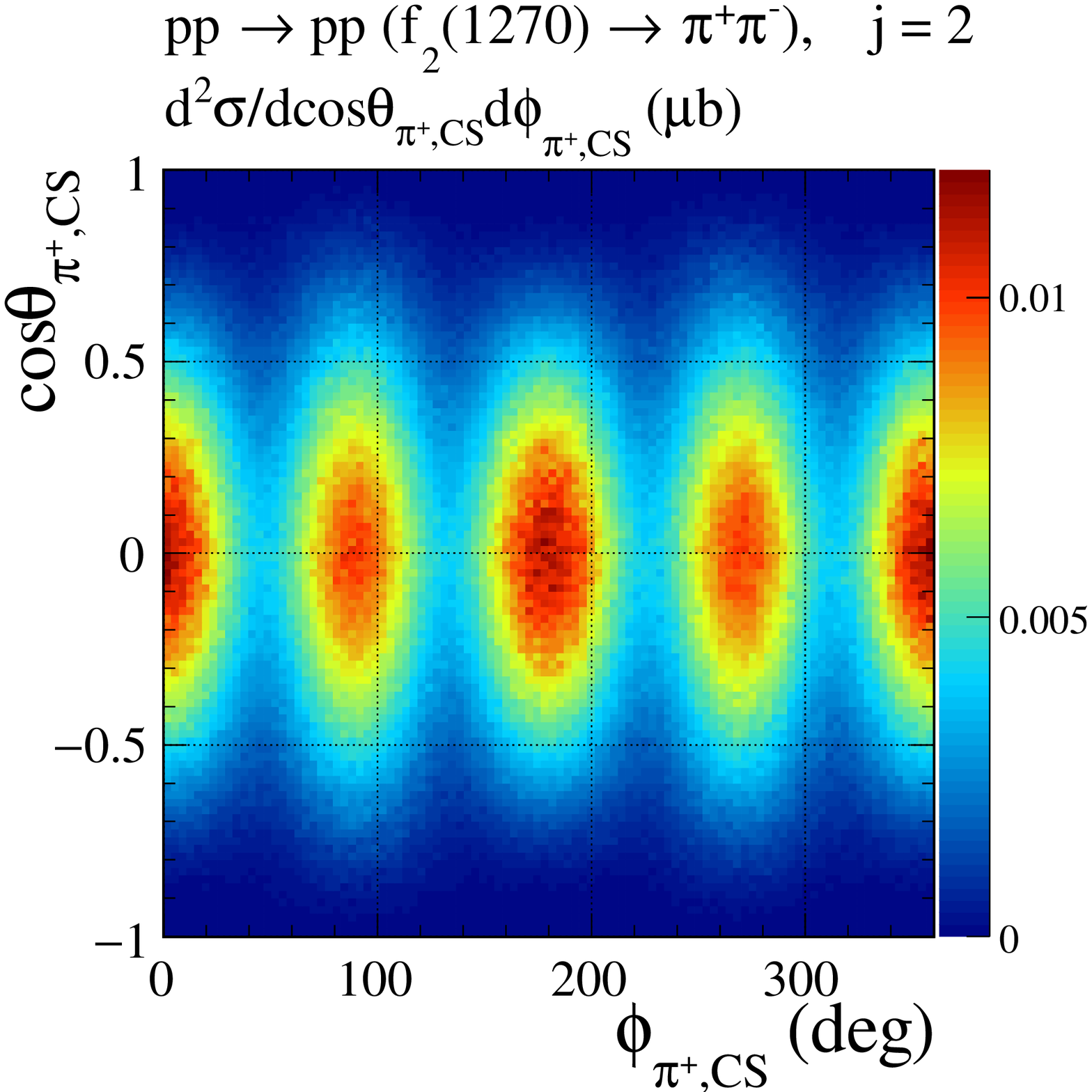}
(c)\includegraphics[width=0.4\textwidth]{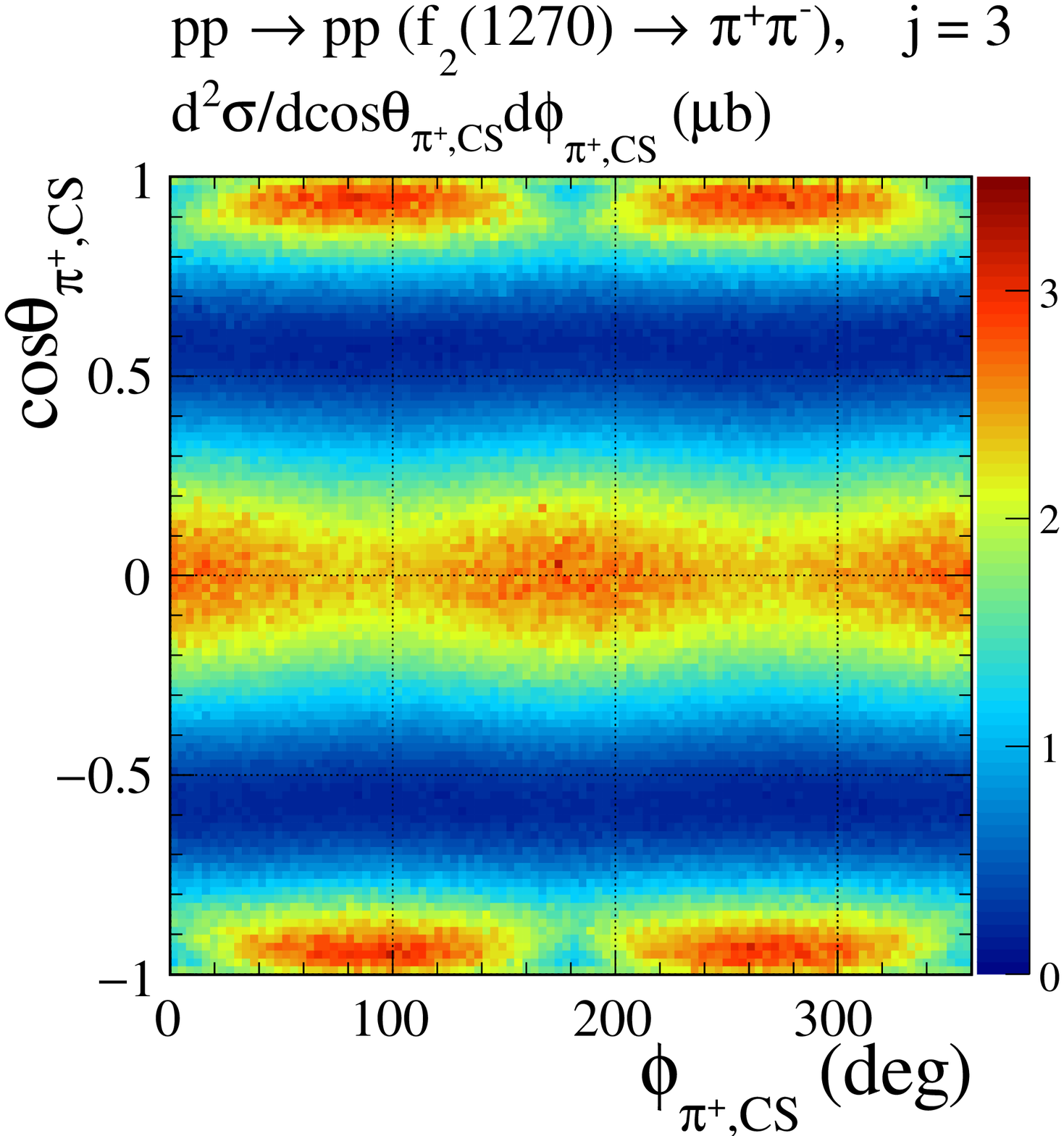}
(d)\includegraphics[width=0.4\textwidth]{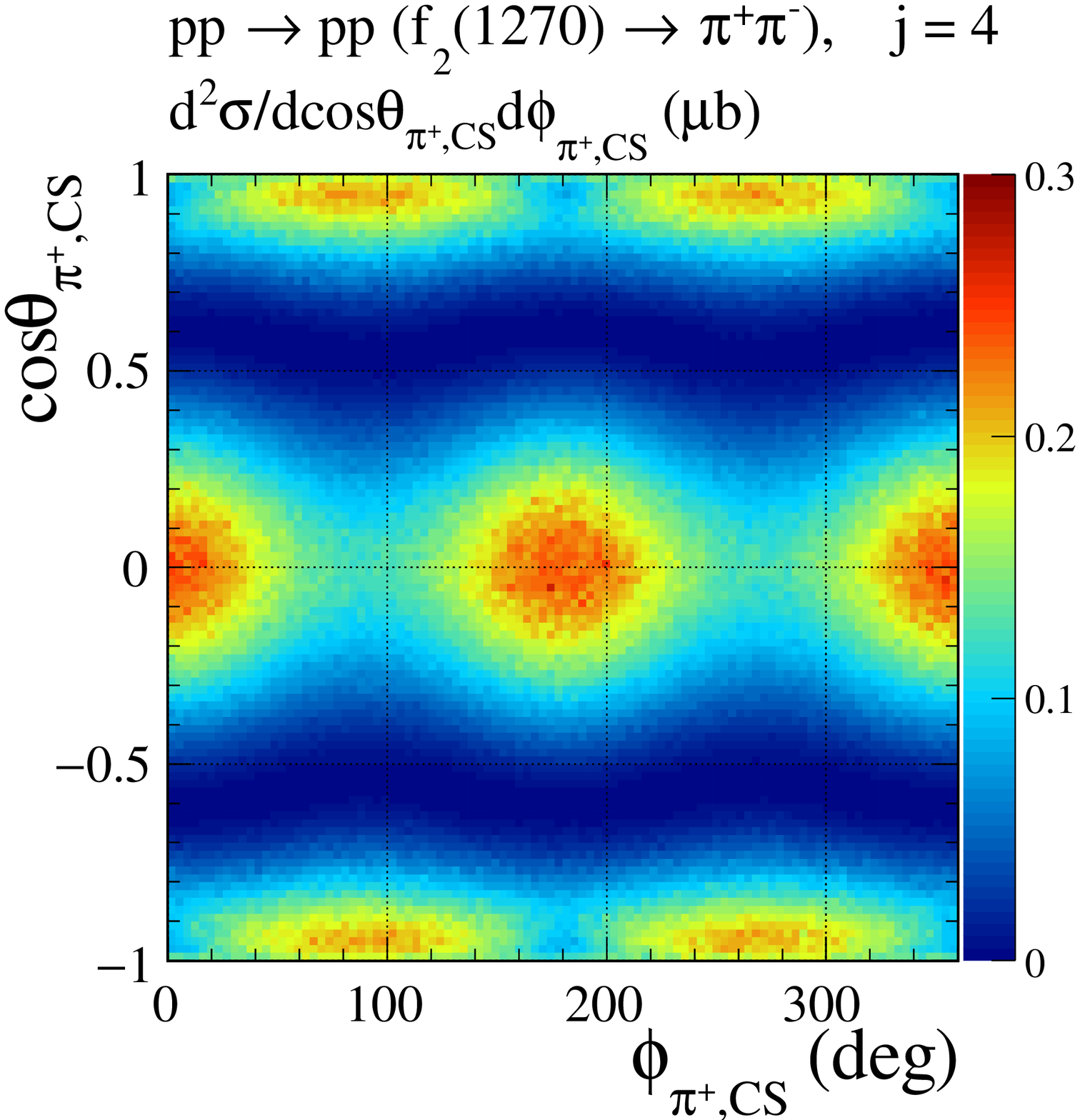}
(e)\includegraphics[width=0.4\textwidth]{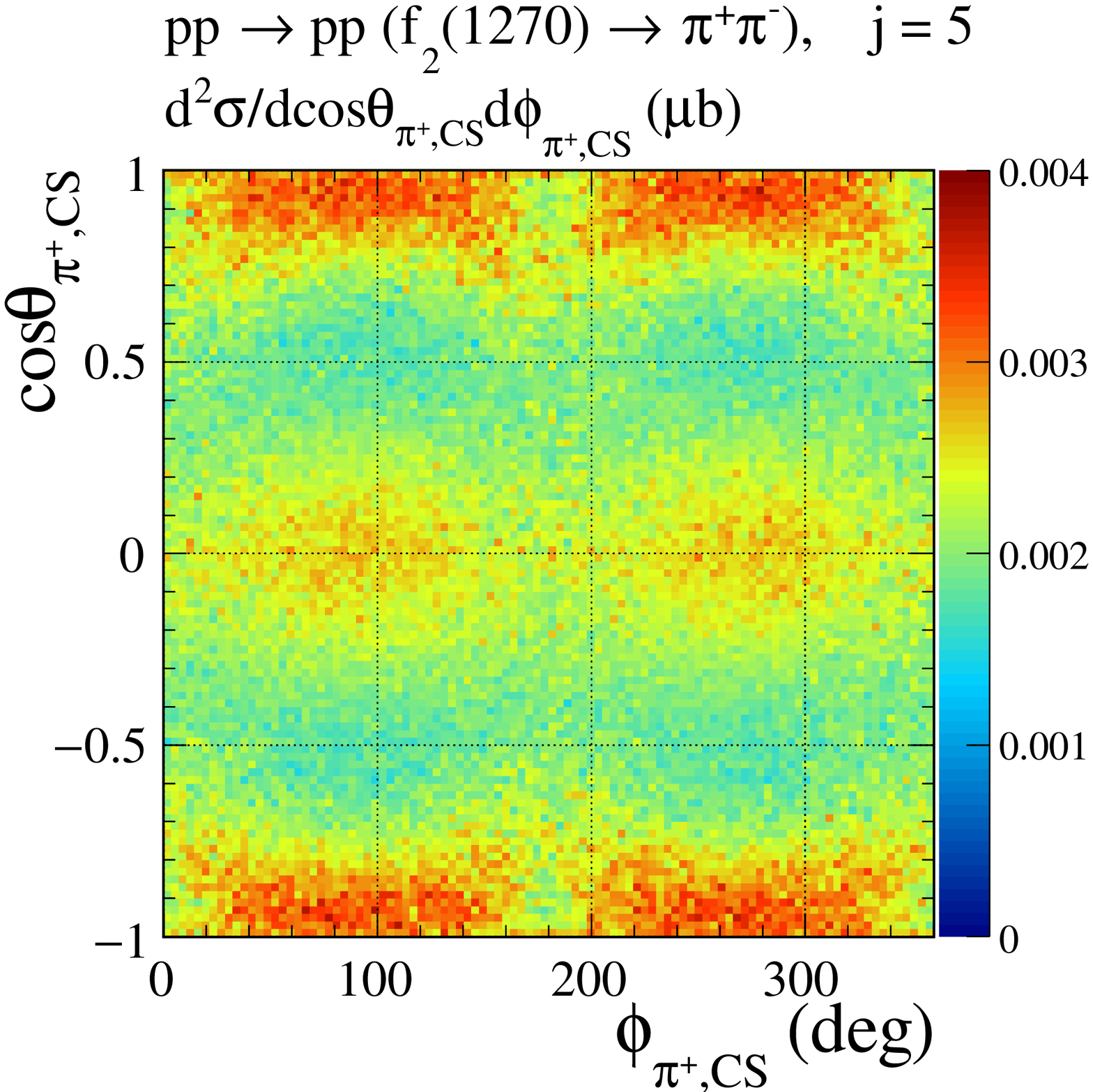}
(f)\includegraphics[width=0.4\textwidth]{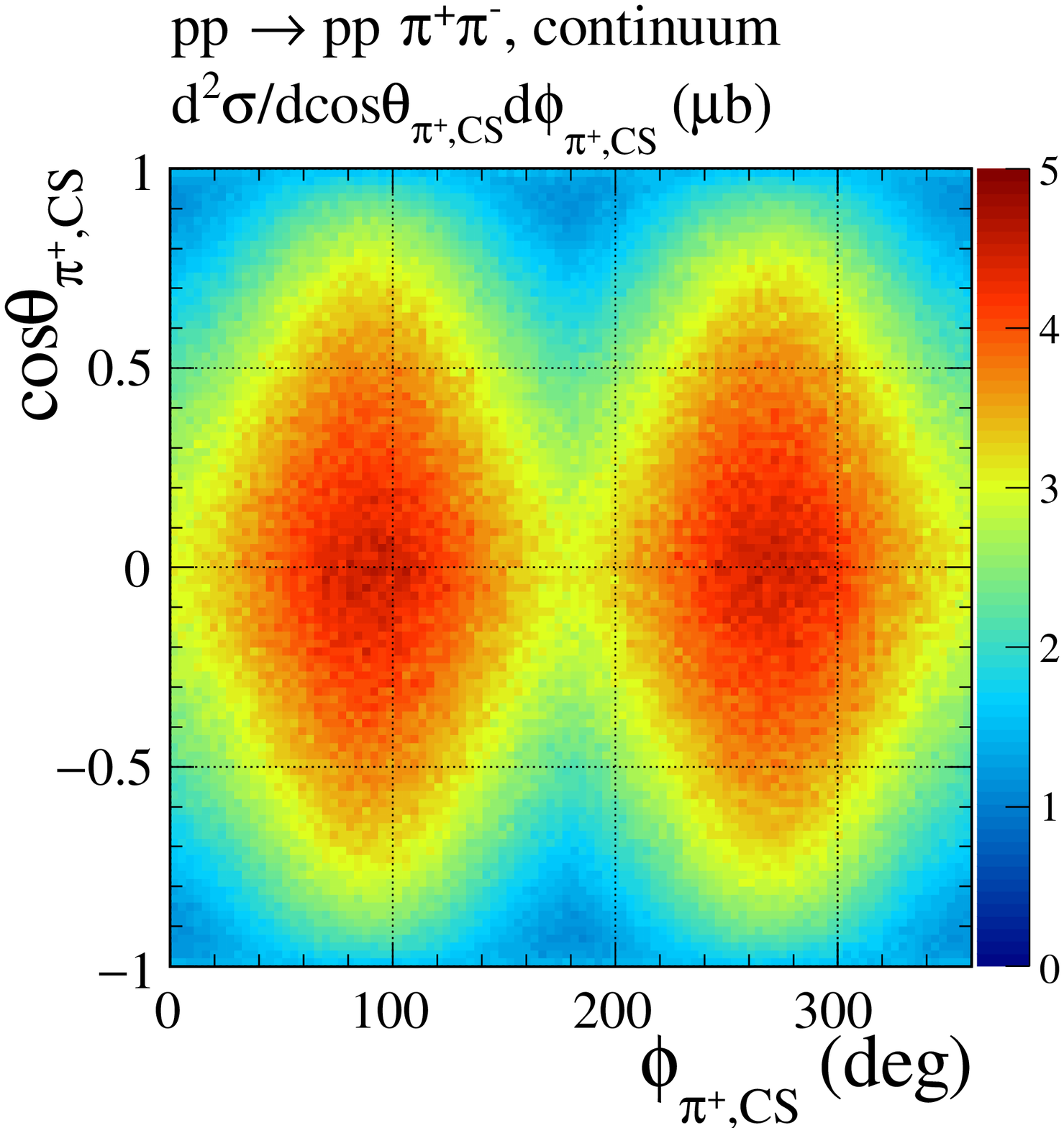}
\caption{\label{fig:3}
\small
The two-dimensional distributions in ($\phi_{\pi^{+},\,{\rm CS}},\cos\theta_{\pi^{+},\,{\rm CS}}$) 
for the $pp \to pp \pi^{+}\pi^{-}$ reaction.
The calculations were done for 
$\sqrt{s} = 13$~TeV and with cuts on $|\eta_{\pi}| < 2.5$.
The individual $f_{2}(1270) \to \pi^{+}\pi^{-}$ contributions 
for five $\Pom \Pom f_{2}$ couplings,
(a)~$j=1$, (b)~$j=2$, (c)~$j=3$, (d)~$j=4$, (e)~$j=5$, 
and (f) the $\pi^{+}\pi^{-}$ continuum term are presented.
The results for $\pi^{+}\pi^{-}$ production via $f_{2}(1270)$ resonance 
were obtained with coupling constants $g_{\Pom \Pom f_{2}}^{(j)} = 1.0$.
No absorption effects were included here.}
\end{figure}

\begin{figure}[!ht]
\includegraphics[width=0.45\textwidth]{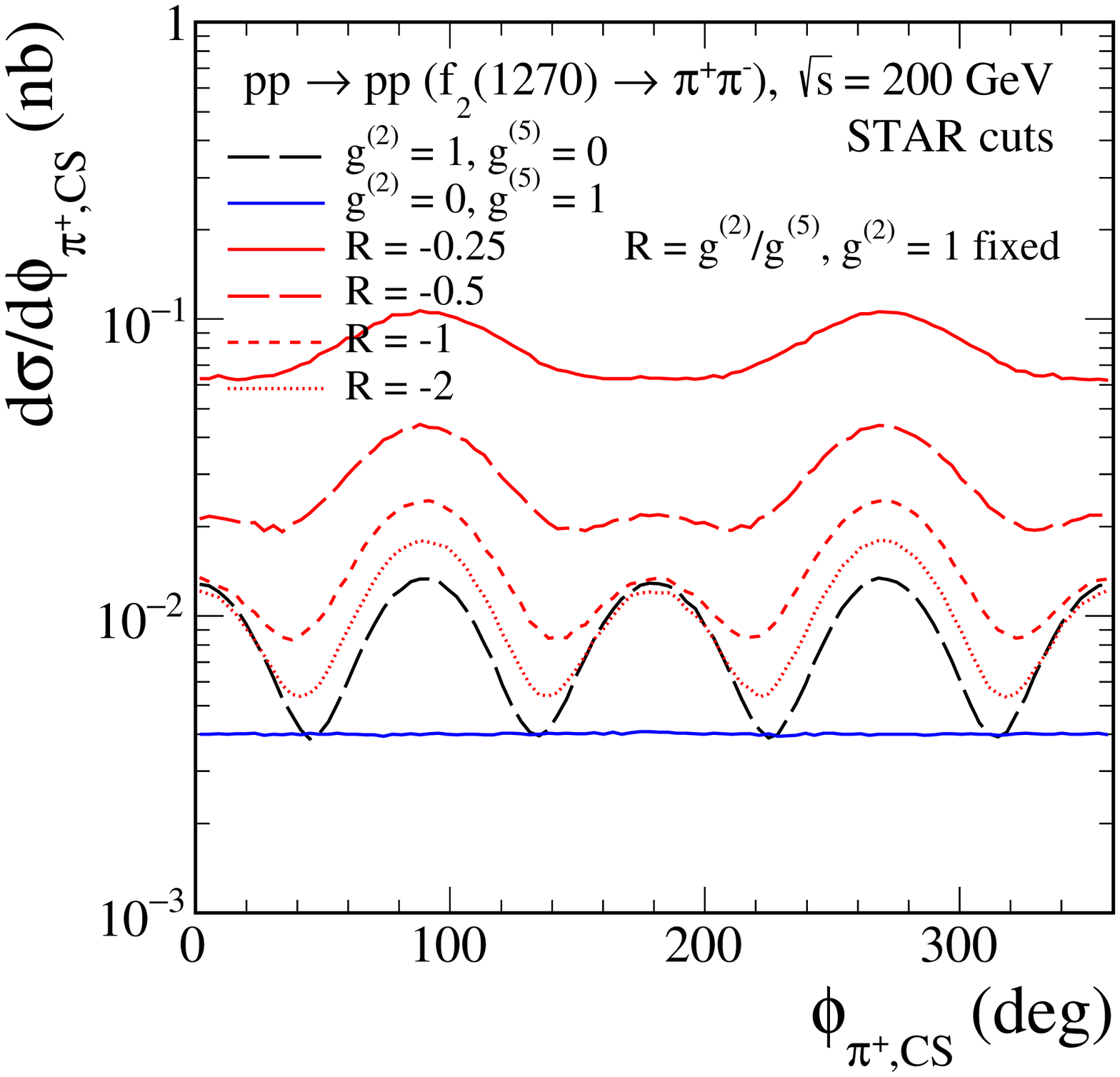}
\includegraphics[width=0.45\textwidth]{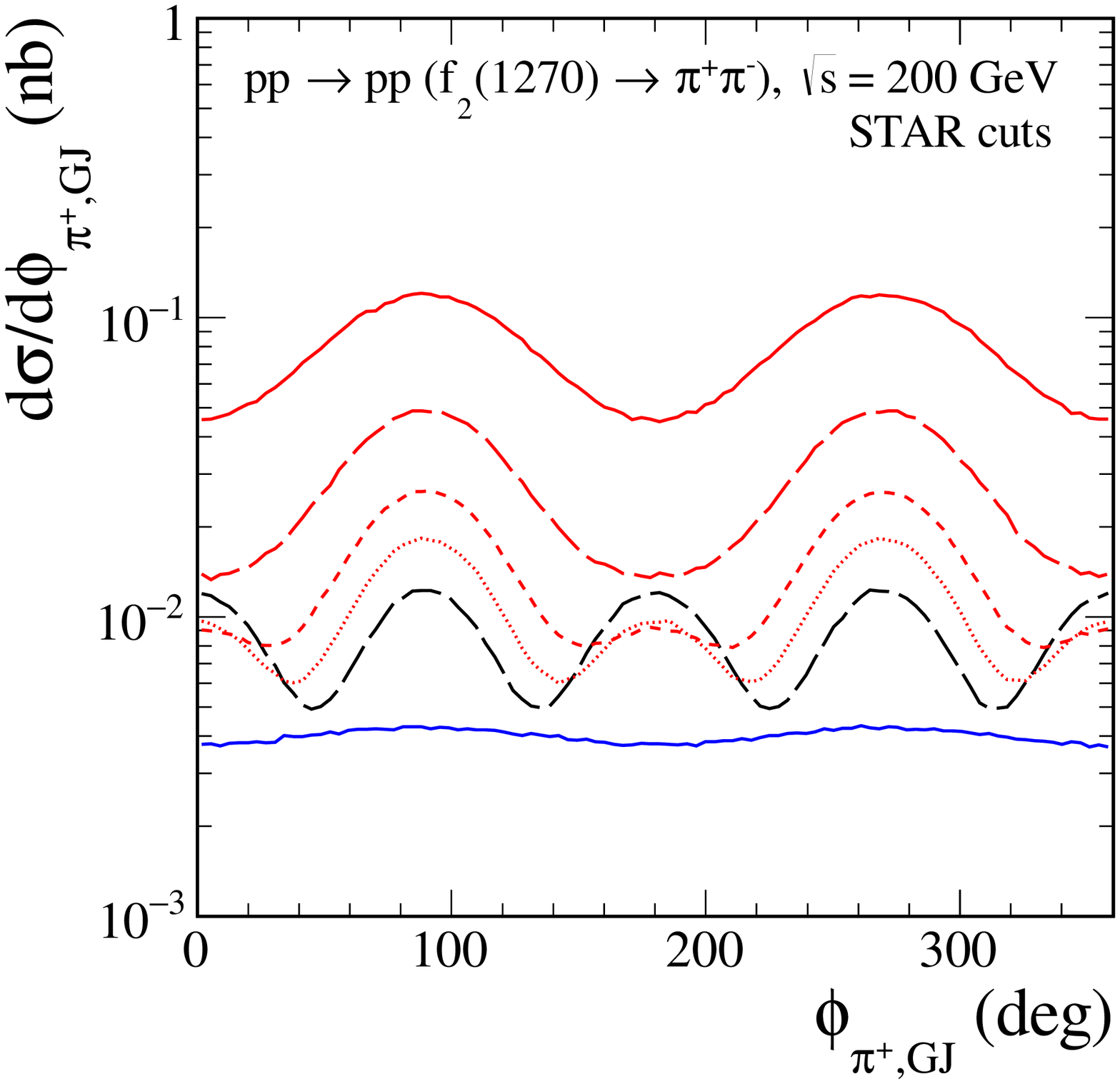}
\includegraphics[width=0.45\textwidth]{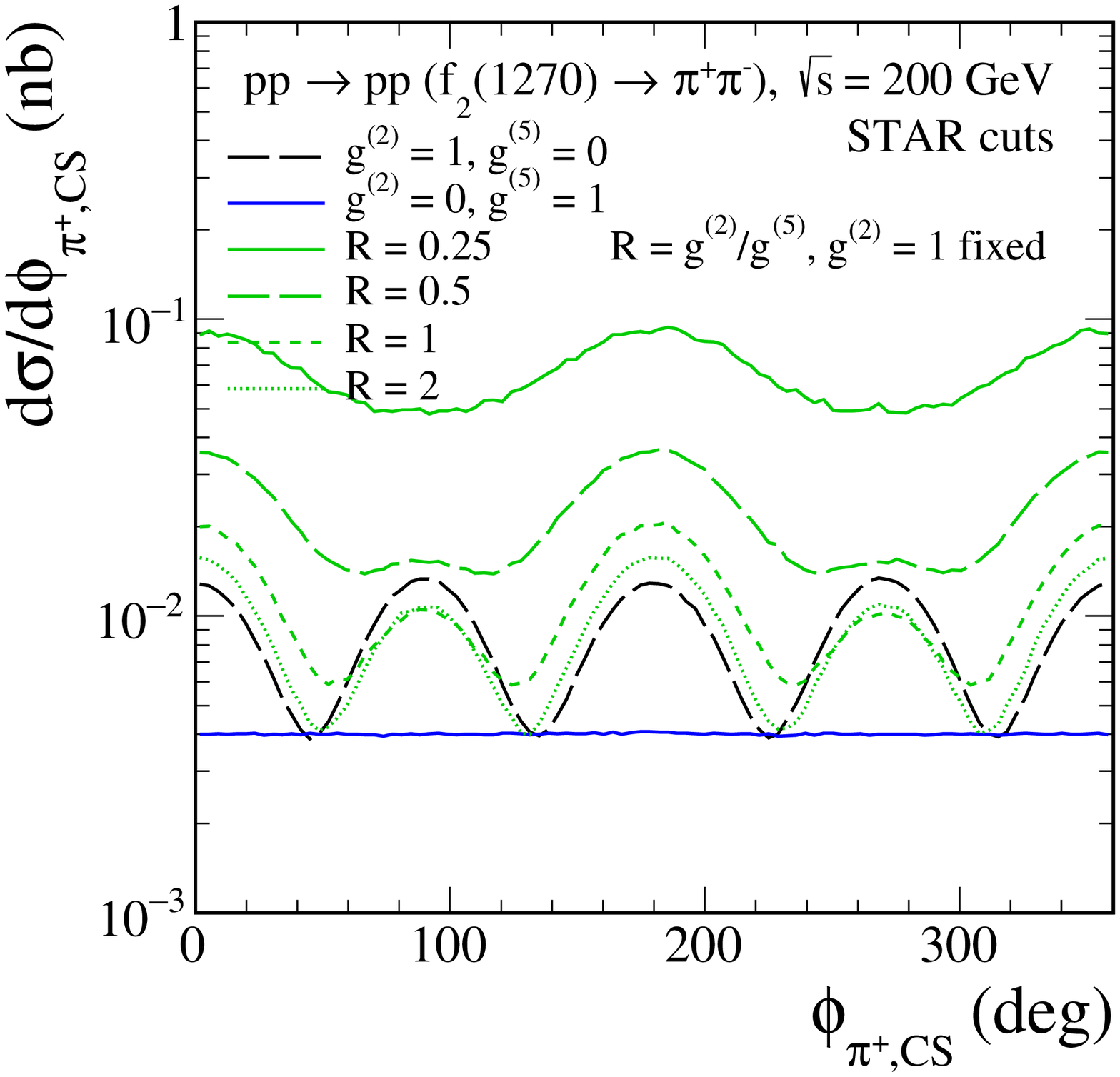}
\includegraphics[width=0.45\textwidth]{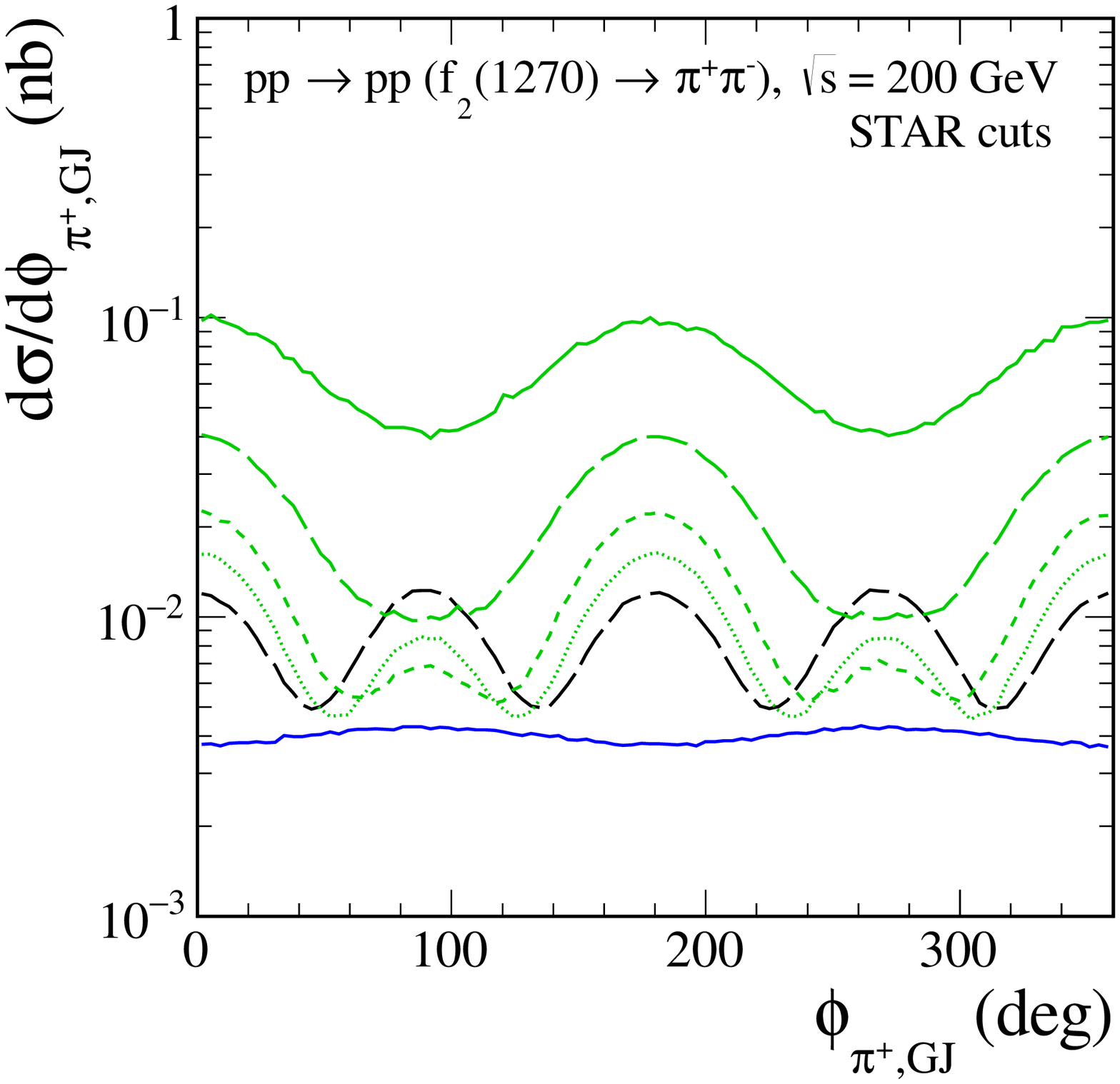}
\caption{\label{fig:STAR_g2g5}
\small
The distributions in azimuthal angle in the $\pi^+ \pi^-$ rest system 
using the CS frame (left panels) and the GJ frame (right panels), respectively.
The calculations were done for the STAR kinematics 
(see the caption of Fig.~\ref{fig:STAR}).
No absorption effects were included here.}
\end{figure}

\begin{figure}[!ht]
\includegraphics[width=0.4\textwidth]{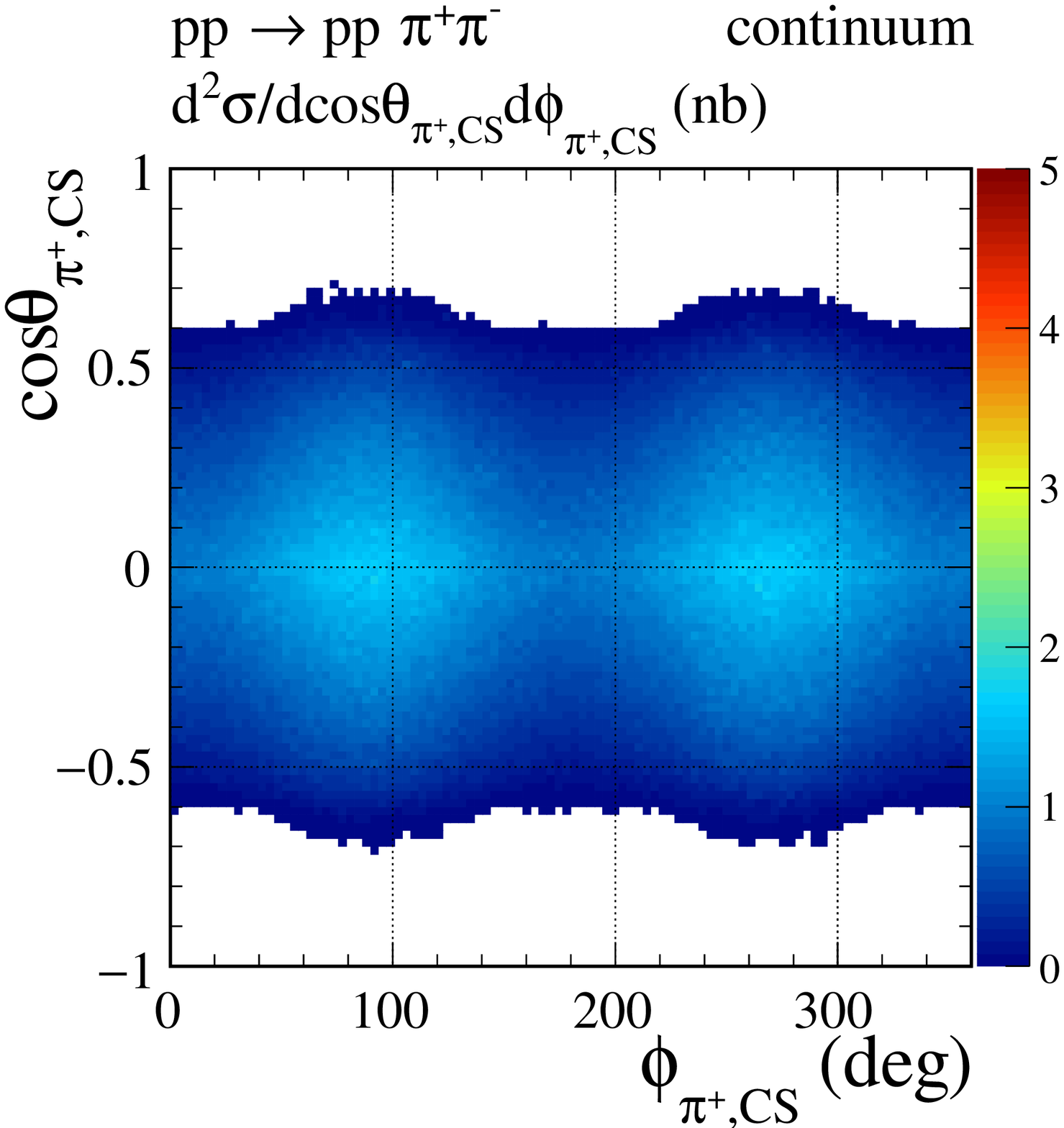}
\includegraphics[width=0.4\textwidth]{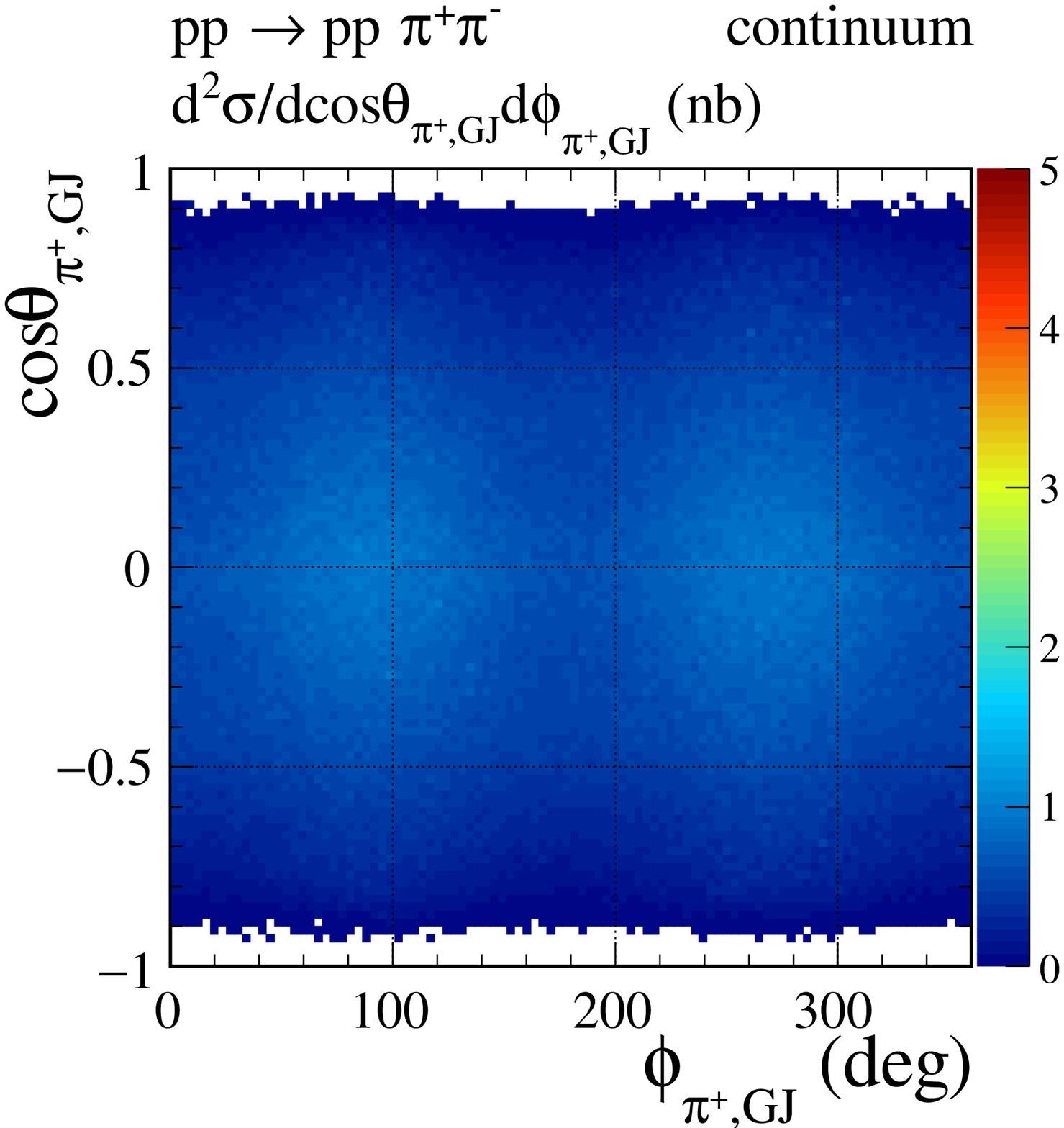}
\includegraphics[width=0.4\textwidth]{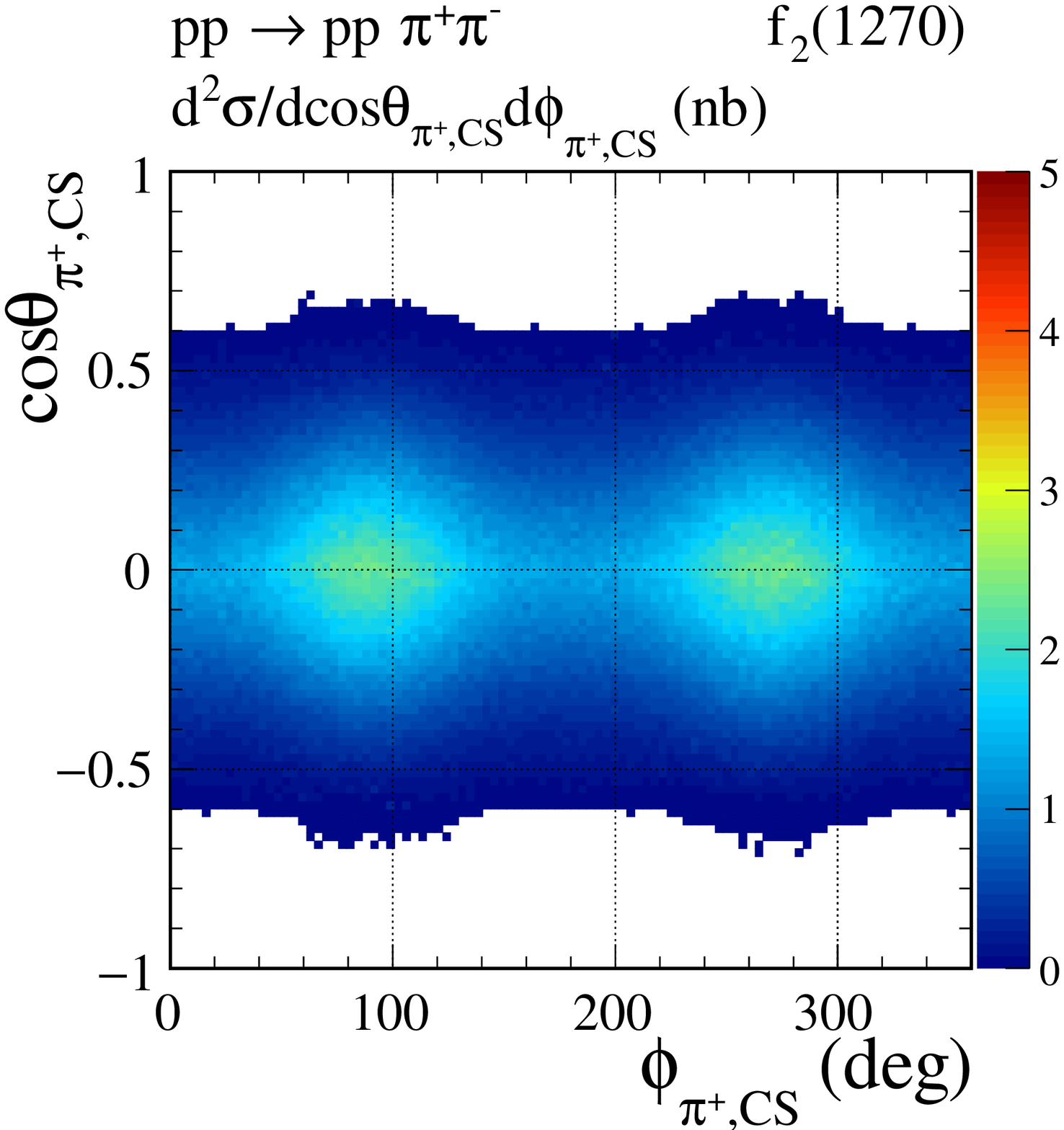}
\includegraphics[width=0.4\textwidth]{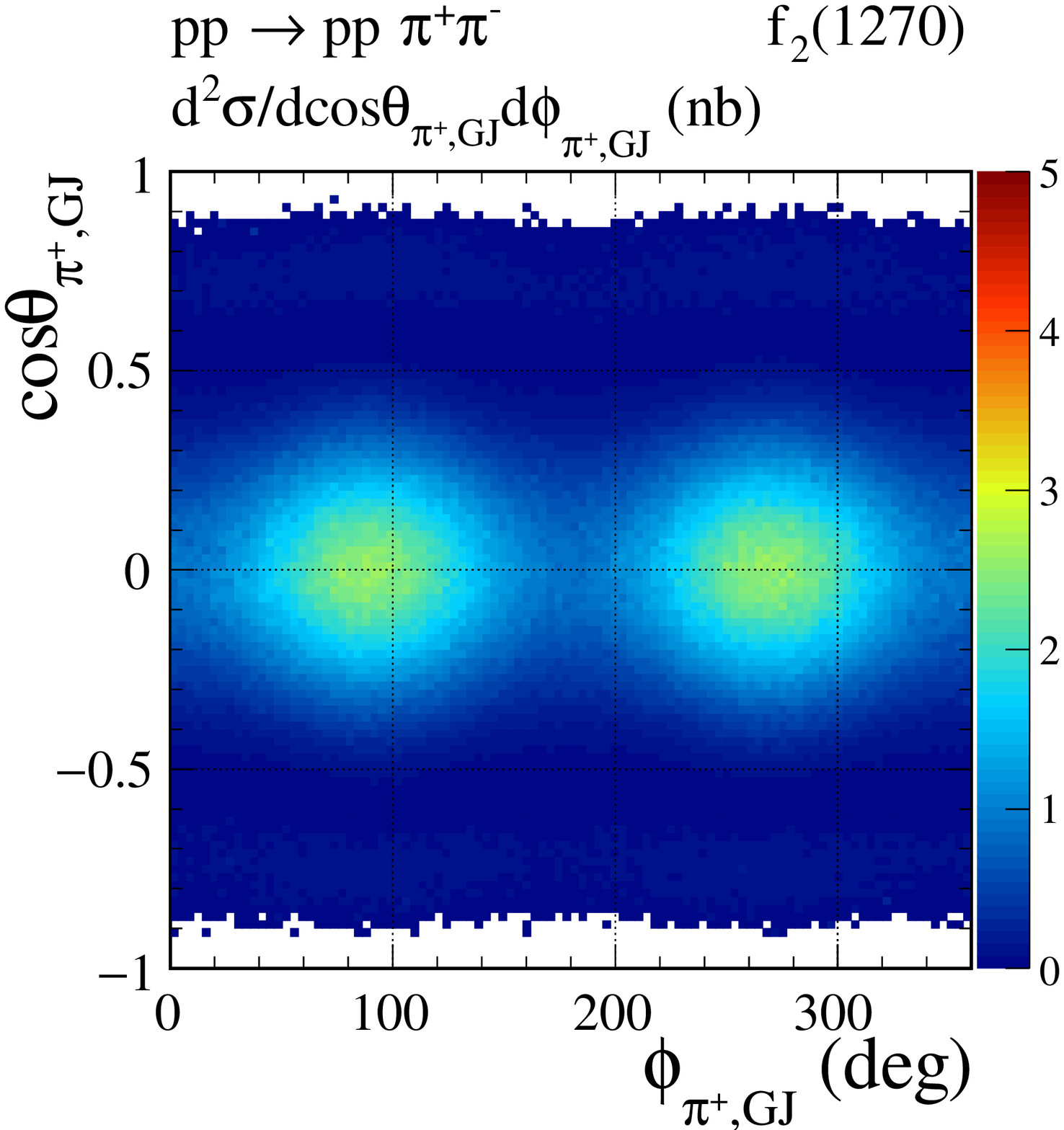}
\includegraphics[width=0.4\textwidth]{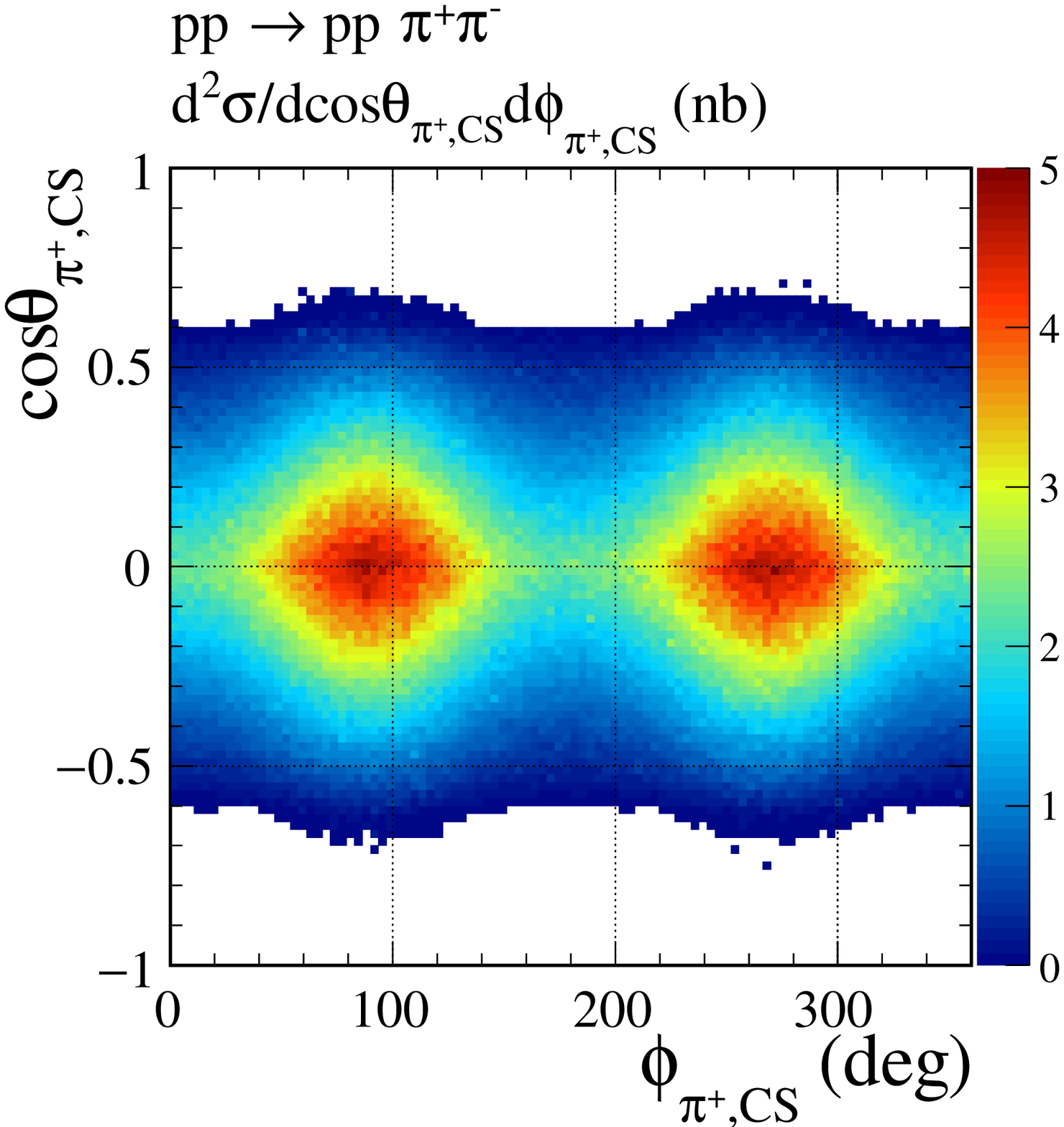}
\includegraphics[width=0.4\textwidth]{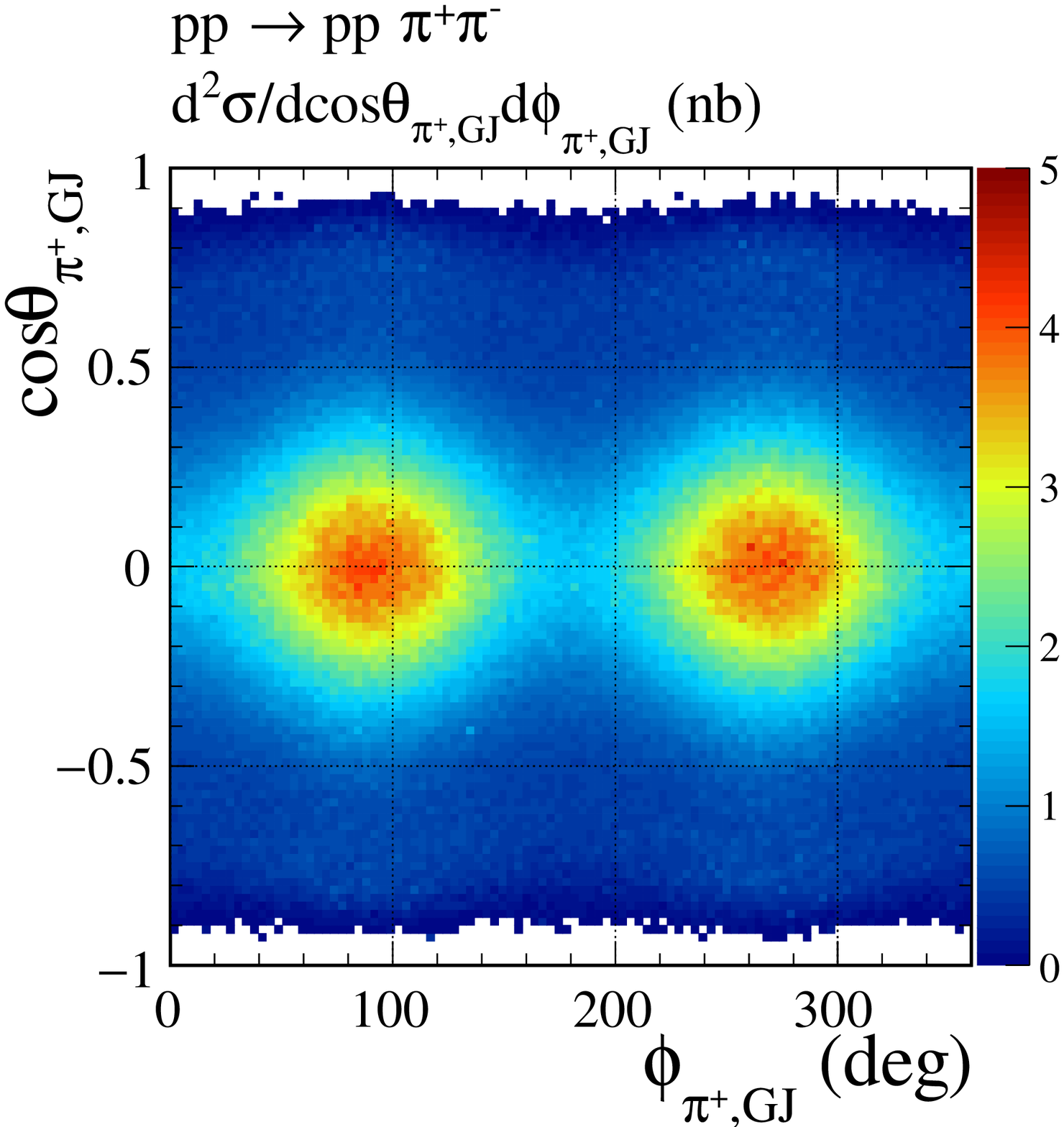}
\caption{\label{fig:1a}
\small
The distributions in ($\phi_{\pi^{+},\,{\rm CS}},\cos\theta_{\pi^{+},\,{\rm CS}}$) (the left panels) 
and in ($\phi_{\pi^{+},\,{\rm GJ}},\cos\theta_{\pi^{+},\,{\rm GJ}}$) (the right panels) for 
the $pp \to pp \pi^{+}\pi^{-}$ reaction.
The calculations were done in the dipion invariant mass region $M_{\pi^{+} \pi^{-}} \in (1.0,1.5)$~GeV
for $\sqrt{s} = 200$~GeV and the STAR experimental cuts from \cite{Sikora:2018cyk}:
$|\eta_{\pi}| < 0.7$, $p_{t, \pi} > 0.15$~GeV, and (\ref{STAR_cuts}).
In the top panels, we show results for the $\pi^{+}\pi^{-}$ continuum term,
in the center panels, for the $f_{2}(1270)$ resonance term (set A), and
in the bottom panels, for both the contributions added coherently.
Here we took $(g_{\Pom \Pom f_{2}}^{(2)}, g_{\Pom \Pom f_{2}}^{(5)}) = (-4.0, 16.0)$
as discussed in the main text.
The absorption effects are included here.}
\end{figure}

\begin{figure}[!ht]
\includegraphics[width=0.4\textwidth]{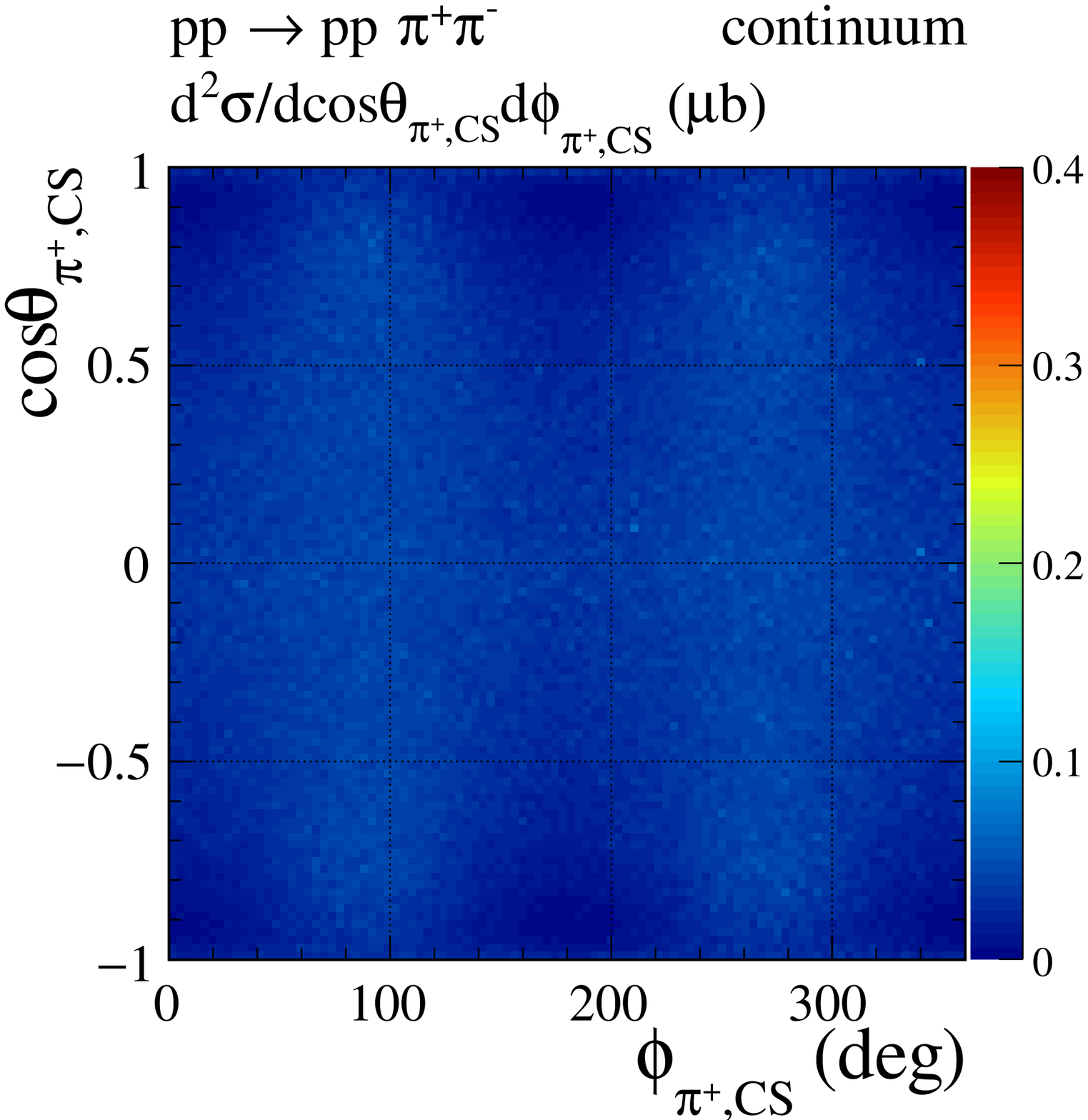}
\includegraphics[width=0.4\textwidth]{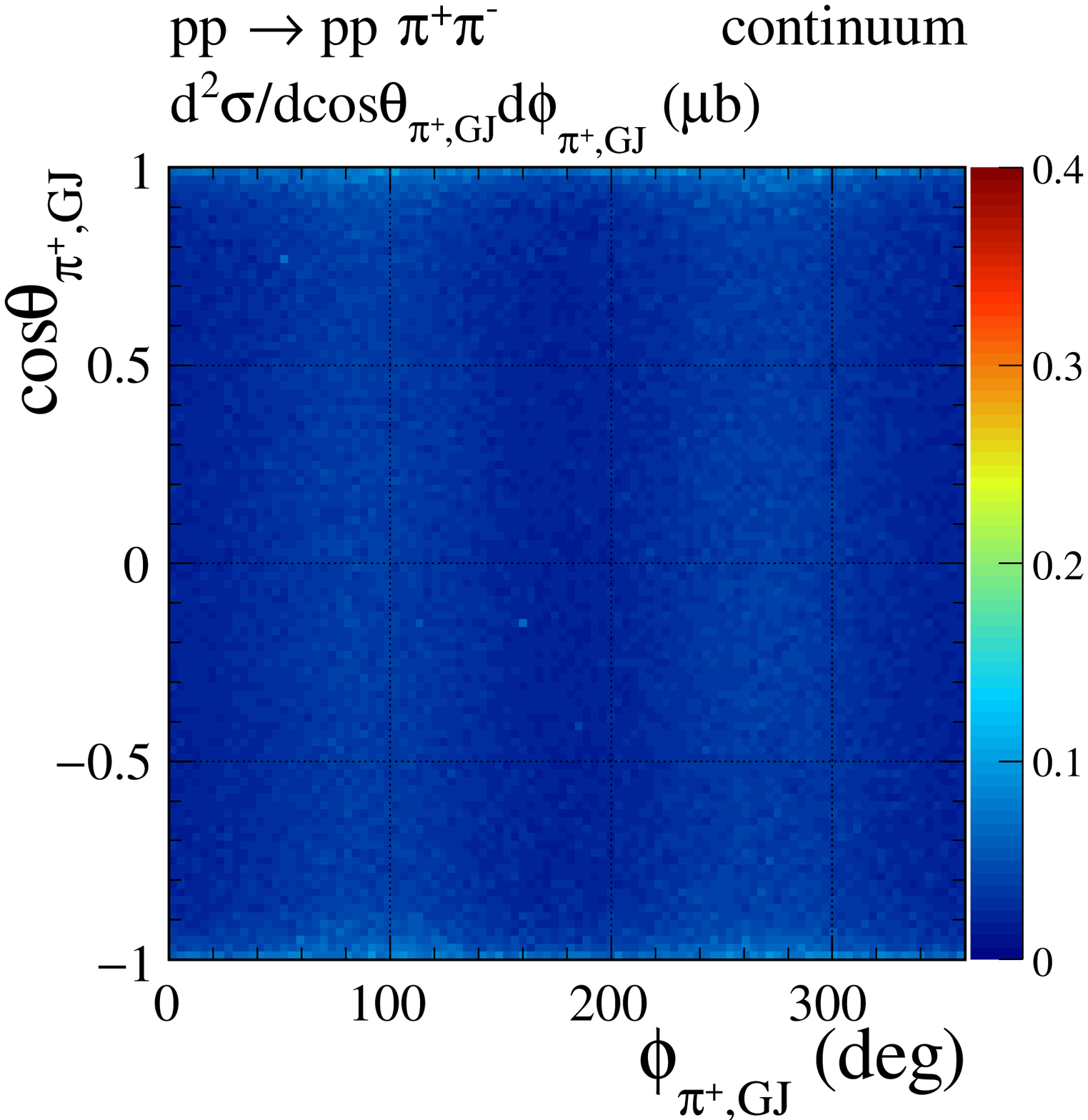}
\includegraphics[width=0.4\textwidth]{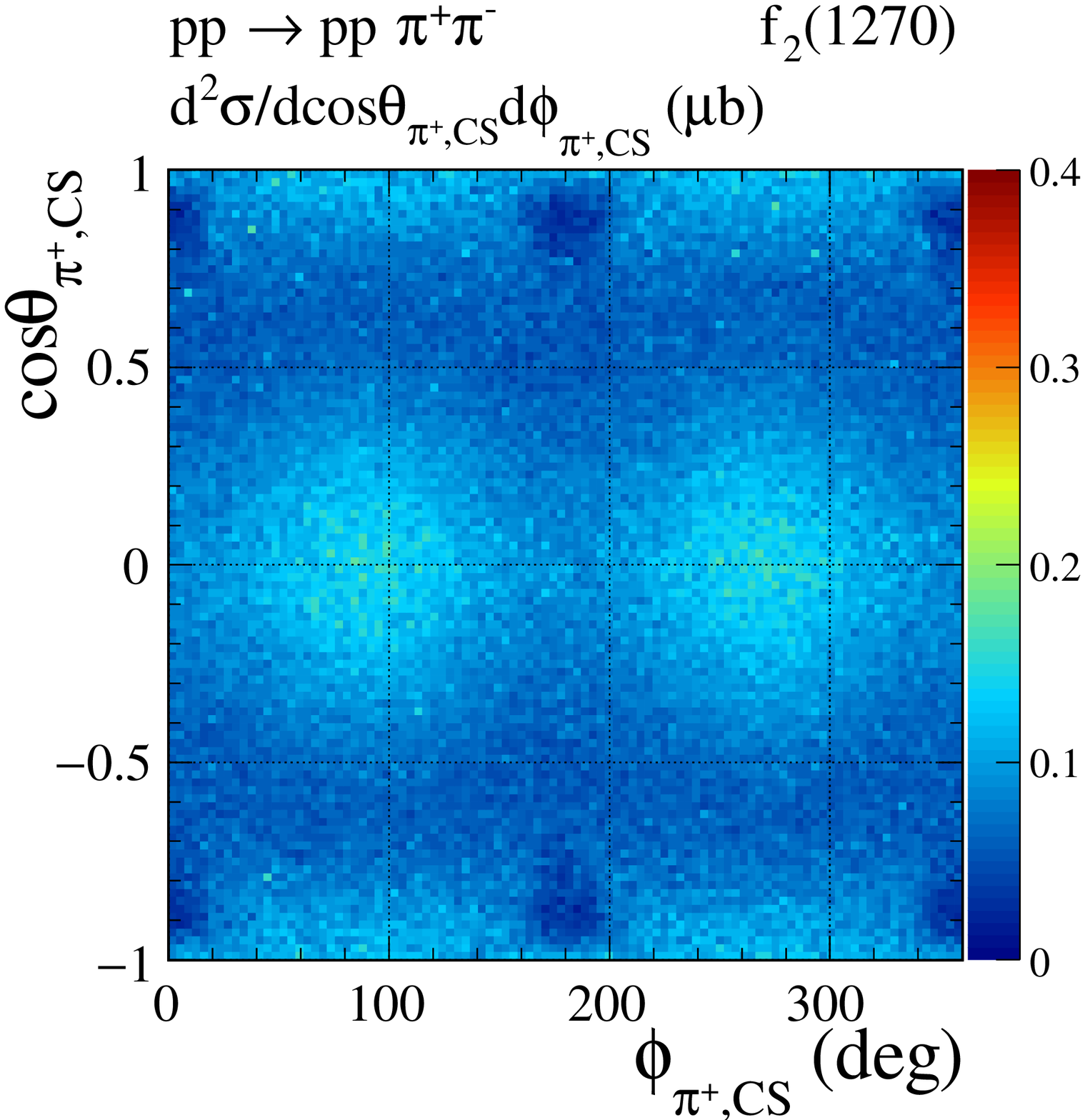}
\includegraphics[width=0.4\textwidth]{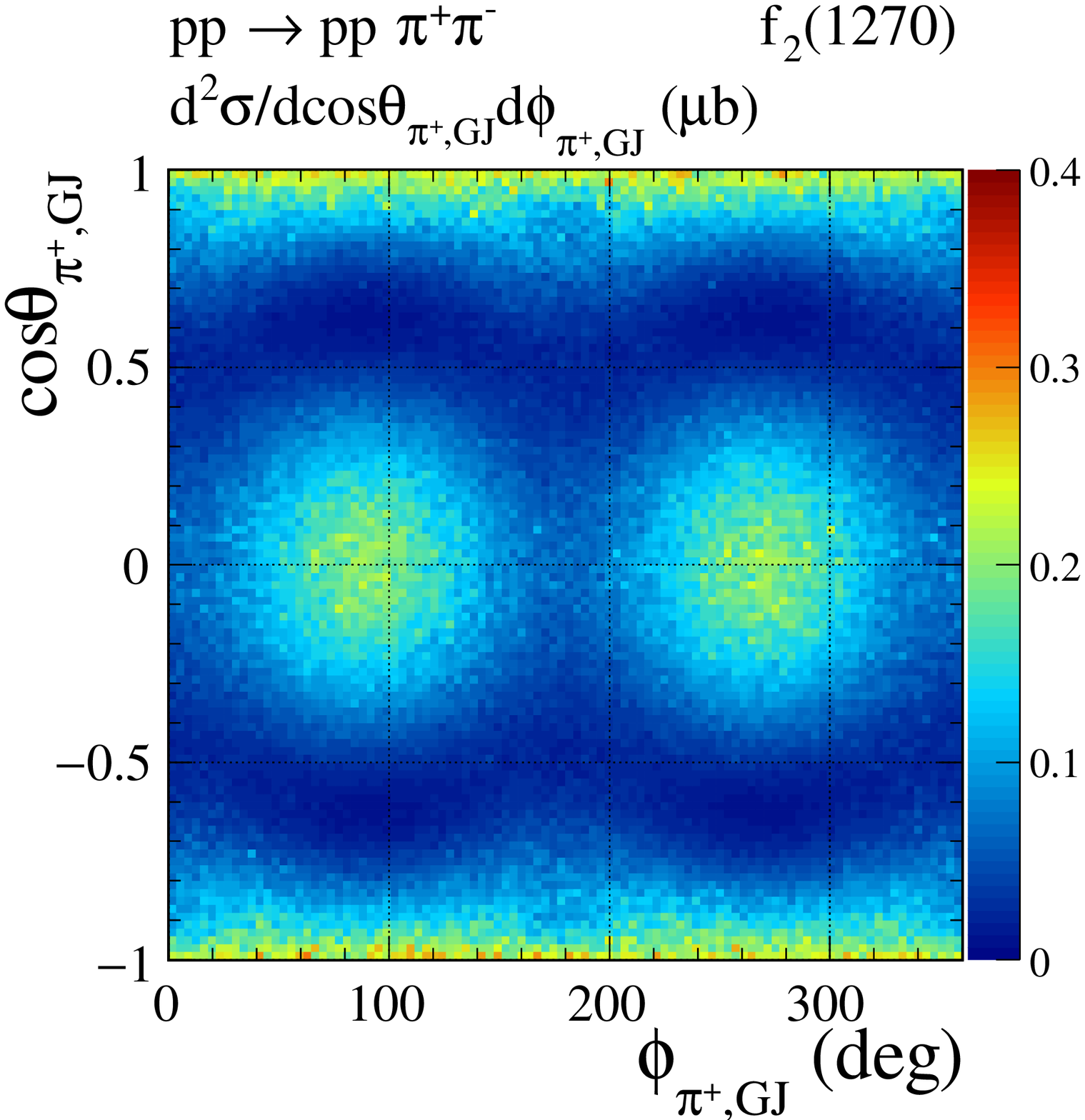}
\includegraphics[width=0.4\textwidth]{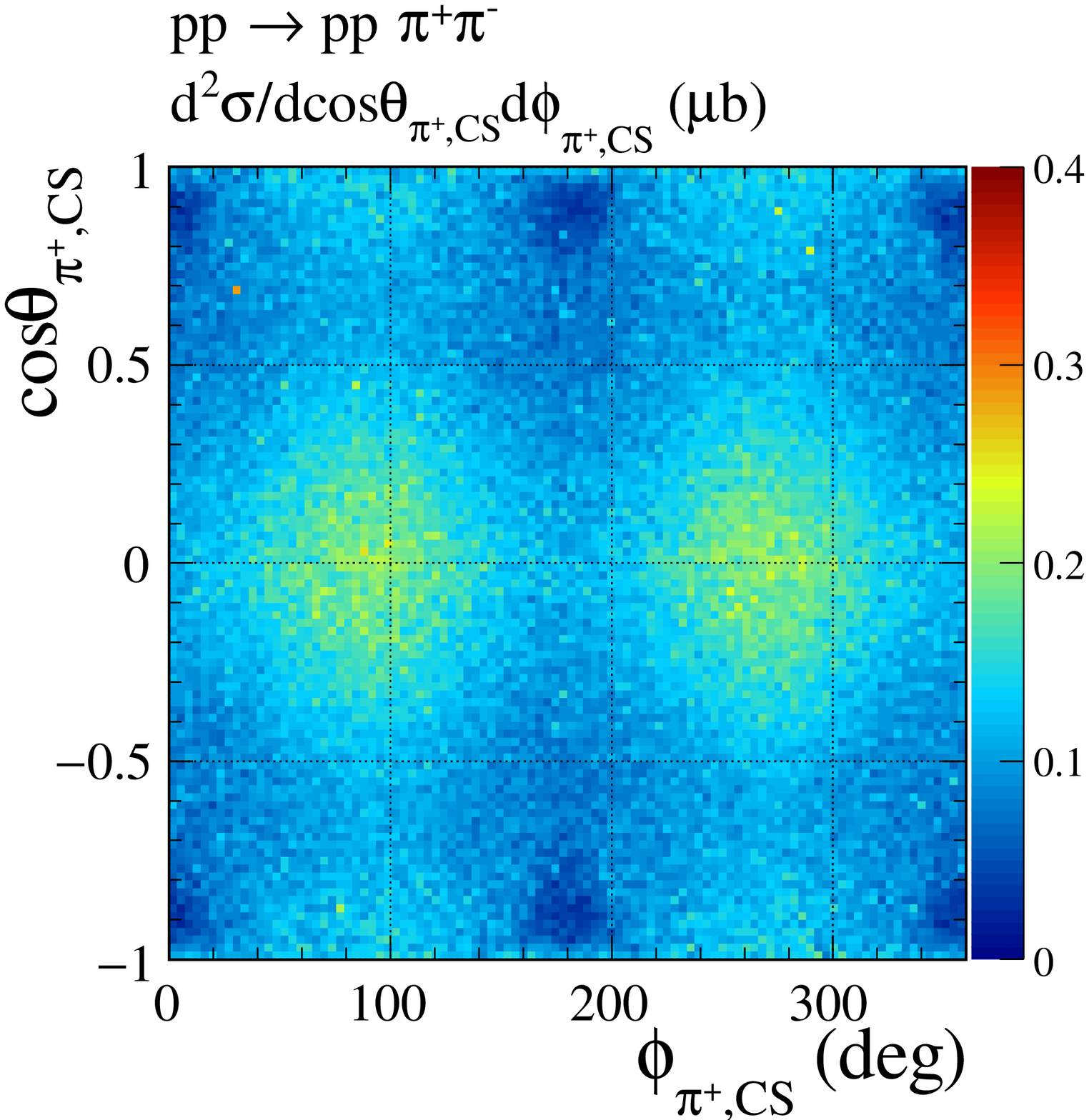}
\includegraphics[width=0.4\textwidth]{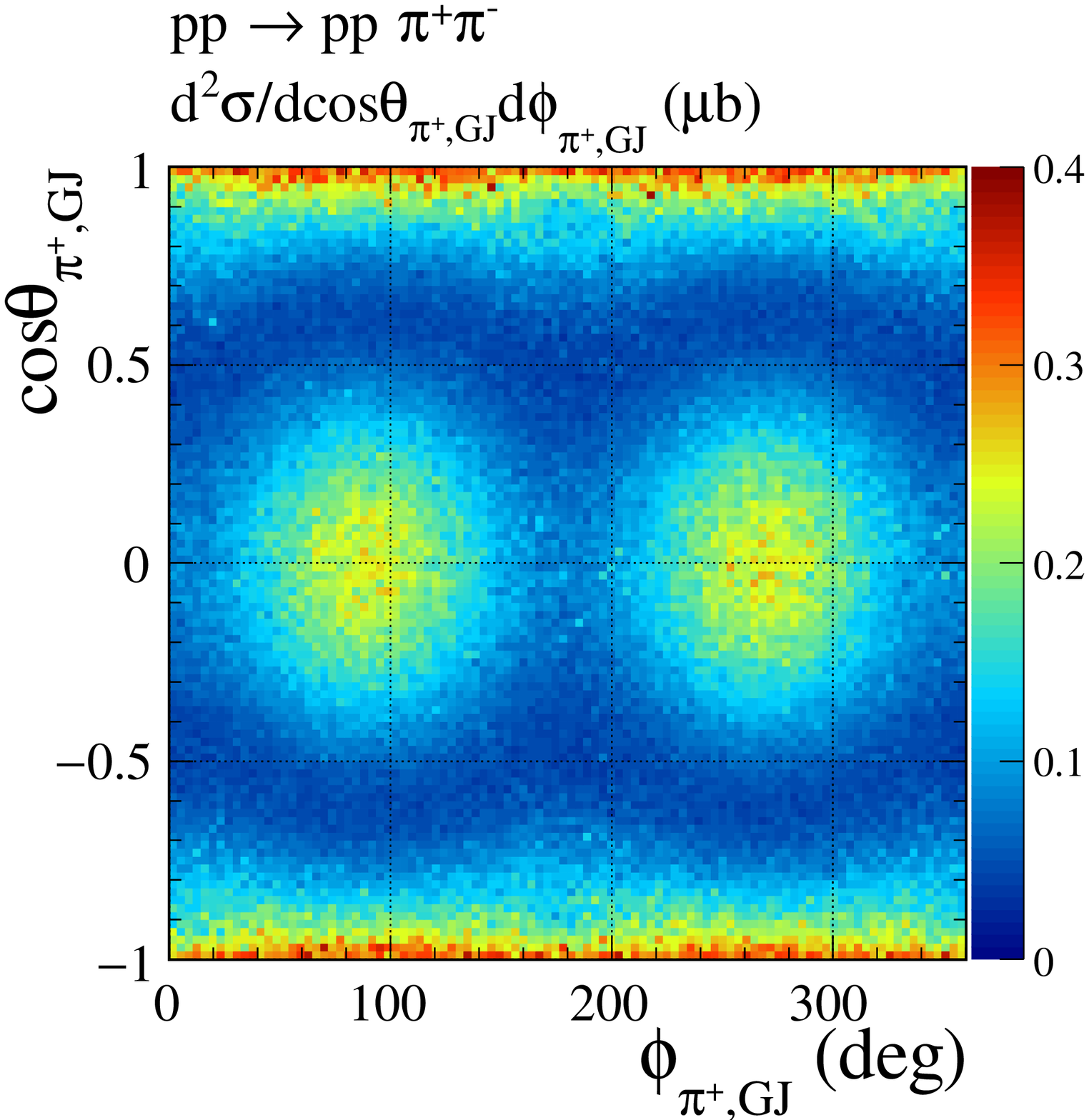}
\caption{\label{fig:1b}
\small
The same as in Fig.~\ref{fig:1a} but for $\sqrt{s} = 13$~TeV
and the ATLAS-ALFA experimental cuts:
$|\eta_{\pi}| < 2.5$, $p_{t, \pi} > 0.1$~GeV, and
0.17~GeV~$< |p_{y,p}| <$~0.50~GeV.
The calculations were done
in the dipion invariant mass region $M_{\pi^{+} \pi^{-}} \in (1.0,1.5)$~GeV.
The absorption effects are included here.}
\end{figure}

\begin{figure}[!ht]
\includegraphics[width=0.45\textwidth]{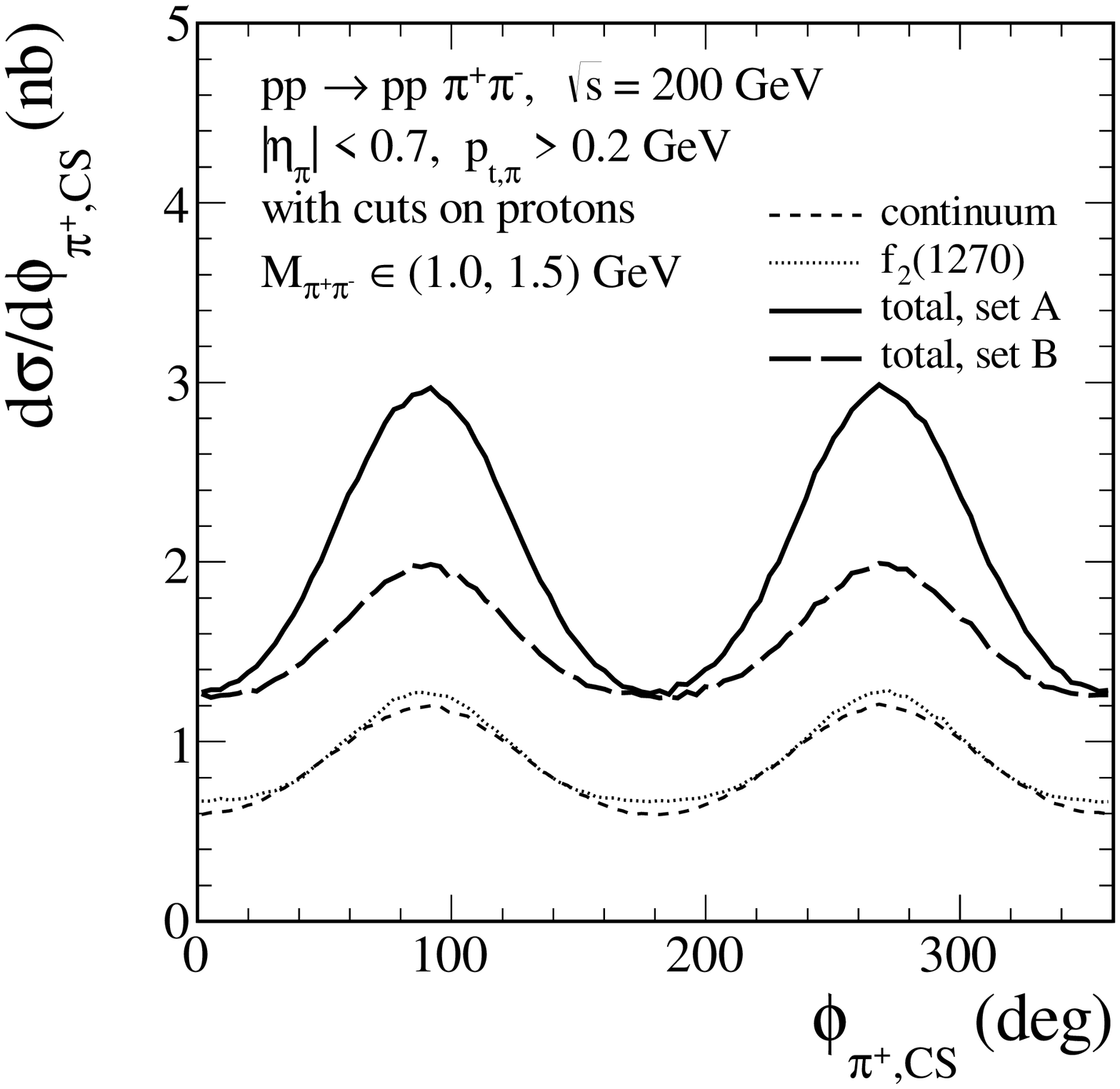}
\includegraphics[width=0.45\textwidth]{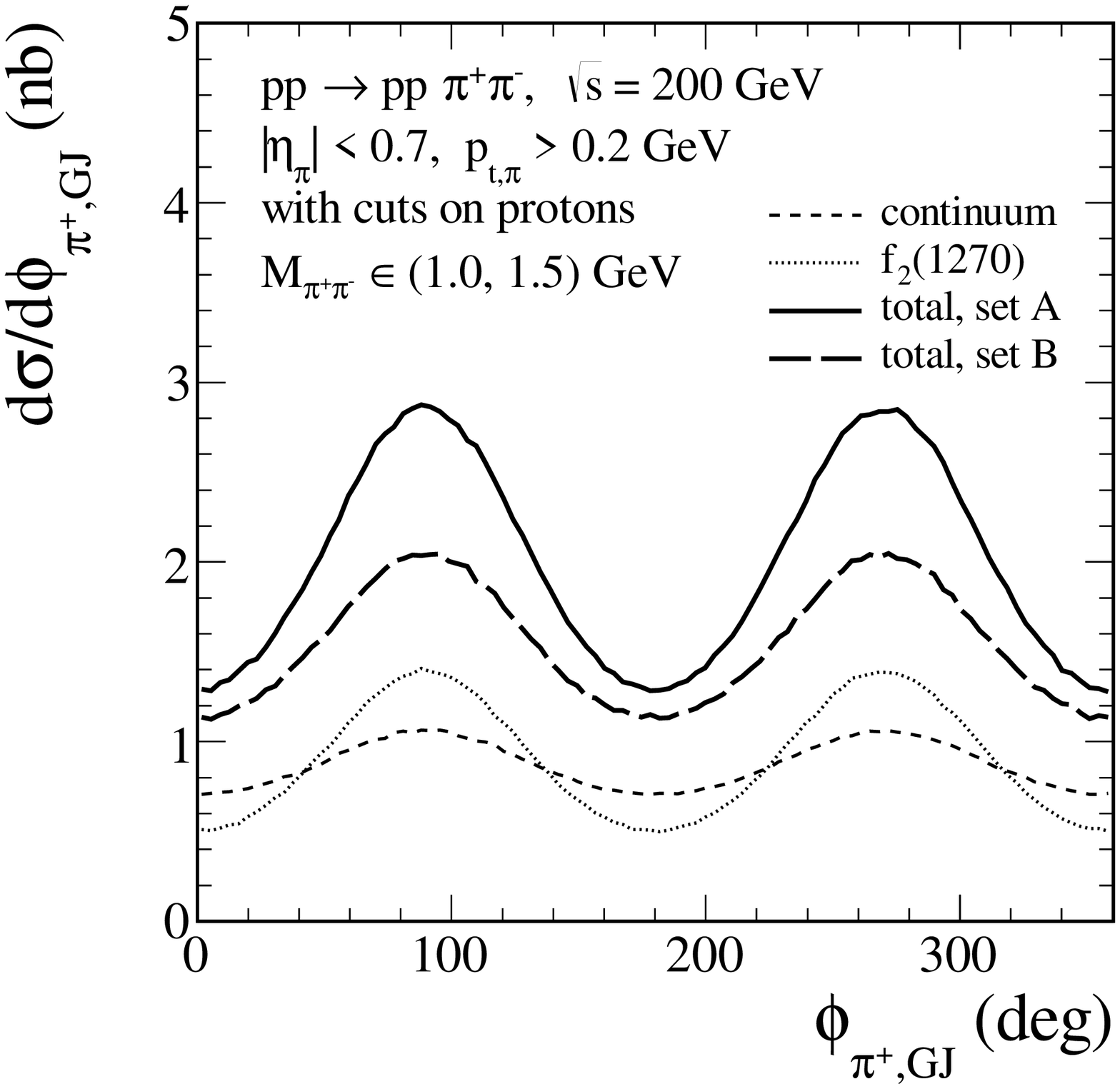}
\includegraphics[width=0.45\textwidth]{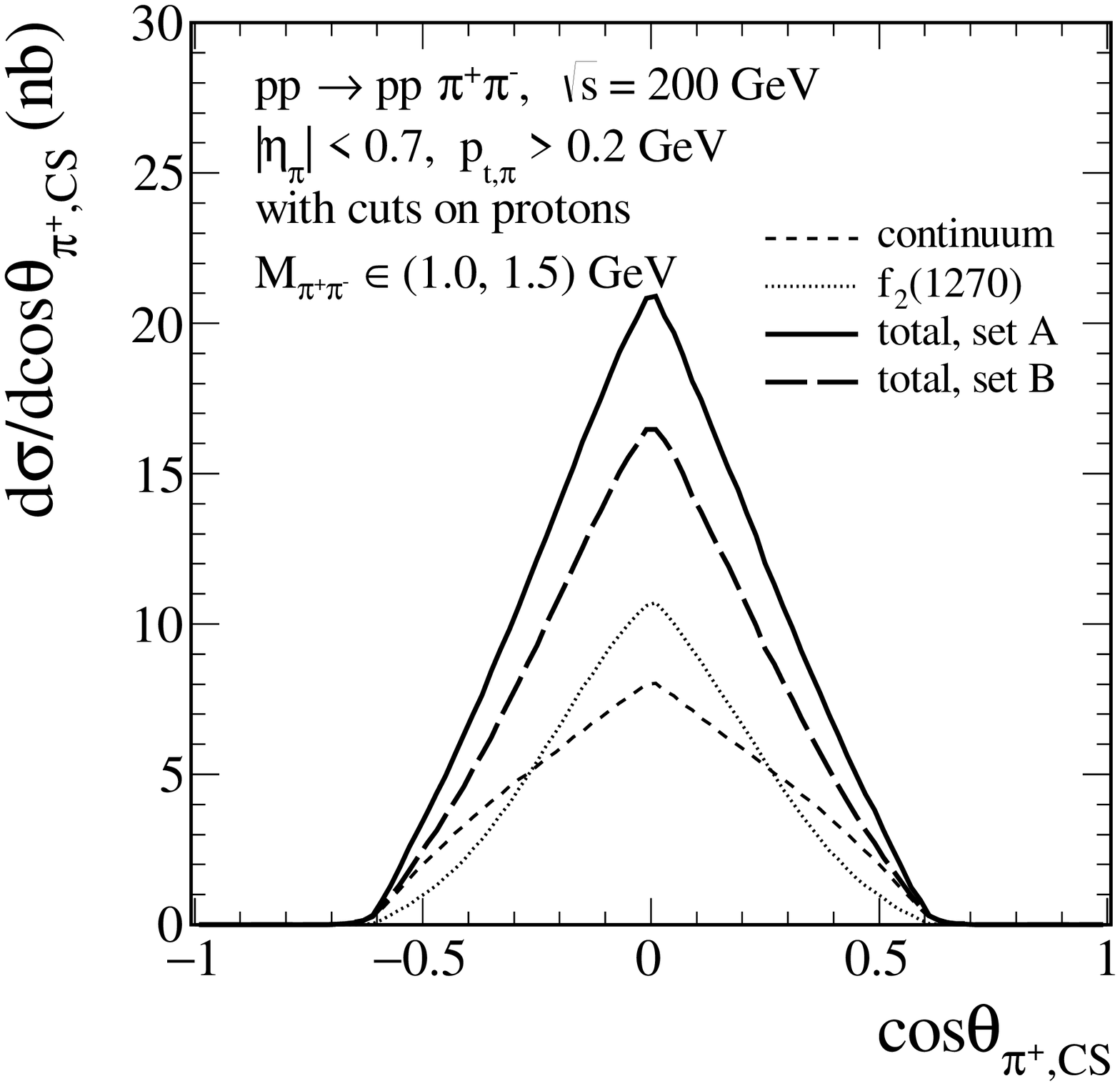}
\includegraphics[width=0.45\textwidth]{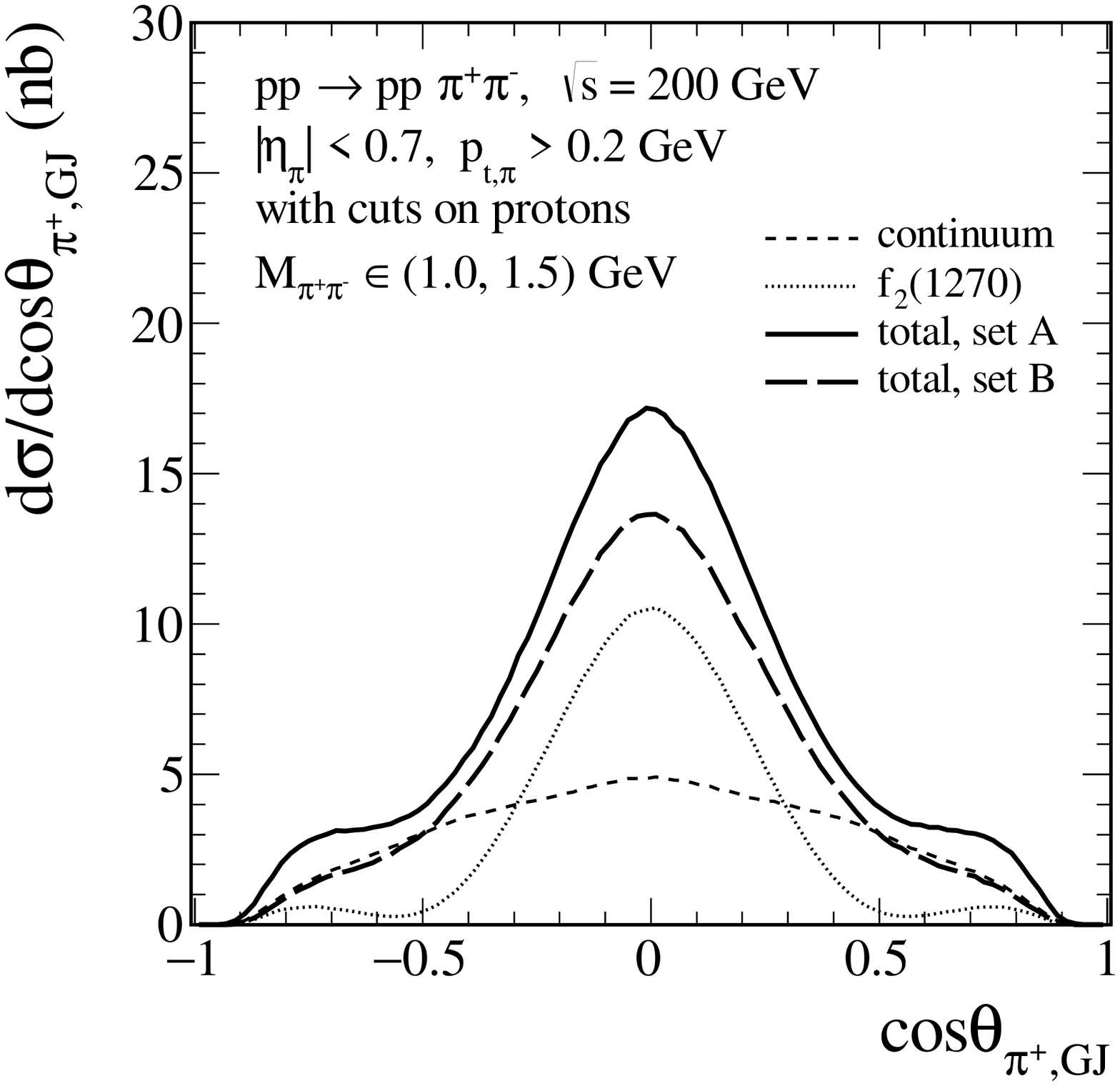}
\caption{\label{fig:2a}
\small
The angular distributions for the $pp \to pp \pi^{+}\pi^{-}$ reaction.
The calculations were done for $\sqrt{s} = 200$~GeV 
in the dipion invariant mass region $M_{\pi^{+} \pi^{-}} \in (1.0,1.5)$~GeV
and for the STAR experimental cuts specified in \cite{Sikora:2018cyk}.
The results for the $\pi^{+}\pi^{-}$ continuum term (the short-dashed line),
for the $f_{2}(1270)$ resonance term (the dotted line), and
for their coherent sum (the solid and long-dashed lines
corresponding to sets A and B, respectively) are presented.
We have taken here A $(g_{\Pom \Pom f_{2}}^{(2)}, g_{\Pom \Pom f_{2}}^{(5)}) = (-4.0, 16.0)$,
and B $(g_{\Pom \Pom f_{2}}^{(2)}, g_{\Pom \Pom f_{2}}^{(5)}) = (4.0, -16.0)$ as parameter sets.
The absorption effects are included here.}
\end{figure}

\begin{figure}[!ht]
\includegraphics[width=0.45\textwidth]{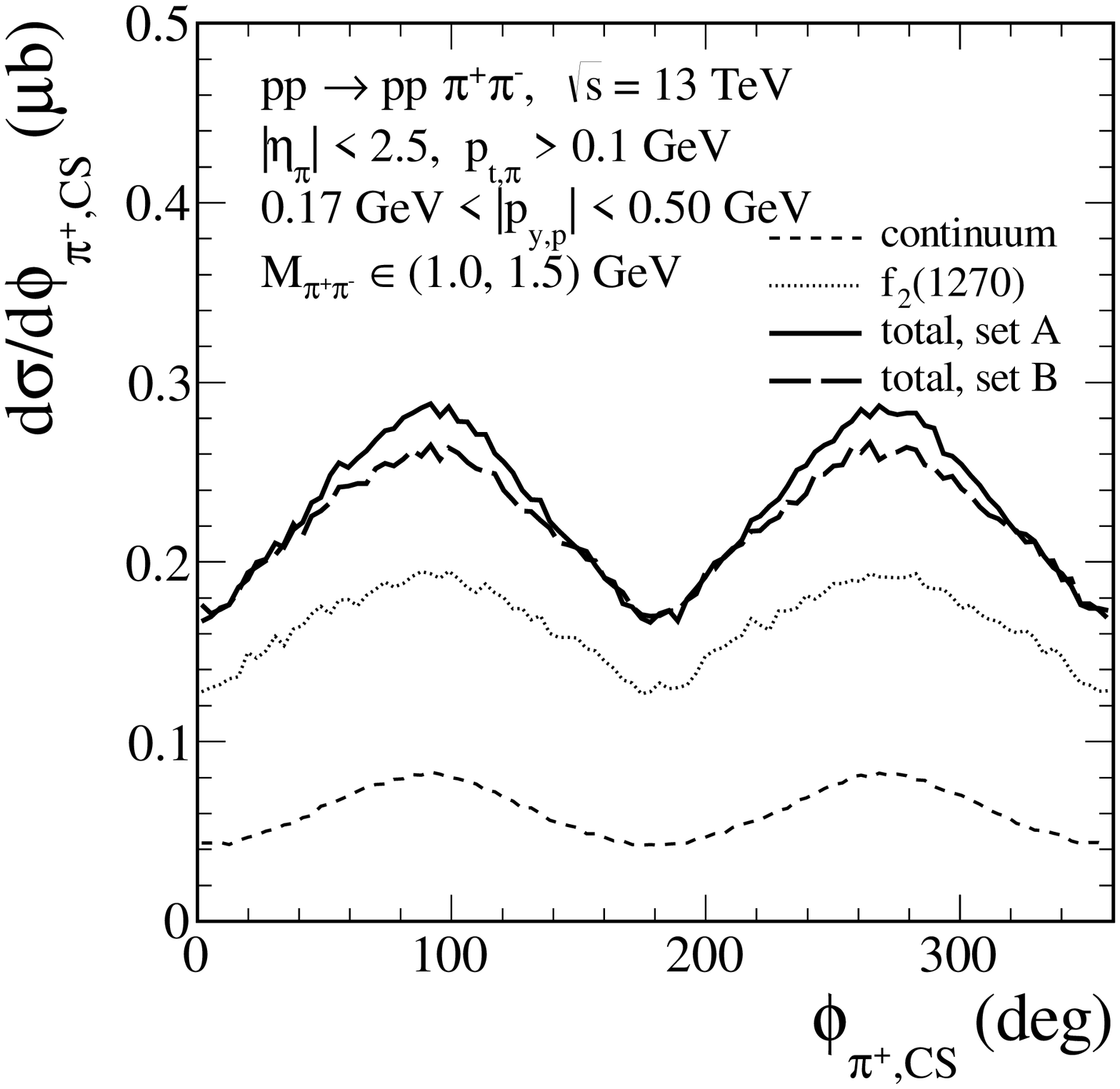}
\includegraphics[width=0.45\textwidth]{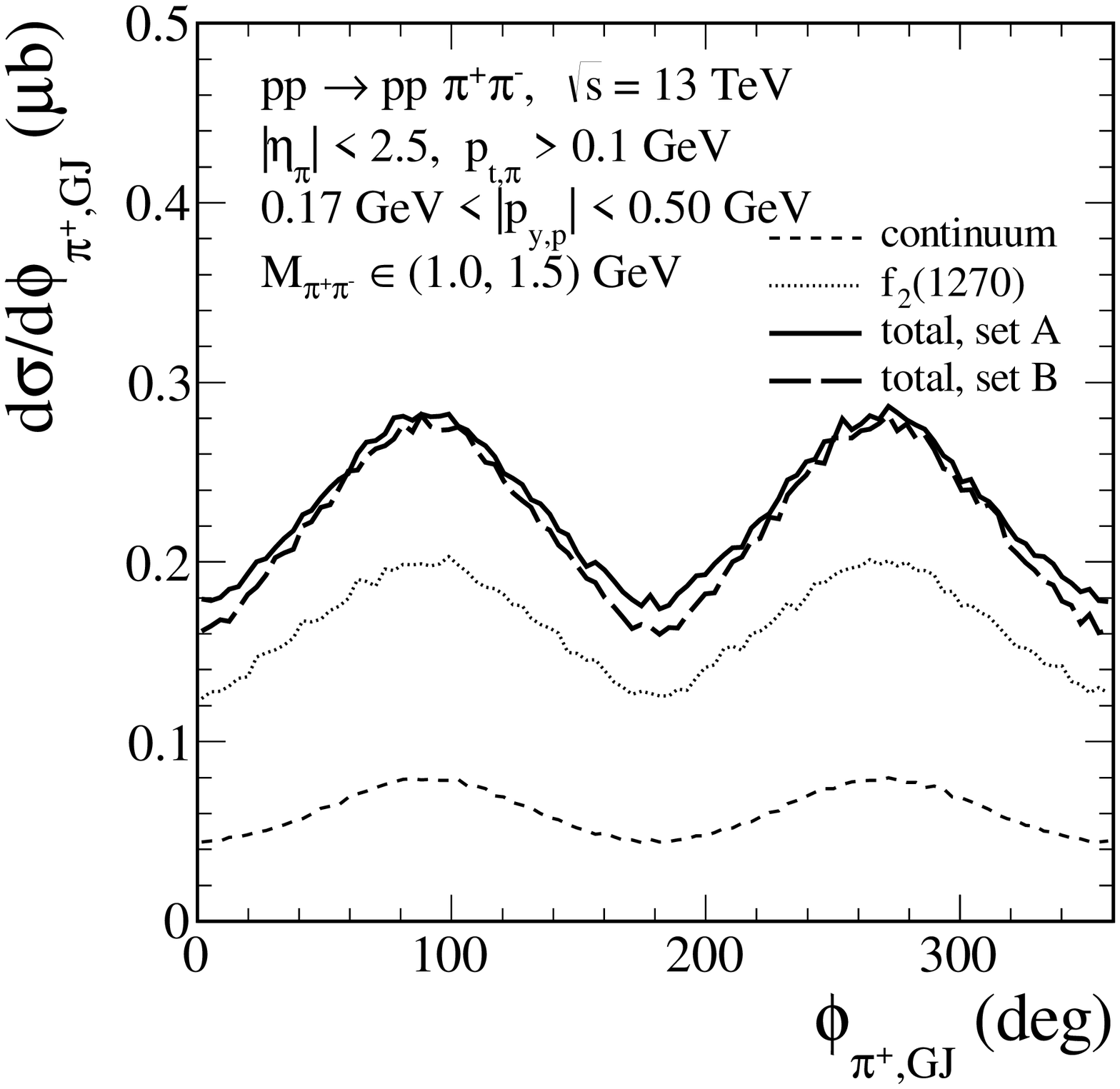}
\includegraphics[width=0.45\textwidth]{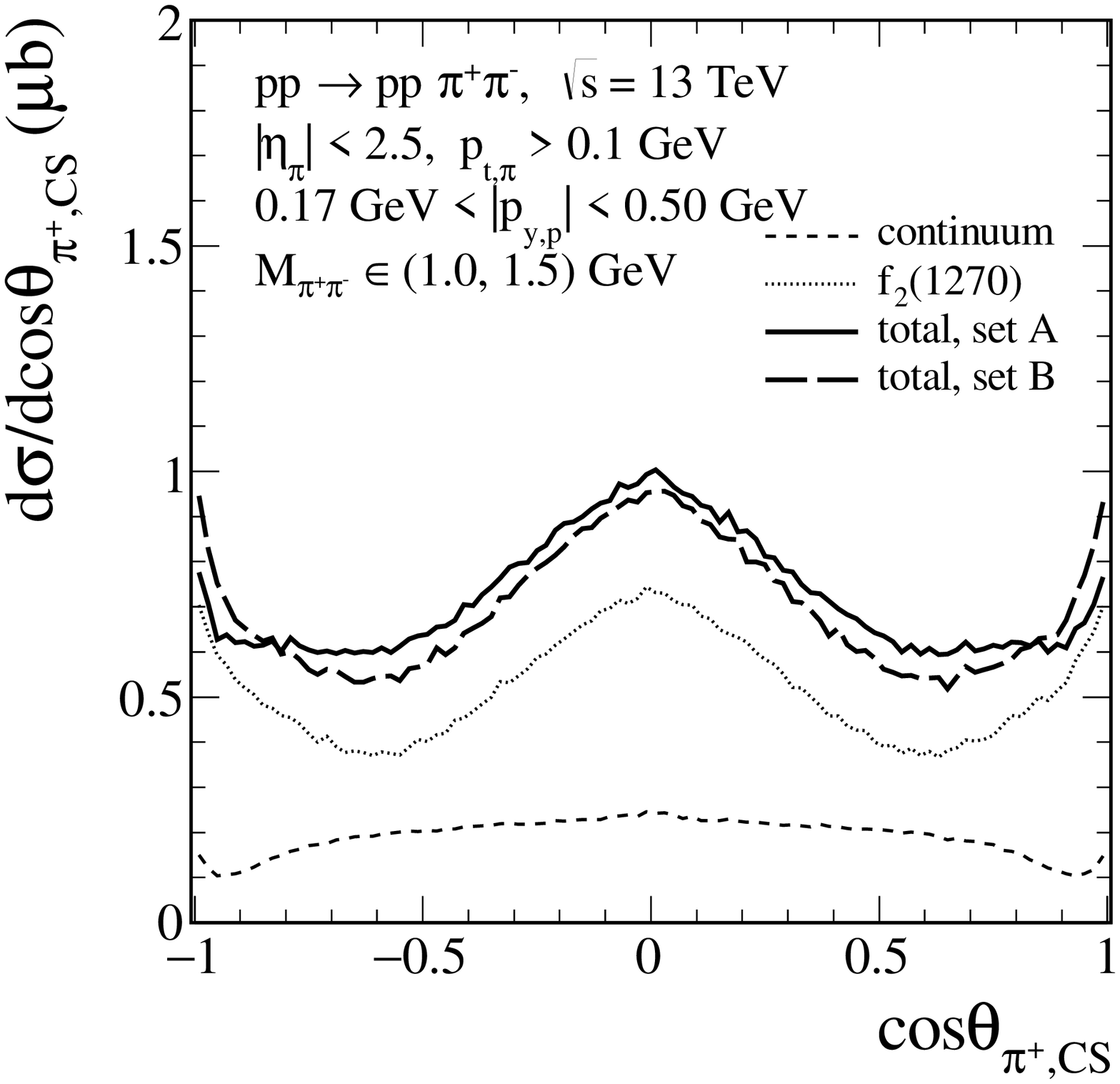}
\includegraphics[width=0.45\textwidth]{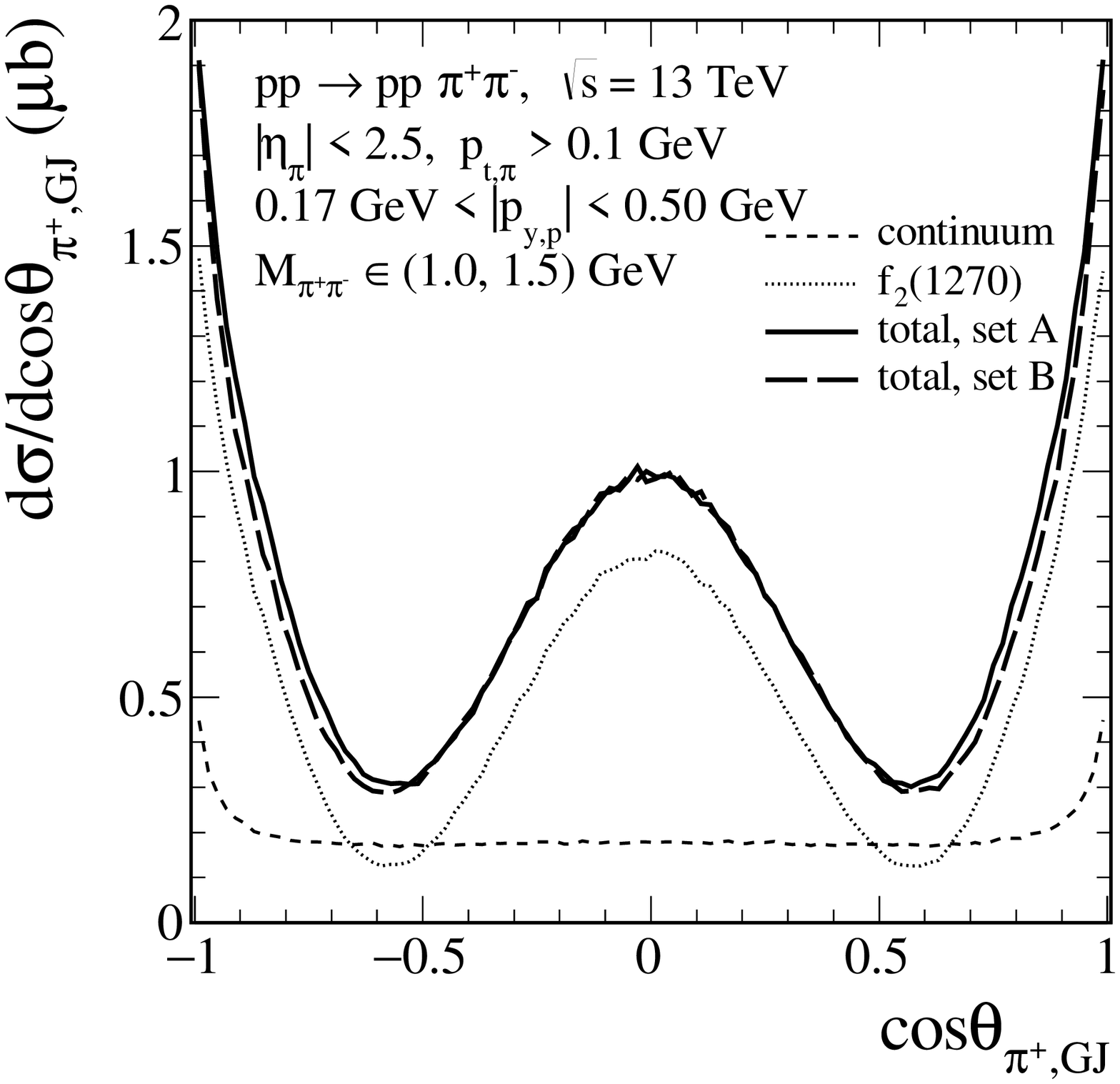}
\caption{\label{fig:2b}
\small
The same as in Fig.~\ref{fig:2a} but for the ATLAS-ALFA kinematics: 
$\sqrt{s} = 13$~TeV, $|\eta_{\pi}| < 2.5$, $p_{t, \pi} > 0.1$~GeV, and
0.17~GeV~$< |p_{y,p}|<$~0.50~GeV.
The calculations were done
in the dipion invariant mass region $M_{\pi^{+} \pi^{-}} \in (1.0,1.5)$~GeV.
The absorption effects are included here.}
\end{figure}

As discussed in the introduction, very good observables which can 
be used for visualizing the role of the $\Pom \Pom f_{2}$ couplings,
given by Eqs.~(\ref{A11})--(\ref{A17}) (cf. also Appendix~A of \cite{Lebiedowicz:2016ioh}),
could be the differential cross sections
$d\sigma/d\cos\theta_{\pi^{+}}$ and $d\sigma/d\phi_{\pi^{+}}$,
both in the CS and the GJ systems of reference;
see (\ref{CS}) and (\ref{GJ}), respectively.
In Figs.~\ref{fig:1}--\ref{fig:STAR_g2g5} and \ref{fig:1a}--\ref{fig:2b} 
we show such angular distributions 
for the $\pi^{+}$ meson in the $\pi^{+}\pi^{-}$ rest frame.

In Fig.~\ref{fig:1} we collected angular distributions for all (seven) 
independent $\Pom \Pom f_{2}(1270)$ couplings
for $\sqrt{s} = 13$~TeV, $p_{t, \pi} > 0.1$~GeV and 
for two different cuts on the pseudorapidities of the pions,
$|\eta_{\pi}| < 1.0$ (the top panels), and $|\eta_{\pi}| < 2.5$ 
(the bottom panels),
that will be measured in the LHC experiments.
In Fig.~\ref{fig:STAR} we show results for the STAR experimental conditions
with extra cuts on the leading protons, specified in \cite{Sikora:2018cyk},
\begin{eqnarray}
&&(p_{x,p} + 0.3 \;{\rm GeV})^{2} + p_{y,p}^{2} < 0.25\;{\rm GeV}^{2} \,, \nonumber \\
&&0.2\;{\rm GeV} < |p_{y,p}| < 0.4\;{\rm GeV}\,, \quad
p_{x,p}>-0.2\;{\rm GeV}\,.
\label{STAR_cuts}
\end{eqnarray}
Quite different distributions are obtained for different couplings.
Note that the shape of the angular distributions
depends on the coverage in $|\eta_{\pi}|$.
From the left top panel in Fig.~\ref{fig:1} we see that
the condition $|\eta_{\pi}| < 1.0$ leads to a reduction of
the cross sections mostly at $\cos\theta_{\pi^{+},\,{\rm CS}} \approx \pm 1$
compared to the results with $|\eta_{\pi}| < 2.5$ shown in the left bottom panel.
To our surprise, particularly interesting are the distributions 
in azimuthal angle.
The distributions for the resonance contribution alone can be approximated as 
\begin{eqnarray}
d\sigma / d\phi_{\pi^{+},\,{\rm CS}} \approx A \pm B \cos(n \,\phi_{\pi^{+},\,{\rm CS}})\,,
\label{dsig_dphi_approx}
\end{eqnarray} 
for $|\cos\theta_{\pi^{+},\,{\rm CS}}| < 0.5$ (as will be shown below),
where $A$ and $B$ depend on experimental conditions.
For most of the couplings $n = 2$ but for the $j = 2$ coupling it is $n = 4$.
The reader is asked to note the different number of oscillations
for the $j = 2$ coupling. 
The shape of $\phi_{\pi^{+},\,{\rm CS}}$ distributions depends also 
on the cuts on $|\eta_{\pi}|$.
Therefore, we expect these differences to be better visible when one compares
the results related to different regions of pion pseudorapidity.
Let us note that the LHCb Collaboration can measure $\pi^{+}\pi^{-}$ production
for $2.0<\eta_{\pi}<4.5$ \cite{Goncerz:2018xcw}.

In Fig.~\ref{fig:3} we show the two-dimensional distributions 
in ($\phi_{\pi^{+},\,{\rm CS}},\cos\theta_{\pi^{+},\,{\rm CS}}$)
for $\sqrt{s} = 13$~TeV and $|\eta_{\pi}| < 2.5$.
We can observe interesting structures for the $pp \to pp \pi^{+}\pi^{-}$ reaction.
We show results for the individual $\Pom \Pom f_{2}(1270)$ coupling terms
and for the continuum $\pi^{+}\pi^{-}$ production.
Different tensorial couplings generate very different patterns
which should be checked experimentally.

Some preliminary low-energy COMPASS results 
\cite{Austregesilo:2013yxa,Austregesilo:2014oxa}
suggest the presence of two maxima in the $\phi_{\pi^{+},\,{\rm GJ}}$ distribution.
So far there are no official analogous data for high-energy scattering
either from STAR or the LHC experiments.
Nevertheless we have asked ourselves the question if and how we can
get a similar structure 
(two maxima at $\phi_{\pi^{+},\,{\rm GJ}} = \pi/2$, $3/2\pi$)
in terms of our $\Pom \Pom f_{2}$ couplings (\ref{A11}) to (\ref{A17}).

In Fig.~\ref{fig:STAR_g2g5} we show the azimuthal angle distributions
using the CS (\ref{CS}) and the GJ (\ref{GJ}) frames.
Here we examine the combination of two $\Pom \Pom f_{2}$ couplings:
$j = 2$ (\ref{A12}) and $j = 5$ (\ref{A15}).
We show results for the individual $j = 2, 5$ coupling terms
and for their coherent sum.
For this purpose, we fixed the $j = 2$ coupling constant to
$g_{\Pom \Pom f_{2}}^{(2)} = 1.0$, and assumed
various values for $g_{\Pom \Pom f_{2}}^{(5)}$.
In the top and bottom panels, the red and green lines
correspond to the results when both couplings have opposite signs 
and the same signs, respectively.
Different interference patterns can be seen there
depending on the ratio of the two couplings,
$R = g_{\Pom \Pom f_{2}}^{(2)}/g_{\Pom \Pom f_{2}}^{(5)}$.

Now we discuss whether the absorption effects (the $pp$-rescattering corrections) 
may change the angular distributions discussed so far in the Born approximation.
We have checked that a slightly different size of absorption effects
may occur for the $j = 1, ..., 7$ resonant terms.
The absorption effects lead to a significant reduction of the cross section. 
However, the shapes of the polar and azimuthal angle distributions are practically not changed.
This indicates that the absorption effects should not disturb the determination
of the type of the $\Pom \Pom f_{2}(1270)$ coupling. 
However, the continuum and the resonant terms may
be differently affected by absorption.
This will have to be taken into account when one tries to extract the strengths
of the couplings from such distributions.

The measurement of forward protons would be useful
to better understand absorption effects.
The {\tt GenEx} Monte Carlo generator \cite{Kycia:2014hea,Kycia:2017ota}
could be used in this context.
We refer the reader to \cite{Kycia:2017iij} where a first calculation
of four-pion continuum production in the $pp \to pp \pi^{+}\pi^{-}\pi^{+}\pi^{-}$ reaction
with the help of the {\tt GenEx} code was performed.

Clearly, by a comparison of our model results to high-energy experimental data
we shall be able to determine or at least set limits on the parameters 
of the $\Pom \Pom f_{2}(1270)$ coupling.
At the moment, however, this is not yet possible since only
some, mostly preliminary, experimental distributions were presented
\cite{Adamczyk:2014ofa,Aaltonen:2015uva,Khachatryan:2017xsi,Sikora:2018cyk,CMS:2019qjb}.

\begin{figure}[!ht]
(a)\includegraphics[width=0.4\textwidth]{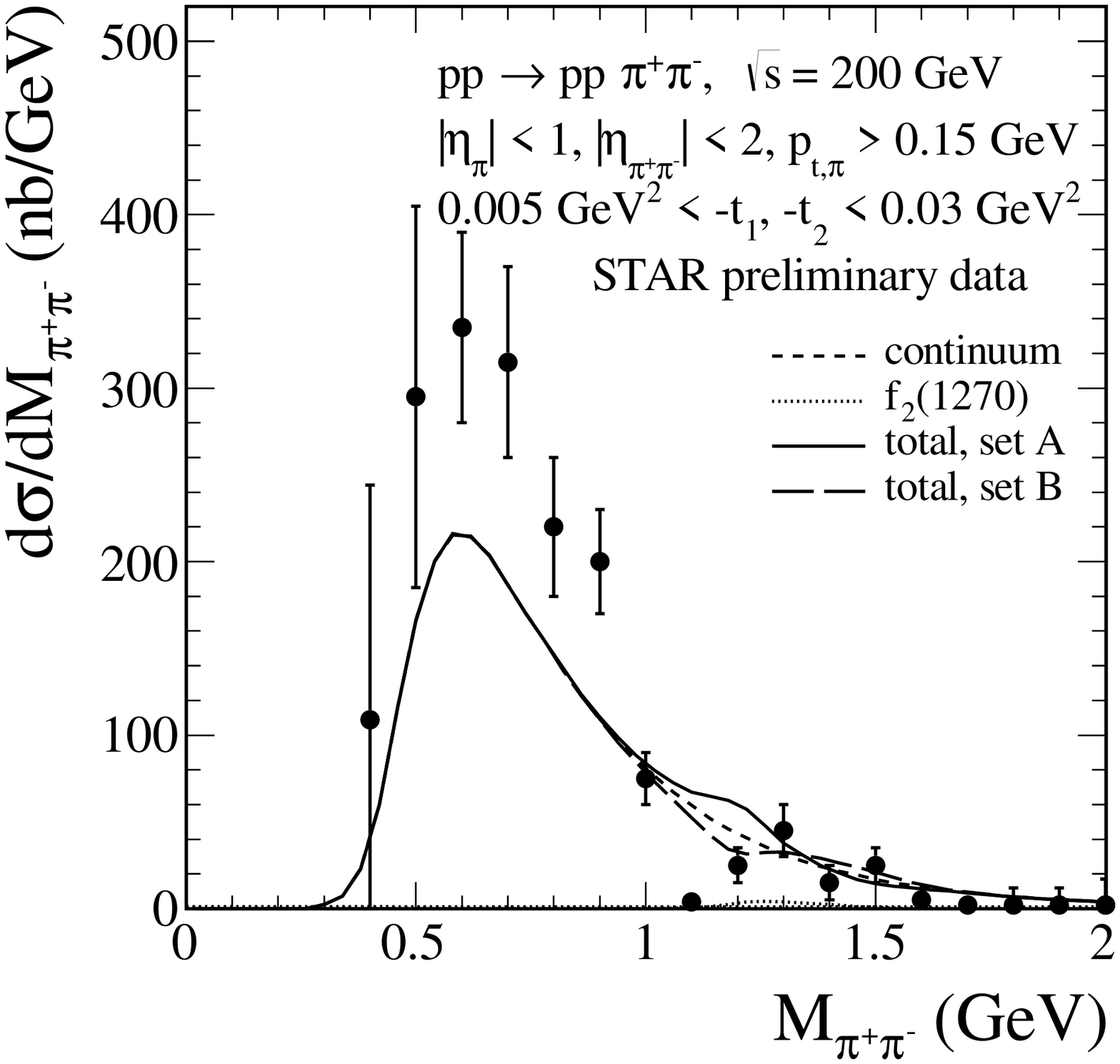}
(b)\includegraphics[width=0.4\textwidth]{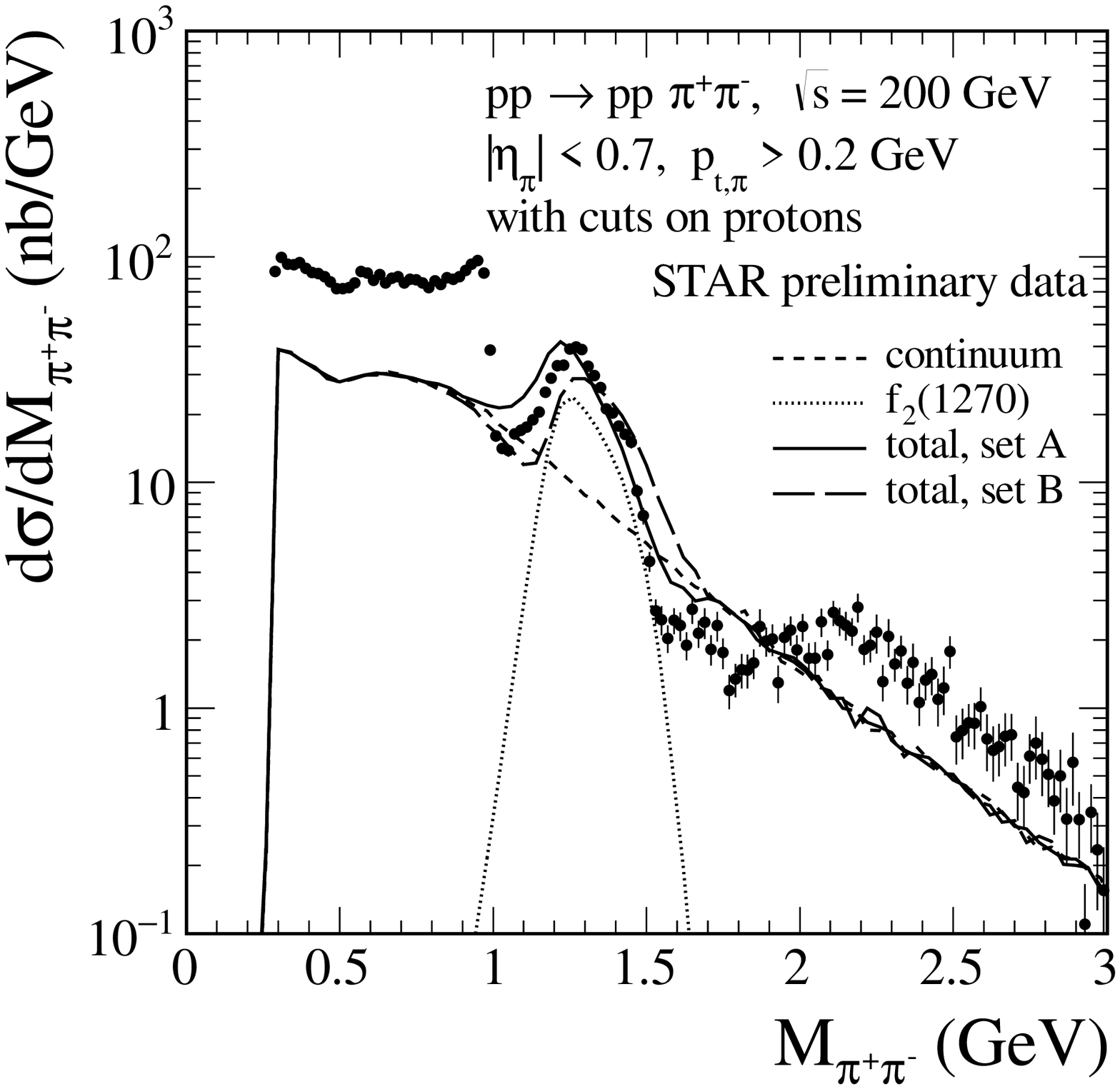}
(c)\includegraphics[width=0.4\textwidth]{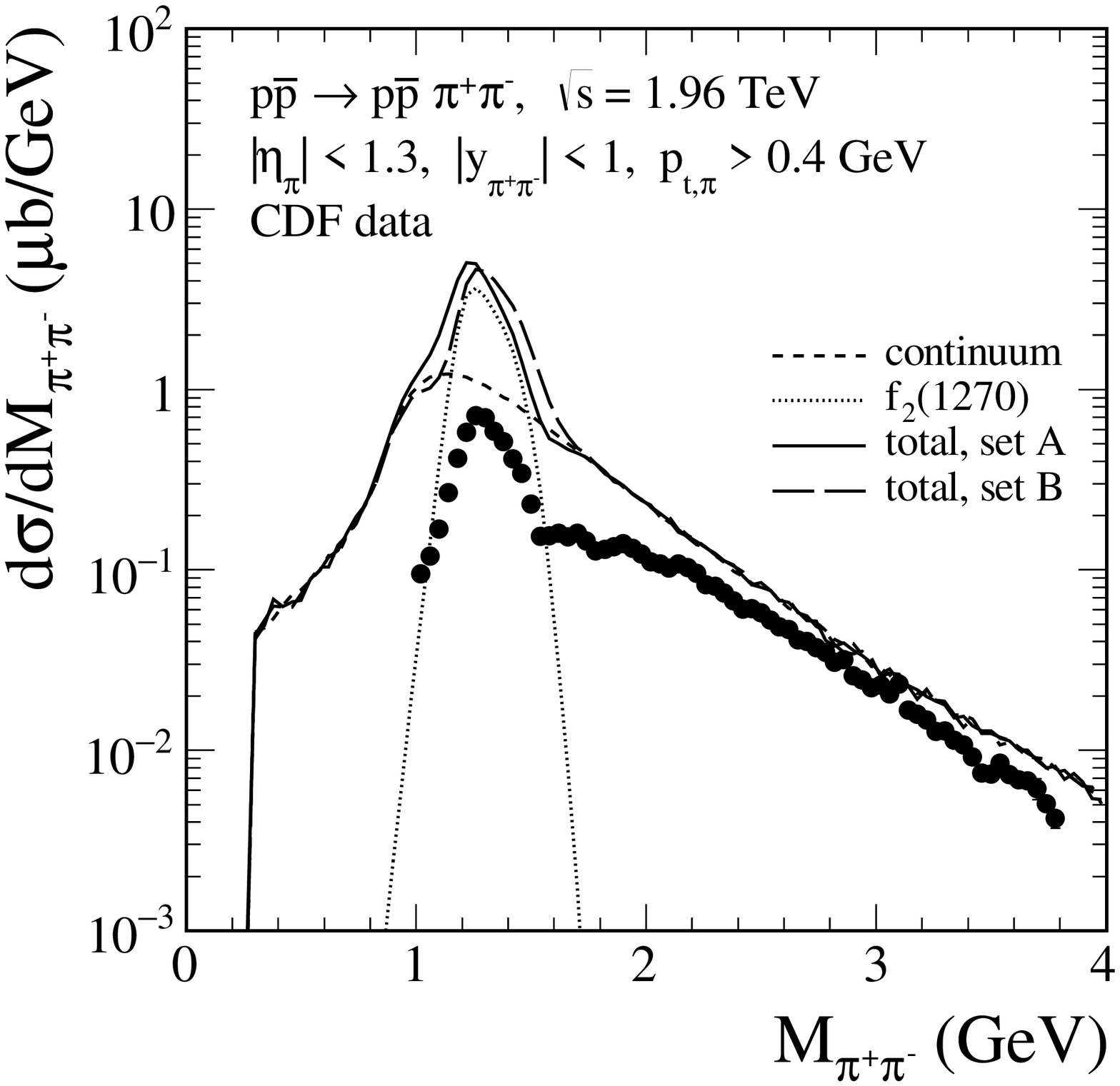}
(d)\includegraphics[width=0.4\textwidth]{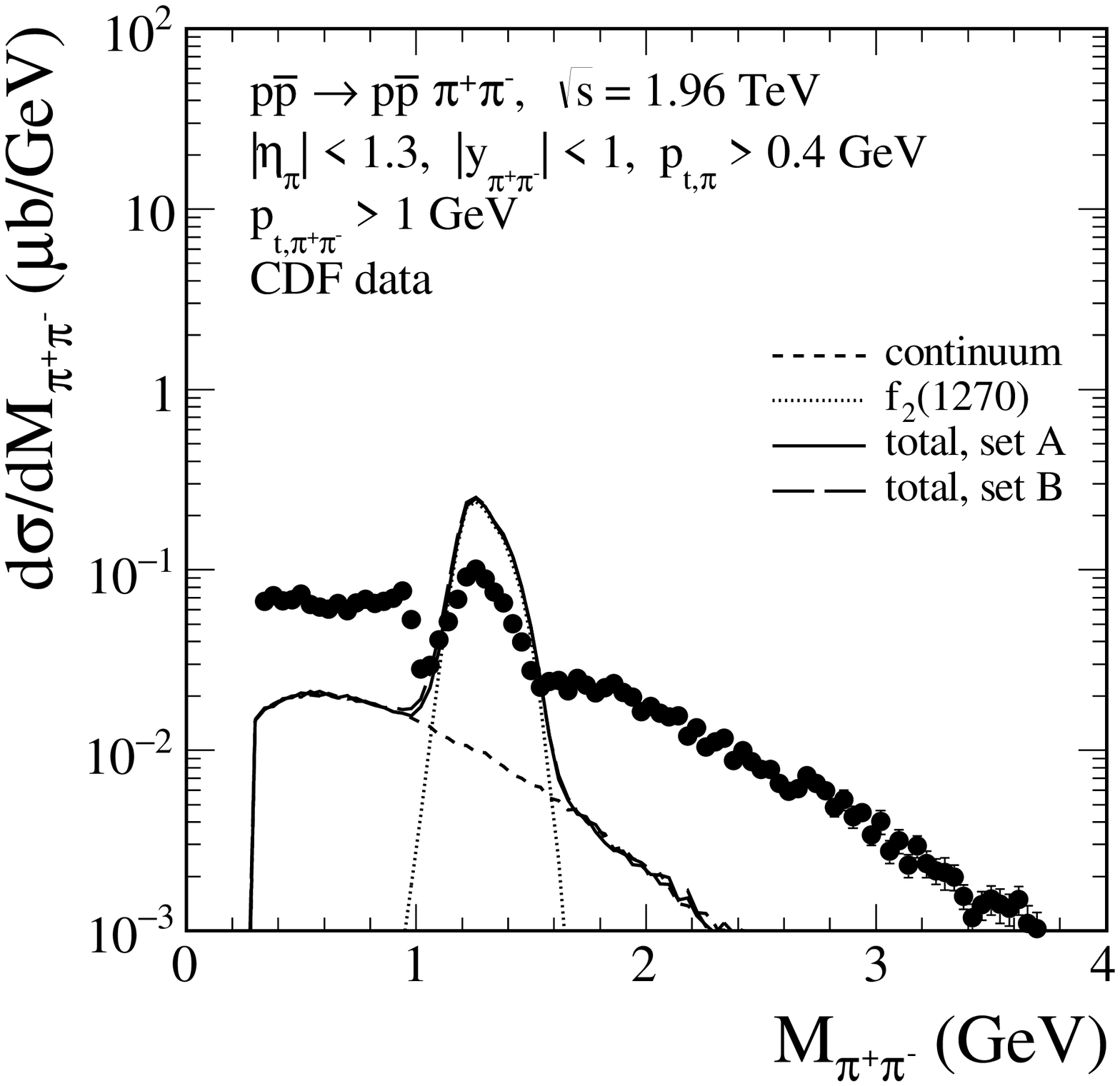}
(e)\includegraphics[width=0.4\textwidth]{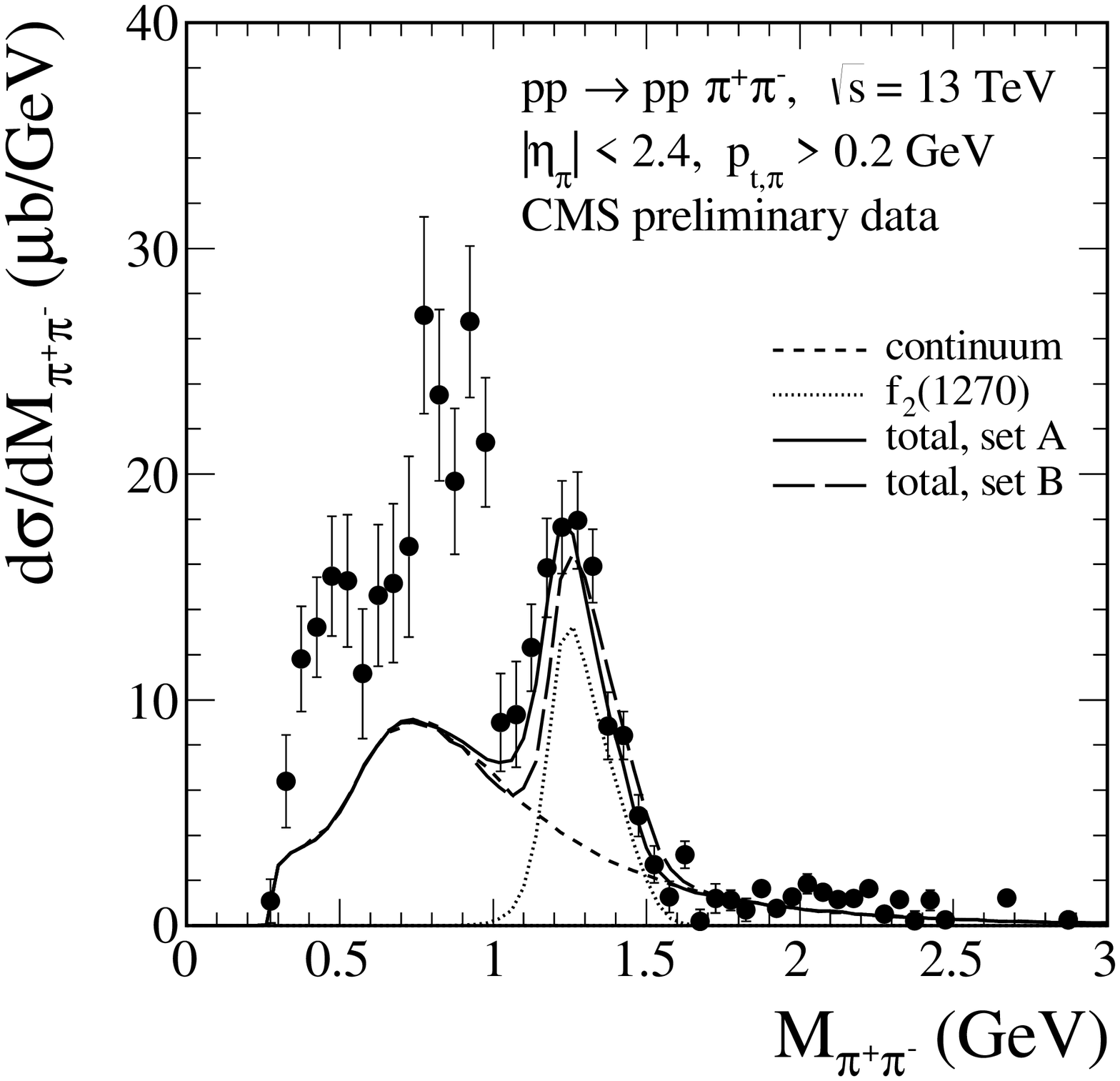}
(f)\includegraphics[width=0.4\textwidth]{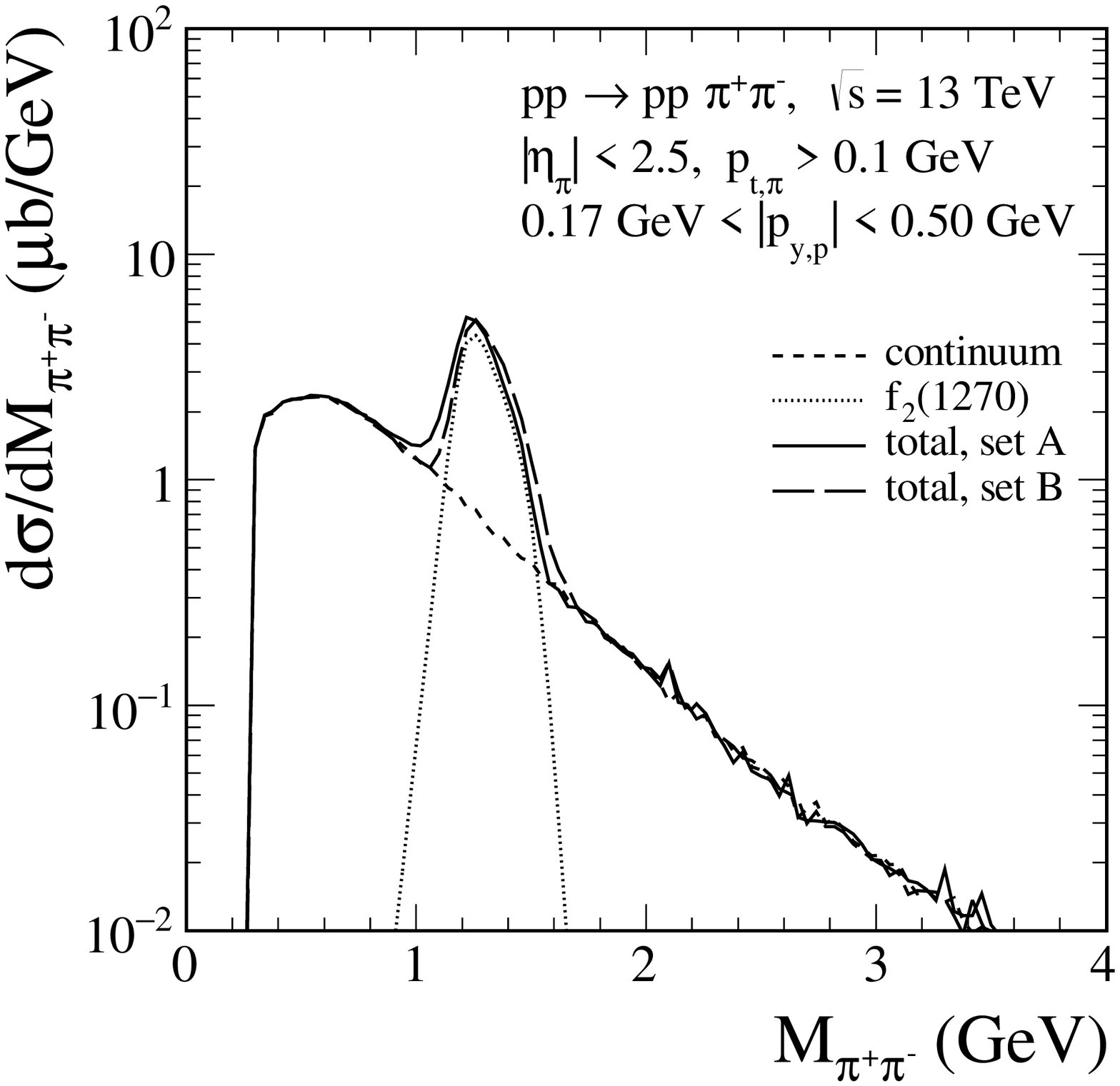}
\caption{\label{fig:exp}
\small
Two-pion invariant mass distributions with the relevant kinematical cuts
for (a), (b) STAR, (c), (d) CDF, (e) CMS, and (f) ATLAS-ALFA experiments.
The STAR preliminary data from \cite{Adamczyk:2014ofa,Sikora:2018cyk},
the CDF data from \cite{Aaltonen:2015uva},
and the CMS preliminary data from \cite{CMS:2019qjb} are shown.
The calculations for the STAR and ATLAS-ALFA experiments were done
with extra cuts on the leading protons.
The short-dashed lines represent the nonresonant continuum contribution,
the dotted lines represent the results for the $f_{2}(1270)$ contribution,
while the solid and long-dashed lines represent their coherent sum 
for the two parameter sets A and B, respectively.
Here we take, in set A $(g_{\Pom \Pom f_{2}}^{(2)}, g_{\Pom \Pom f_{2}}^{(5)}) = (-4.0, 16.0)$
and, in set B $(g_{\Pom \Pom f_{2}}^{(2)}, g_{\Pom \Pom f_{2}}^{(5)}) = (4.0, -16.0)$; 
see (\ref{Fpompommeson_tensor}), (\ref{A12}), (\ref{A15}).
The absorption effects are included here.}
\end{figure}

In Fig.~\ref{fig:exp} we show the dipion invariant mass distributions
for different experimental conditions specified in the legend.
One can see the recent high-energy data from the STAR, CDF, and CMS experiments,
as well as predictions of our model.
Panels (a) and (b) show the preliminary STAR data from \cite{Adamczyk:2014ofa} 
and \cite{Sikora:2018cyk}, respectively.
Panels (c) and (d) show the CDF experimental data from \cite{Aaltonen:2015uva}.
Panel (e) shows a very recent result obtained by the CMS Collaboration \cite{CMS:2019qjb}.
In the calculations we include both the nonresonant continuum and $f_{2}(1270)$ terms.
The panels (b) and (f) show the results including extra cuts on the outgoing protons.
For the STAR experiment we take the cuts (\ref{STAR_cuts})
and for the ATLAS-ALFA experiment we take 0.17~GeV~$< |p_{y,p}|<$~0.50~GeV.
The absorption effects (the $pp$-rescattering corrections only) 
were taken into account at the amplitude level.
The two-pion continuum was fixed by using 
the monopole form of the off-shell pion form factor
with the cut-off parameter $\Lambda_{\rm{off,M}} = 0.8$~GeV;
see (3.18) of \cite{Lebiedowicz:2016ioh}.
For the $f_{2}(1270)$ contribution, 
in order to get distinct maxima at $\phi_{\pi^{+},\,{\rm GJ}} = \pi/2$, $3/2\pi$,
we take a combination of two $\Pom \Pom f_{2}$ couplings,
$(g_{\Pom \Pom f_{2}}^{(2)}, g_{\Pom \Pom f_{2}}^{(5)}) = (-4.0, 16.0)$ (set A)
and $(4.0, -16.0)$ (set B), which correspond to the solid and long-dashed lines, respectively.
The complete results indicate an interference effect 
of the continuum and the $f_{2}(1270)$ term.
For comparison we show also the contributions of the individual terms separately.

We can see from Fig.~\ref{fig:exp}, that in the $f_{2}$ mass region 
we describe fairly well the preliminary STAR and CMS data 
but we overestimate the CDF data \cite{Aaltonen:2015uva}.
In the CMS and CDF measurements there are possible contributions 
of proton dissociation.
The continuum contribution
underestimates the data in the region $M_{\pi^{+}\pi^{-}} < 1$~GeV, however,
there are also possible other processes
e.g. from $f_{0}(500)$, $f_{0}(980)$, and $\rho^{0}$ production
not included in the present analysis; 
see e.g. \cite{Lebiedowicz:2014bea,Lebiedowicz:2016ioh}.
Also other effects, such as the rescattering corrections discussed in
\cite{Lebiedowicz:2011nb,Lebiedowicz:2015eka,Ryutin:2019khx},
can be very important there.

We emphasize, that in our calculation of the $\pi^{+}\pi^{-}$-continuum term
we include not only the leading pomeron exchanges ($\Pom \Pom \to \pi^{+}\pi^{-}$)
but also the $\Pom f_{2 \Reg}$, $f_{2 \Reg} \Pom$, 
and $f_{2 \Reg} f_{2 \Reg}$ exchanges.
There is interference between the corresponding amplitudes.
Their role is very important especially at low energies (COMPASS, WA102, ISR) 
but even for the STAR kinematics their contribution is not negligible.
Adding the $f_{2 \Reg}$ reggeon exchanges
increases the cross section by 56\% and 45\% for the kinematical conditions 
shown in Figs.~\ref{fig:exp}~(a) and (b), respectively.
A similar role of secondary reggeons can be expected for the production of resonances.
This means that our results for the $f_{2}(1270)$ resonance
(roughly matched to the STAR data) should be treated rather as an upper estimate.
This may be the reason why our result for $f_{2}(1270)$ is well above the CDF data.

We summarize this part by the general observation that it is very difficult to
describe all available data with the same set of parameters.
High-energy central exclusive data expected from CMS-TOTEM and ATLAS-ALFA
will allow a better understanding of the diffractive production mechanisms.

In Figs.~\ref{fig:1a} and \ref{fig:1b} 
we show the two-dimensional angular distributions 
for the STAR and ATLAS-ALFA kinematics, respectively.
In the left panels the results for the CS system
and in the right panels for the GJ system are presented.
In the top panels we show results for the continuum term,
in the center panels for the $f_{2}(1270)$ term,
and in the bottom panels for their coherent sum.
Here we take the set~A with the $\Pom \Pom f_{2}$ coupling parameters
$(g_{\Pom \Pom f_{2}}^{(2)}, g_{\Pom \Pom f_{2}}^{(5)}) = (-4.0, 16.0)$.
Figures~\ref{fig:2a} and \ref{fig:2b} 
show that the complete results indicate an interference effect 
of the continuum and the $f_{2}(1270)$ term calculated for the sets A and B,
see the solid and long-dashed lines, respectively.
The interference effect depends crucially on the choice of 
the $\Pom \Pom f_{2}(1270)$ coupling.
A combined analysis of the $M_{\pi^{+}\pi^{-}}$ and angular distributions 
in the $\pi^{+} \pi^{-}$ rest frames would,
therefore, help to pin down the underlying reaction mechanism.

\section{Conclusions}
In the present work we have considered the possibility
to extract the $\Pom \Pom f_{2}(1270)$ couplings from the analysis
of pion angular distributions in the $\pi^+ \pi^-$ rest system,
using the Collins-Soper (CS) and the Gottfried-Jackson (GJ) frames.
We have considered the tensor-pomeron model for which
there are 7 possible $\Pom \Pom f_{2}(1270)$ couplings; 
see Eqs.~(\ref{A11})--(\ref{A17}) and Appendix~A~of~\cite{Lebiedowicz:2016ioh}.
We have shown that the shape of such distributions strongly 
depends on the functional form of the $\Pom \Pom f_{2}(1270)$ coupling.
In particular, we have shown that the azimuthal angle distributions 
may have different numbers of oscillations.
The corresponding distributions can be approximately represented 
by the formula (\ref{dsig_dphi_approx}):
$A \pm B \cos(n \, \phi_{\pi^{+},\,{\rm CS}})$, where $n = 2,4$.
Two-dimensional distributions in the CS system
($\phi_{\pi^{+},\,{\rm CS}},\cos\theta_{\pi^{+},\,{\rm CS}}$),
($M_{\pi^{+}\pi^{-}},\phi_{\pi^{+},\,{\rm CS}}$),
($M_{\pi^{+}\pi^{-}},\cos\theta_{\pi^{+},\,{\rm CS}}$),
and respectively in the GJ system,
will give even more information
and could also be useful in understanding the role of experimental cuts.
Can such distributions be used to fix the $\Pom \Pom f_{2}(1270)$ coupling?
The answer will require dedicated experimental studies by 
the STAR, ALICE, ATLAS-ALFA, CMS-TOTEM, and LHCb Collaborations.
This requires comparisons of our model results 
with precise 'exclusive' experimental data
simultaneously in several differential observables.

We have shown how to select linear combinations of the different
$\Pom \Pom f_{2}(1270)$ coupling constants to get two maxima
in $\phi_{\pi^{+},\,{\rm GJ}}$ (or $\phi_{\pi^{+},\,{\rm CS}}$)
as observed at low energies by the COMPASS Collaboration;
see \cite{Austregesilo:2013yxa,Austregesilo:2014oxa}.

In the diffractive process considered the $f_{2}(1270)$ resonance 
cannot be completely isolated from the continuum background as the
corresponding amplitudes strongly interfere~\cite{Lebiedowicz:2016ioh}.
We have discussed how the interference of the resonance 
and the continuum background may change the angular distributions
$d\sigma/d\cos\theta_{\pi^{+},\,{\rm CS}}$ and $d\sigma/d\phi_{\pi^{+},\,{\rm CS}}$.
The absorption effects change the overall normalization of such distributions but leave 
the shape essentially unchanged.
This is in contrast to the $d\sigma/d\phi_{pp}$ distributions
where absorption effects considerably modify the corresponding 
shapes; see e.g. \cite{Lebiedowicz:2015eka,Lebiedowicz:2016zka}.

In the present analysis we have concentrated on the pronounced
$f_2(1270)$ resonance, clearly seen in the $\pi^+ \pi^-$ channel. 
We have discussed methods how to pin down the pomeron-pomeron-$f_{2}(1270)$ coupling.
The analysis presented may be extended
also to other resonances seen in different final state channels.
We strongly encourage experimental groups to start such analyses.
We think that this will bring in a new tool
for analysing exclusive diffractive processes
and will provide new inspirations in searching for more exotic states
such as glueballs, for instance. The exclusive diffractive processes 
were always claimed to be a good area to learn about the physics of glueballs.
The extension of our methods to the production of glueballs, 
to be identified in suitable decay channels, 
should shed light on the pomeron-pomeron-glueball couplings.
These represent very interesting quantities: 
the coupling of three (mainly) gluonic objects.

\appendix
\section{Remarks on the transformation from the c.m. to the $\pi \pi$ rest system}
\label{sec:appendixA}

In this section we discuss the relation between quantities in the c.m. system
and the $\pi \pi$ rest system.
Momenta in the c.m. system will be denoted by $p_{\rm c.m.}$, $k_{\rm c.m.}$, etc.,
momenta in the $\pi \pi$ rest system by $p_{\rm R}$, $k_{\rm R}$, etc.
We assume that the transformation from the c.m. to the $\pi \pi$ rest system
is made by a boost, that is, by a rotation free Lorentz transformation
\begin{eqnarray}
\Lambda(-\bp_{\textbf{34},\,\rm c.m.}) &=&
\Bigl( \Lambda^{\mu}_{\;\;\;\nu}(-\bp_{\textbf{34},\,\rm c.m.}) \Bigr) \nonumber\\
&=&
\left( \begin{array}{cl}
\dfrac{ p^{0}_{34,\,\rm c.m.}}{M_{\pi\pi}} & \;\; 
\dfrac{-p^{j}_{34,\,\rm c.m.}}{M_{\pi\pi}}
\\ 
\dfrac{-p^{i}_{34,\,\rm c.m.}}{M_{\pi\pi}} & \;\; 
\delta^{ij}+\Bigl( \dfrac{p^{0}_{34,\,\rm c.m.}}{M_{\pi\pi}}-1 \Bigr)
\dfrac{p^{i}_{34,\,\rm c.m.}
       p^{j}_{34,\,\rm c.m.}}{(\bp_{\textbf{34},\,\rm c.m.})^{2}} \\
\end{array} \right) \,,
\label{AA1}
\end{eqnarray} 
where $i,j \in \left\lbrace 1,2,3 \right\rbrace$.

We have then for any four vector $l = (l^{\mu})$
\begin{eqnarray}
\Lambda(-\bp_{\textbf{34},\,\rm c.m.}) \,l_{\rm c.m.} = l_{\rm R}\,.
\label{AA2}
\end{eqnarray} 

The reverse transformation is 
$\Lambda^{-1}(-\bp_{\textbf{34},\,\rm c.m.}) = \Lambda(\bp_{\textbf{34},\,\rm c.m.})$
\begin{eqnarray}
\Lambda^{-1}(-\bp_{\textbf{34},\,\rm c.m.}) \,l_{\rm R} =
\Lambda(\bp_{\textbf{34},\,\rm c.m.}) \,l_{\rm R} = l_{\rm c.m.}\,.
\label{AA3}
\end{eqnarray} 

In particular we get
\begin{eqnarray}
\Lambda(-\bp_{\textbf{34},\,\rm c.m.}) \,p_{34,\,\rm c.m.} =
p_{34,\,\rm R} =
\left( \begin{array}{c}
M_{\pi \pi} \\
0 \\
\end{array} \right)\,,
\label{AA4}
\end{eqnarray} 
\begin{eqnarray}
\Lambda(-\bp_{\textbf{34},\,\rm c.m.}) \,\frac{M_{\pi \pi}}{\sqrt{s}}(p_{a}+p_{b})_{\rm c.m.} =
                                         \frac{M_{\pi \pi}}{\sqrt{s}}(p_{a}+p_{b})_{\rm R} =
\left( \begin{array}{c}
 p^{0}_{34,\,\rm c.m.} \\
-\bp_{\textbf{34},\,\rm c.m.} \\
\end{array} \right)\,,
\label{AA5}
\end{eqnarray} 
\begin{eqnarray}
q_{1, \, \rm R} &=&
\Lambda(-\bp_{\textbf{34},\,\rm c.m.}) \,q_{1, \, \rm c.m.} \nonumber \\
&=&
\left( \begin{array}{l}
\dfrac{(p_{34} \cdot q_{1})}{M_{\pi \pi}} 
\\
\bq_{\textbf{1},\, \rm c.m.}+\dfrac{\bp_{\textbf{34},\,\rm c.m.}}{M_{\pi \pi}}
\left(-q^{0}_{1, \, \rm c.m.} + (p^{0}_{34,\,\rm c.m.} - M_{\pi \pi})
\dfrac{(\bp_{\textbf{34},\,\rm c.m.} \cdot \bq_{\textbf{1},\,\rm c.m.})}{(\bp_{\textbf{34},\,\rm c.m.})^{2}}
\right) \\
\end{array} \right)\,.
\label{AA6}
\end{eqnarray} 

With these relations we can now express the unit vectors
of the Gottfried-Jackson (GJ) system of (\ref{GJ}) entirely by vectors defined in
the $\pi \pi$ rest system. We have $p_{34} = q_{1} + q_{2}$
and therefore from (\ref{AA5}) and (\ref{AA6})

\begin{eqnarray}
\bq_{\textbf{1},\,\rm c.m.} \times \bq_{\textbf{2},\,\rm c.m.} =
\bq_{\textbf{1},\,\rm c.m.} \times \bp_{\textbf{34},\,\rm c.m.} =
-\frac{M_{\pi\pi}}{\sqrt{s}} \bq_{\textbf{1},\,\rm R} \times  (\bpa + \bpb)_{\rm R}
\,;
\label{AA7}
\end{eqnarray} 
\begin{equation}
\begin{split}
& \be_{\textbf{3},\,\rm GJ} = \frac{\bq_{\textbf{1},\,\rm R}}{|\bq_{\textbf{1},\,\rm R}|}\,,\\
& \be_{\textbf{2},\,\rm GJ} = - \frac{\bq_{\textbf{1},\,\rm R} \times (\bpa + \bpb)_{\rm R}}
                                    {|\bq_{\textbf{1},\,\rm R} \times (\bpa + \bpb)_{\rm R}|}\,,\\
& \be_{\textbf{1},\,\rm GJ} = \be_{\textbf{2},\,\rm GJ} \times \be_{\textbf{3},\,\rm GJ}\,.
\end{split}
\label{AA8}
\end{equation}

Note that for setting up this GJ system only 
the momentum of one of the outgoing
protons in the reaction (\ref{2to4_reaction}) 
has to be measured, plus, of course the momenta
of $\pi^{+}$ and $\pi^{-}$ giving $p_{34}$.

\acknowledgments
We are indebted to Leszek Adamczyk and Rafa{\l} Sikora for useful discussions.
This work was partially supported by
the Polish National Science Centre Grant
No. 2018/31/B /ST2/03537 and by the Center for Innovation and Transfer of Natural Sciences 
and Engineering Knowledge in Rzesz\'ow.

\bibliography{refs}

\end{document}